\documentclass[aps,reprint,showpacs,onecolumn,superscriptaddress,pra,tightenlines]{revtex4-2}

\usepackage[dvips]{graphicx}
\usepackage{amsmath,amssymb,amsthm,mathrsfs,amsfonts,dsfont}
\usepackage{subfigure, epsfig}
\usepackage{braket}
\usepackage{bm}
\usepackage{enumerate}
\usepackage{color}
\usepackage{qcircuit}
\usepackage{graphicx}
\usepackage{algorithm}
\usepackage{algpseudocode}
\usepackage{ulem}
\usepackage{lineno}

\newcommand{\floor}[1]{\left\lfloor #1\right \rfloor }

\usepackage[utf8]{inputenc}
\usepackage{hyperref}
\hypersetup{colorlinks=true, linkcolor=blue, citecolor=blue, urlcolor=black }

\newcommand{\tr}{\mathrm{Tr}}

\newcommand{\blue}[1]{\textcolor{black}{#1}}
\newcommand{\sun}[1]{\textcolor{black}{#1}}

\newcommand{\mol}[1]{\mathrm{#1}}

\newcommand{\pbra}[1]{\left( #1 \right)}
\newcommand{\cbra}[1]{\left\{ #1 \right\}}

\newcommand{\var}{\operatorname{Var}}

\newcommand{\Kcal}{\mathcal{K}}
\newcommand{\Scal}{\mathcal{S}}

\newcommand{\Omat}{\bm{\mathrm O}}
\newcommand{\Pmat}{\bm{\mathrm P}}
\newcommand{\Qmat}{\bm{\mathrm Q}}
\newcommand{\Olmat}{\bm{\mathrm O}_l}
\newcommand{\Olpmat}{\bm{\mathrm O}_{l'}}
\newcommand{\omat}{\bm{\mathrm o}}

\renewcommand{\vec}[1]{\boldsymbol{#1}}

\DeclareUnicodeCharacter{FFFC}{-}
\usepackage{varioref}
\labelformat{equation}{(#1)}

\labelformat{section}{#1}

\labelformat{figure}{#1}

\labelformat{proposition}{#1}

\labelformat{lemma}{#1}

\labelformat{theorem}{#1}

\labelformat{observation}{#1}

\labelformat{definition}{#1}

\newcommand{\adda}{Hefei National Research Center for Physical Sciences at the Microscale and School of Physical Sciences, University of Science and Technology of China, Hefei 230026, China}
\newcommand{\addb}{Shanghai Research Center for Quantum Science and CAS Center for Excellence in Quantum Information and Quantum Physics, University of Science and Technology of China, Shanghai 201315, China}
\newcommand{\addc}{Hefei National Laboratory, University of Science and Technology of China, Hefei 230088, China}
\newcommand{\addd}{Center on Frontiers of Computing Studies, Peking University, Beijing 100871, China}
\newcommand{\adde}{Clarendon Laboratory, University of Oxford, Parks Road, Oxford OX1 3PU, United Kingdom}
\newcommand{\addf}{School of Computer Science, Peking University, Beijing 100871, China}
\newcommand{\addg}{QOLS, Blackett Laboratory, Imperial College London, London SW7 2AZ, United Kingdom}

\begin{document}

\title{Experimental quantum computational chemistry with optimised unitary coupled cluster ansatz}

% \title{\yx{Experimental realisation of quantum computational chemistry with superconducting qubits}}

% ------------  AUTHORS AND AFFILIATIONS ----------
\author{Shaojun Guo}
\thanks{These authors contributed equally to this work.}
\affiliation{\adda}
\affiliation{\addb}
\affiliation{\addc}
\author{Jinzhao Sun}
\thanks{These authors contributed equally to this work.}
% \thanks{Present address: Blackett Laboratory, Imperial College London, London SW7 2AZ, United Kingdom}
\affiliation{\addd}
\affiliation{\addg}
\affiliation{\adde}

\author{Haoran Qian} 
\thanks{These authors contributed equally to this work.}
\affiliation{\adda}
\affiliation{\addb}
\affiliation{\addc}
\author{Ming Gong} 
\thanks{These authors contributed equally to this work.}
\affiliation{\adda}
\affiliation{\addb}
\affiliation{\addc}
\author{Yukun Zhang}
\affiliation{\addd}
\affiliation{\addf}
\author{Fusheng Chen} 
\affiliation{\adda}
\affiliation{\addb}
\affiliation{\addc}
\author{Yangsen Ye} 
\affiliation{\adda}
\affiliation{\addb}
\affiliation{\addc}
\author{Yulin Wu}
\affiliation{\adda}
\affiliation{\addb}
\affiliation{\addc}
\author{Sirui Cao}
\affiliation{\adda}
\affiliation{\addb}
\affiliation{\addc}
\author{Kun Liu}
\affiliation{\addd}
% \affiliation{\addf}
\author{Chen Zha} 
\affiliation{\adda}
\affiliation{\addb}
\affiliation{\addc}
\author{Chong Ying} 
\affiliation{\adda}
\affiliation{\addb}
\affiliation{\addc}
\author{Qingling Zhu} 
\affiliation{\adda}
\affiliation{\addb}
\affiliation{\addc}
\author{He-Liang Huang} 
\affiliation{\adda}
\affiliation{\addb}
\affiliation{\addc}
\author{Youwei Zhao} 
\affiliation{\adda}
\affiliation{\addb}
\affiliation{\addc}
\author{Shaowei Li} 
\affiliation{\adda}
\affiliation{\addb}
\affiliation{\addc}
\author{Shiyu Wang} 
\affiliation{\adda}
\affiliation{\addb}
\affiliation{\addc}
\author{Jiale Yu} 
\affiliation{\adda}
\affiliation{\addb}
\affiliation{\addc}
\author{Daojin Fan} 
\affiliation{\adda}
\affiliation{\addb}
\affiliation{\addc}
\author{Dachao Wu} 
\affiliation{\adda}
\affiliation{\addb}
\affiliation{\addc}
\author{Hong Su} 
\affiliation{\adda}
\affiliation{\addb}
\affiliation{\addc}
\author{Hui Deng} 
\affiliation{\adda}
\affiliation{\addb}
\affiliation{\addc}
\author{Hao Rong} 
\affiliation{\adda}
\affiliation{\addb}
\affiliation{\addc}
\author{Yuan Li}
\affiliation{\adda}
\affiliation{\addb}
\affiliation{\addc}
\author{Kaili Zhang} 
\affiliation{\adda}
\affiliation{\addb}
\affiliation{\addc}
\author{Tung-Hsun Chung} 
\affiliation{\adda}
\affiliation{\addb}
\affiliation{\addc}
\author{Futian Liang} 
\affiliation{\adda}
\affiliation{\addb}
\affiliation{\addc}
\author{Jin Lin} 
\affiliation{\adda}
\affiliation{\addb}
\affiliation{\addc}
\author{Yu Xu} 
\affiliation{\adda}
\affiliation{\addb}
\affiliation{\addc}
\author{Lihua Sun} 
\affiliation{\adda}
\affiliation{\addb}
\affiliation{\addc}
\author{Cheng Guo} 
\affiliation{\adda}
\affiliation{\addb}
\affiliation{\addc}
\author{Na Li} 
\affiliation{\adda}
\affiliation{\addb}
\affiliation{\addc}
\author{Yong-Heng Huo} 
\author{Cheng-Zhi Peng}
\affiliation{\adda}
\affiliation{\addb}
\affiliation{\addc}
\author{Chao-Yang Lu} 
\affiliation{\adda}
\affiliation{\addb}
\affiliation{\addc}
\author{Xiao Yuan}
\affiliation{\addd}
\affiliation{\addf}
\author{Xiaobo Zhu} 
\affiliation{\adda}
\affiliation{\addb}
\affiliation{\addc}
\author{Jian-Wei Pan}
\affiliation{\adda}
\affiliation{\addb}
\affiliation{\addc}

\date{\today}

% --------------------  ABSTRACT  --------------------

\begin{abstract}

\textbf{
Quantum computational chemistry has emerged as an important application of quantum computing~\cite{aspuru2005simulated,lanyon2010towards,arguello2019analogue,reiher2017elucidating,Cao2019,mcardle2018quantum,bauer2020quantum}.
Hybrid quantum-classical computing methods, such as variational quantum eigensolvers (VQE)~\cite{Alan14,cerezo2021variational,Bharti2021RMP}, have been designed as promising solutions to quantum chemistry problems, yet
% show promise as a near-term application,
challenges due to theoretical complexity~\cite{lee2022there,stilck2021limitations,lee2021even,dalton2022variational} and experimental imperfections~\cite{wang2021noise} hinder progress in achieving reliable and accurate results.
Experimental works for solving electronic structures  are consequently still restricted to nonscalable (hardware efficient)~\cite{kandala2017hardware,kandala2019error} or classically simulable (Hartree-Fock) ansatz~\cite{arute2020hartree}, or limited to a few qubits with large errors~\cite{o2016scalable,PhysRevA.95.020501,PhysRevX.8.031022,colless2018computation,nam2020ground}. 
% \sun{Maybe we should leave this sentence to emphasise the experimental difficulty?}
% Significant challenges both in theory and experiments must be overcome to enable the solution to quantum computational chemistry. 
The experimental realisation of scalable and high-precision quantum chemistry simulation remains elusive.
% Overcoming these challenges, both in theory and experiments, is essential for enabling practical quantum computational chemistry.
Here, we address the critical challenges {associated with} solving molecular electronic structures using noisy quantum processors.
Our protocol presents significant improvements in the circuit depth and running time, key metrics for chemistry simulation. Through systematic hardware enhancements and the integration of error mitigation techniques~\cite{PhysRevA.103.042605,Czarnik_2021,lowe2021unified,PhysRevA.103.042605,PhysRevLett.58.83,claudino2021improving}, we 
push forward the limit of experimental quantum computational chemistry
and 
successfully scale up the implementation of VQE with an optimised unitary coupled-cluster ansatz to 12 qubits. We produce high-precision results of the ground-state energy for molecules with error suppression by around two orders of magnitude. 
We achieve chemical accuracy for H$_2$ at all bond distances and LiH at small bond distances in the experiment, even beyond the two recent concurrent works~\cite{obrien_purification-based_2023,2022arXiv221202482Z}.
Our work demonstrates a feasible path towards a scalable solution to electronic structure calculation, validating the key technological features and identifying future challenges for this goal.
}

\end{abstract}

\maketitle

% \linenumbers

%\section{Introduction}

% , and is often considered to be the first instance towards practical quantum computing
% Hybrid quantum-classical computing solutions, although seemingly promising, entail prohibitive resources in terms of practical computation. 
% As a high-potential application of quantum computing, extensive research in both theory and experiments has been conducted to study quantum chemistry simulation. 
Quantum computational chemistry, a high-potential application of quantum computing, has prompted extensive research in both theory and experiments~\cite{aspuru2005simulated,lanyon2010towards,arguello2019analogue,reiher2017elucidating,Cao2019,mcardle2018quantum,bauer2020quantum}.
% move forwards towards this goal \sun{This sentence may look a bit weak}. 
Among the various approaches explored, variational quantum algorithms~\cite{Alan14,cerezo2021variational,Bharti2021RMP} emerge as a promising near-term solution with their potential for executing the task using shallow circuits.
% that is believed to have near-term applications. 
% due to the inherent complex nature of molecules, quantum chemistry problems pose unique challenges for practical quantum computation, especially with current limited noisy quantum processors. 
% However, the inherent complexity of molecular systems \sout{also} presents challenges for practical quantum computation~\cite{lee2022there}, particularly considering the limitations of current noisy quantum processors \sun{in terms of noise and measurement costs}~\cite{stilck2021limitations}.
% Despite the theoretical potential, \sun{considering the finite resources and gate fidelities}, resources required for solutions to practical quantum chemistry problems generally remain prohibitive~\cite{dalton2022variational,lee2021even}. 
Despite the theoretical potential, due to the inherent complexity of molecular systems, quantum resources required for practical computation in terms of the gate count, measurement numbers and total running time are prohibitively large~\cite{dalton2022variational,bittel2021training,Gonthier_Measurements_2022,yen_deterministic_2023}.
% \sun{JS:  total running time or measurement cost, which is better? If total running time, then we may need to explain what it is.}
Taking into account the noise levels (above error-correction thresholds) and limited allowed gate counts (a few hundred), achieving quantum chemistry simulation with noisy quantum processors remains elusive~\cite{stilck2021limitations,wang2021noise}.
Experiments in quantum chemistry are thus primarily restricted to {proof-of-principle} demonstrations with either small-scale~\cite{o2016scalable,PhysRevX.8.031022,colless2018computation,nam2020ground,PhysRevA.95.020501} or limited circuit ansatz~\cite{kandala2017hardware,arute2020hartree,kandala2019error}. 
% Although quantum chemistry is considered to be the first instance towards practical quantum computing, 
A notable gap exists between theoretical quantum chemistry algorithms and experimental realisations. 

% Bridging this gap requires further research and advancements to overcome the challenges posed by complex molecular systems and limited hardware capabilities.

% It remains elusive if scalable and high-precision experimental quantum chemistry simulation can be achieved with noisy quantum processors.

% Significant challenges both in theory and experiments must be overcome to enable the solution to quantum computational chemistry. Challenges are closely intertwined. These challenges create a pessimistic outlook for achieving realistic implementations

% Quantum chemistry simulation requires high accuracy. However, the accuracy of chemistry simulation greatly suffered from decoherence and readout errors.
% We realised QEM in mitigating decoherence and readout errors in chemistry by realising QEM., and mitigated the crucial errors in chemistry simulation.

% The pursuit of quantum chemistry simulation involves addressing 

% The main challenges of \sun{quantum chemistry simulation} are the theoretical complexity and implementation errors. 
{
The pursuit of quantum computational chemistry involves addressing critical challenges related to excessive demands due to the complexity of algorithms and experimental errors.}
Achieving high-accuracy simulation entails the use of a more complex and expressive ansatz, which, however, introduces more errors in practice and leads to less accurate results. Many existing quantum chemistry experiments  are consequently restricted to either nonscalable (hardware-efficient)~\cite{kandala2017hardware,kandala2019error} or classically simulable (Hartree-Fock) ansatz~\cite{arute2020hartree}, which, in theory, cannot be useful for practical problems. 
{The experimental realisation of quantum chemistry simulations is further complicated by the presence of noise~\cite{quek2022exponentially,wang2021noise}, resulting in a sense of pessimism.}
% The experimental realisation of quantum chemistry simulations with expressive ansatz is indeed challenging with noisy quantum hardware. 
% introduces more challenges and a sense of pessimism due to the presence of noise. 
Even small-scale demonstrations~\cite{o2016scalable,PhysRevA.95.020501,PhysRevX.8.031022} face significant difficulties in achieving the desired level of precision known as chemical accuracy (1.6 milli-Hartree), which is vital for reliable calculations and predictions in quantum chemistry.
Significant efforts are required in protocol optimisation, hardware improvements, and effective error mitigation in order to enable scalable and high-precision experimental quantum chemistry simulation with noisy quantum processors.

% Experimental realisation makes it much pessimistic due to noise from decoherence and readout. The hardware limitations make it hard even for small-scale demonstrations to achieve the so-called chemical accuracy (1.6 milli-Hartree), an essential requirement for reliable calculations or predictions in quantum computational chemistry.
% Quantum chemistry simulation poses significant challenges due to the need for high precision, specifically achieving what is known as "chemical accuracy" (1.6 milli-Hartree). This level of precision is essential for reliable calculations and predictions in quantum computational chemistry. 

% QUantum Chemistry simulation is particularly difficult since it requires high precision, which ofter requires to achieving the so-called chemical accuracy (1.6 milli-Hartree), an essential requirement for reliable calculations or predictions in quantum computational chemistry. Even for small-scale demonstrations, errors from decoherence and readout will make the result far away from realistic requirement.

% The hardware limitations restrict high-precision results even for small-scale demonstrations, casting doubt on the feasibility of practical implementations.  

In this work, we address those challenges associated with quantum chemistry simulation, and realise variational quantum
eigensolver (VQE) for $\mol{H_2}$, $\mol{LiH}$, and $\mol{F_2}$ on a noisy intermediate-scale superconducting quantum processor. {We consider a multi-reference initial state and develop an optimised UCC ansatz~\cite{romero2018strategies,anand2022quantum,ucc-2} by selecting symmetry-conserving and dominant terms.}
% These theoretical advancements 
We demonstrate significant improvements in critical metrics for quantum chemistry simulation, including the circuit depth, measurement cost and total running time. 
% Our results largely alleviate the theoretical complexity of VQE and enable the implementation of an optimised unitary coupled cluster  ansatzz~\cite{romero2018strategies,anand2022quantum,ucc-2} with significantly reduced circuit depth and running time.
% We demonstrate significant improvements in the circuit depth and running time, key metrics for chemistry simulation. 
% Comparing the three molecules with increasing system size, 
% With our method, we find that the resource costs (average running time and measurement shots) increase nearly linearly. 
% instead of hardware-efficient ansatz, with the current generation of quantum devices. 
On the experimental side, we developed a systematic approach aimed at optimising hardware tailored for quantum chemistry simulations. We achieved high-fidelity parallel gates leveraging the advantage of the flip-chip structure and tunable couplers, which suppress the crosstalk significantly. {We introduced a calibrated implementation of basic operations in UCC-type ansatz to improve the circuit accuracy.} The correlated readout error in measurements is significantly suppressed with the optimisation of readout parameters, which is verified by random state measurement. Furthermore, we designed and implemented four quantum error mitigation (QEM) protocols~\cite{PhysRevA.103.042605,Czarnik_2021,lowe2021unified,PhysRevA.103.042605,PhysRevLett.58.83,claudino2021improving} to mitigate decoherence, readout, and algorithmic errors accordingly, which is crucial to obtain reliable and accurate results for chemistry applications. The efficacy of QEM in mitigating different types of errors is  validated, which 
enables two orders of magnitude improvement of the calculation accuracy for the three molecules studied in this work.
With these improvements, we can finally push forward the implementation of VQE with chemically inspired  UCC ansatz and scale up the error-mitigated simulation to 12 qubits. We report that {our solution} allows the saturation of chemical accuracy for $\mol{H_2}$ at all bond distances and $\mol{LiH}$ at small bond distances, {with relative errors compared to initial energies less than 1.00$\%$ for all these molecules.}
Our work explores the potential way for scalable and reliable quantum simulation, which is a pressing and unavoidable question that arises when utilising noisy processors.

\begin{figure*}[t]
\centering
\includegraphics[width=0.7\textwidth]{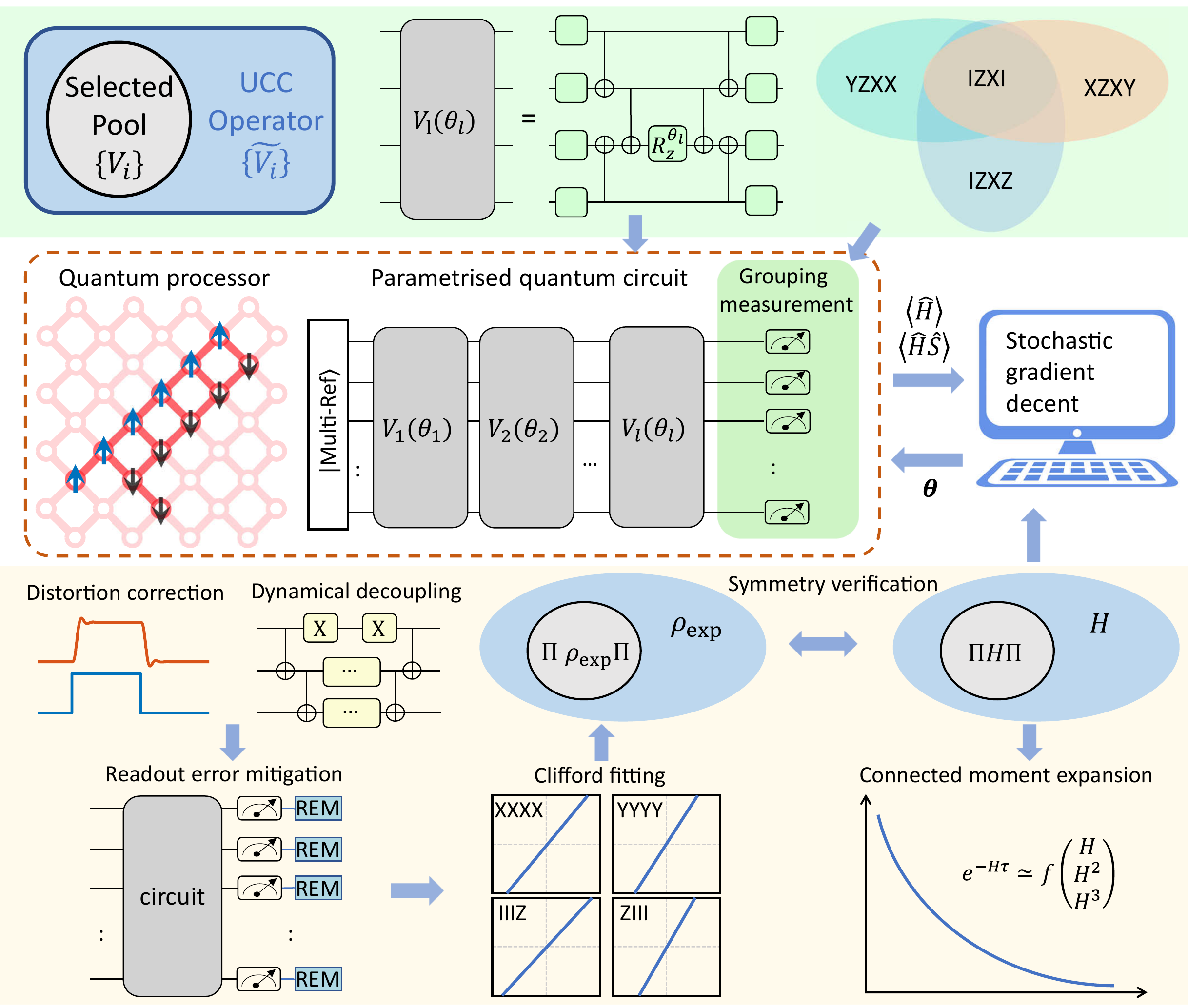}
\caption{\textbf{A diagrammatic scheme for the variational quantum eigensolver on a superconducting quantum processor: device, quantum circuit, measurement, and error mitigation.}
Middle: The experimental circuit for 12-qubit  VQE of $\mol{F_2}$ over a $2 \times 6$ qubit grid on the ``Zuchongzhi 2.0'' quantum processor. Middle left: Topology of qubits on the quantum processor. The twelve selected qubits  used in the experiments {with the encoding of spin up and spin down} are coloured.
Middle right: Schematic diagram of the VQE process.
Top left: Selection of the operators from the UCC operator pool that contribute dominantly to the energy decrease.
Top middle: The elementary multi-qubit Pauli rotation gate in UCC, which is composed of CNOT gates, single-qubit Pauli rotation gates, and single-qubit Clifford gates. 
The single-qubit Pauli rotation is decomposed into a sequence of single-qubit gates $\mathrm{R_Y}$-$\mathrm{R_Z}$-$\mathrm{R_Y}$ in our experiment.
Top right: An example of the overlapped grouping method for measuring observables, which exploits the qubit-wise compatibility of observables.
Bottom: Hardware optimisation and error mitigation techniques integrated into the experiment.
Distortion correction and dynamical decoupling are first applied to improve the fidelity of the elementary operators and suppress dephasing during idle times, respectively. 
Then  readout error mitigation, Clifford circuit fitting, symmetry verification by parity $\hat{S}$, and connected moment expansion 
are  applied sequentially to calculate the error-mitigated  expectation values of observables.
}\label{fig1}
\end{figure*}

% The development of quantum algorithms is going faster and indeed further beyond that of experimental hardware.
% An emergent question is what currently available hardware can produce for quantum chemistry  simulation, intertwining both theoretical and experimental advancements. 

%\section{Building blocks of the algorithm}

We briefly summarise the basic building blocks of our algorithm, which are outlined in Fig.~\ref{fig1}, and refer to Supplementary Information for  detailed discussions.
% \sout{
% We consider second-quantisation under the Born-Oppenheimer approximation and apply Jordan-Wigner transformation to obtain the Hamiltonian $\hat H$.}
{We consider the molecular Hamiltonian under the Born-Oppenheimer approximation in second-quantisation and apply the Jordan-Wigner transformation to obtain its qubit form $\hat H$.}
% \sun{SUN: To xiao: $\hat H$, to be more accurate, is expressed in an operator form, not a qubit form in conventional QC paper. However, I prefer to adopt the operator form here to make it more natural to build a connection with UCC ansatz, and the discussion in SM, although it is in a qubit form now - this could be reasonable since Pauli operator is an operator as well
% }
We employ variational quantum eigensolvers to find the ground state of $\hat H$.
The main idea is to prepare a parametrised quantum state $\ket{\Psi(\Vec{\theta})}$ by a quantum processor and update the parameters classically.
A chemically inspired choice for~$\ket{\Psi(\Vec{\theta})}$  is 
the UCC ansatz $
    \ket{\Psi(\Vec{\theta})} = \exp{(\hat{T}(\vec \theta) - \hat{T}^{\dagger}(\vec \theta))}|\Psi_{0}\rangle
$, which effectively considers excitations and de-excitations
above a reference state~$|\Psi_{0}\rangle$~\cite{nam2020ground,PhysRevA.95.020501,PhysRevX.8.031022,colless2018computation}.
Here $\hat{T}(\vec \theta)$ is the truncated cluster operator concerning first- and second-order excitations from occupied orbitals to virtual orbitals.
However, direct implementation of UCC on a quantum computer requires CNOT gates scaling as $\mathcal{O} (N(N-\eta)^2 \eta^2)$ with system size $N$ and electron number $\eta$, which goes beyond the limit of current technology.

To reduce the gate count, we design the circuit by selecting the operators from the UCC operator pool, which satisfies the symmetry constraint imposed by the selection rule~\cite{stanton1991direct,cao2021larger}.
Moreover, since the operators in $\hat T$ have different effects on the ground state~\cite{fan2021circuit}, we select the qubit operators $\hat V$ that  contribute dominantly to the ground state energy decrease (see the top left column of Fig.~\ref{fig1} and Supplementary Section II). 
For instance, the total gate count is reduced by two orders of magnitude for $N=12$. 
% This circuit reduction approach is shown to be computationally efficient.
Furthermore, we compile the circuit according to the 2D topology structure of the superconducting processor to reduce the circuit depth. 
The gate count and circuit depth reduction strategies are scalable and essential for efficient implementations of UCC ansatz.
% ----- New -----
With the ansatz state $\ket{\Psi(\Vec{\theta})}$, we now aim to 
% the ground state is approximated by 
solve a minimisation problem, $\min_{\vec\theta}\braket{\hat{H}}_{\vec\theta}$, with $\braket{\hat{H}}_{\vec\theta}=\braket{\Psi(\Vec{\theta})|\hat{H}|\Psi(\Vec{\theta})}$.
We consider a classical optimiser  of stochastic  gradient descent with analytical gradients obtained via a linear combination of $\braket{\hat{H}}_{\vec\theta}$ with different choices of parameters $\vec\theta$~\cite{PhysRevA.99.032331}. 
% For the $k$-th iteration with parameters $\vec \theta^{k}$, 
% we obtain the gradient element $\vec g^k_j(\vec \theta^{k})$ by the parameter shift rule, which requires the measurement of 
% $\braket{\hat{H}}_{\vec\theta^{\pm}} 
% $
% with
% $
%     \vec\theta^{\pm}= \vec \theta^{k} \pm \pi \Delta^k_j \vec e_j/2
% $ and \blue{define DELTA} unit vector $\vec e_i$~\cite{PhysRevA.99.032331}.
% We stochastically choose the parameters to be optimised to reduce the sample cost for the estimation of gradients. 
% The gradient estimation is unbiased, and \sun{has a smaller variance} compared with the finite difference method.
% The parameters are updated by
% $
%     \vec \theta^{k+1}= \vec \theta^{k} + \alpha^{{k}} \vec g^k(\vec \theta^{k})
% $ with the learning rate $\alpha^k$ until convergence.
Each $\braket{\hat{H}}_{\vec\theta}$ generally consists of $\mathcal{O} (N^4)$ terms, whose measurement cost could be prohibitively large in practice~\cite{Gonthier_Measurements_2022, Wecker_Progress_2015}.
However, as many of the observables have small coefficients and are qubit-wise compatible (hence could be measured simultaneously),  more efficient measurement schemes could be exploited to alleviate the measurement cost~\cite{wu2021overlapped,huang2021efficient}. We develop an optimised overlapped grouping measurement scheme, illustrated in the top right column of Fig.~\ref{fig1}, which reduces the required measurement count and hence the running time significantly (for instance, by two orders of magnitude for $\mol{F_2}$ compared to importance sampling).

\begin{figure}[t!]
\centering
\includegraphics[width=0.8\textwidth]{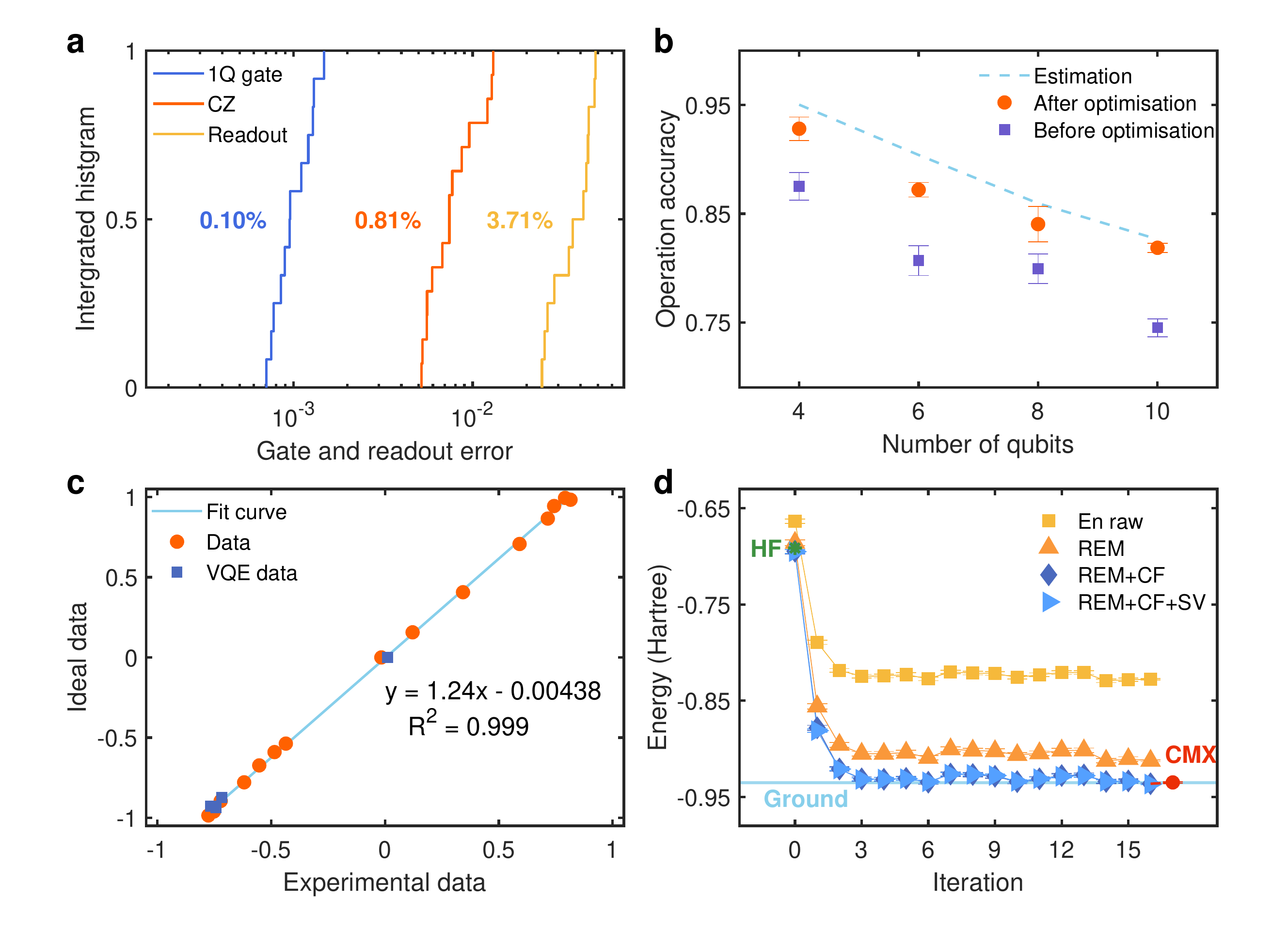}
\caption{\textbf{
Experimental optimisation and implementation of the algorithm.}
(a) Accumulated distribution of single-qubit gate (blue), CZ gate (red) and readout (yellow) errors. The number by the curve is the error on average. 
(b) Accuracy of multiple-qubit $\hat{V}(\vec \theta)$ rotation gate versus the number of qubits. The operation accuracy is obtained by repeatedly executing different numbers of $\hat{V} \hat{V}^{\dag}$ pairs, measuring the $Z$ projection of all qubits, and fitting the data with an exponential function. Under the depolarising error model, the operation accuracy equals the fidelity of $\hat{V}(\vec \theta)$ (see Supplementary Section I). Direct estimation (blue dashed line) is obtained as a product of the fidelities of all  quantum gates applied. The operation accuracy is closer to the direct estimation result after experimental optimisations, in which we used the Ry($-\pi/2$)-Rz($\theta$)-Ry($\pi/2$) gate sequence to replace the amplitude-based Rx($\theta$) gate and corrected the pulse distortion on relative couplers. 
(c) {An example of Clifford fitting for $\mol{H_2}$ with observable $\rm XXYY$. The orange dots correspond to the ideal-noisy expectation values for Clifford-analogue circuits.
The blue solid line is the linear fit of the data to learn the function, which is used to mitigate gate noise.}
The blue squares correspond to the ideal-noisy VQE data, which aligns well with the linear fit. 
(d) The optimisation procedure for H$_2$ with bond distance R~$=2.6$. 
The ground state energy with chemical accuracy (blue region) is calculated by exact diagonalisation as a benchmark. The yellow square, orange triangle, blue diamond, and grey right triangle are the energies without error mitigation, with REM, with REM + CF, and with REM + CF + SV,
respectively. The red dot is the final result of the ground state energy with CMX applied.
}
\label{fig2}
\end{figure}

\begin{figure*}[htb!]
\centering
\includegraphics[width=0.88\textwidth]{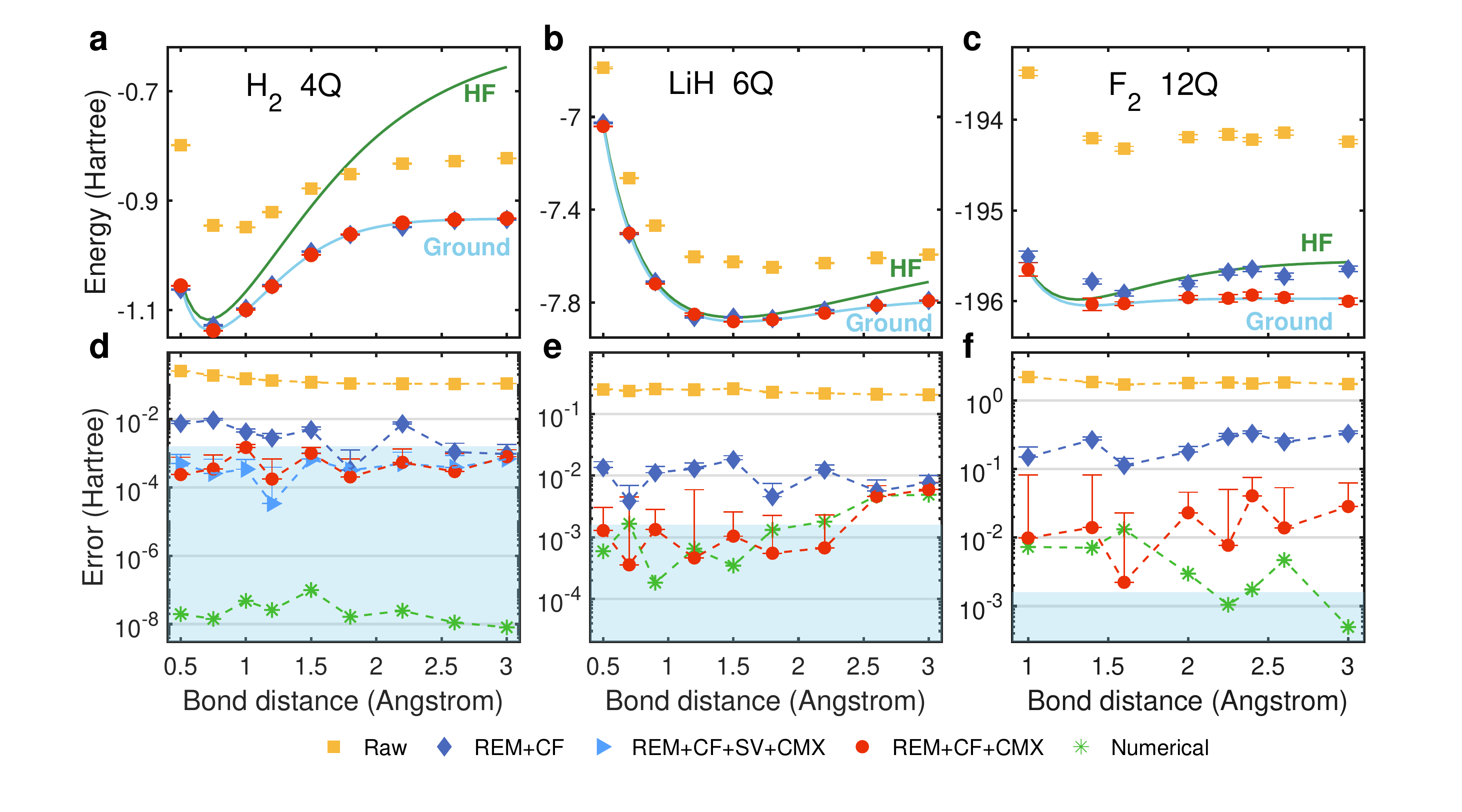}
\caption{
\textbf{The VQE simulations for  potential energy curves for different molecules.}
 (a-c) Potential energy curves as functions of the bond distance for $\mol{H_2}$ (4 qubits), LiH (6 qubits) and $\mol{F_2}$ (12 qubits) molecules with various error mitigation strategies. (d-f) Absolute errors are compared to the exact results. We compare the raw data (yellow squares) with the application of REM and CF (deep blue diamonds), SV (grey-blue triangles) and CMX (red circles). The results marked by green asterisks are energies calculated with the variational parameters searched in the experiment. The ground state energy with chemical accuracy (blue region) is calculated by exact diagonalisation as a reference.
}
\label{fig3}
\end{figure*}

In order to  obtain reliable experimental results, we need to optimise the quantum processor and mitigate experimental errors. In our work, 
various strategies are employed to improve gate operations and  mitigate errors, illustrated in the bottom column of Fig.~\ref{fig1}. 
We adopt dynamical decoupling~\cite{gustavsson2012dynamic} sequences to suppress dephasing errors for idle qubits and introduce a calibrated implementation of  basic operations in UCC to improve the circuit fidelity. 
We further integrate error mitigation methods for the remaining measurement and gate errors. 
Readout error mitigation~(REM) is applied to the classical measurement outcomes to mitigate measurement errors~\cite{PhysRevA.103.042605}.
% For measurement errors, we apply {readout} error mitigation (REM) to the classical measurement outcomes~\cite{PhysRevA.103.042605}.
For gate errors, we {apply} three types of error mitigation methods sequentially to obtain the error-mitigate expectation values of observables. {Considering particle number conservation, we apply an effective symmetry verification (SV) of the particle number parity~\cite{PhysRevA.98.062339};} To mitigate dominant two-qubit gate errors, we apply Clifford fitting (CF) by learning the noise model from noisy and ideal Clifford circuits~\cite{Czarnik_2021,lowe2021unified}; Finally, we consider connected moment expansions (CMX), which effectively implement imaginary time evolution using Hamiltonian moments up to the third order, to correct the energy deviation due to incapable circuit ansatz and implementation errors~\cite{PhysRevLett.58.83,claudino2021improving}. 
These techniques are essential for accurate experimental implementations of variational quantum algorithms.
% We illustrate the procedures in Fig.~\ref{fig1}.

%\section{Experimental realisation and results}

The experiment is conducted on a 66-qubit superconducting quantum processor ``Zuchongzhi 2.0''~\cite{wu2021strong}. We select 12 high-quality qubits arranged in a two-dimensional array, as shown in the middle column of Fig.~\ref{fig1}. We first optimise single-qubit gates, CZ gates and readout performance, following a series of optimisations. {The readout fidelity is $96.3\%$ on average, and the correlated readout error is verified to be negligible, characterised by random state measurements. This is achieved by optimising readout pulse and qubit frequencies, and suppressing the residual ZZ coupling (see Supplementary Information for the methods and results). }
% {The readout fidelity is $96.3\%$ on average and the correlated readout error is verified to be negligible, which is achieved by optimising readout pulse and qubit frequencies, and suppressing the residual ZZ coupling. The correlation is characterised by random state measurements (see Supplementary Information for the methods and results).}
The remaining uncorrelated readout errors are further alleviated by REM in our experiment. 
{The parallel single-qubit} and CZ gate fidelities are $99.90\%$ and $99.19\%$ on average, respectively (see Fig.~\ref{fig2}a), characterised by cross-entropy benchmarking~\cite{Arute2019}. {The high-fidelity parallel gates} are achieved with the advantage of the flip-chip structure and tunable couplers, which suppresses the cross-talk error significantly. 
% {We note that the system's reliability is not only the gate and readout fidelities but also the degree of agreement with the ideal noisy quantum simulator. 
% There experimental imperfections, including crosstalk, distortion, etc., can cause such disagreement. To overcome this issue,
We further optimise the elementary operations (consisting of multiple CZ gates and single-qubit gates) in the UCC circuit by correcting the $Z$ pulse distortion on couplers and replacing the amplitude-based Rx($\theta$) gate. As shown in Fig.~\ref{fig2}b, the operation accuracy notably increases after optimisation and matches the product of the fidelity of individual quantum gates, indicating its scalability for larger systems.
% \sout{
% All these hardware optimisations are crucial for our subsequent implementation of reliable and accurate simulations.
% }
{
We also highlight that the gate optimisation technique developed in this work holds potential for the preparation of multi-qubit Pauli rotation gates, which play a vital role in simulating systems with long-range interactions.
}
{All these hardware optimisations developed in our experiment are crucial for the reliable implementation of our algorithm.}

% \sun{To xiao:}
% [Sun: the concept of physically reliable.
% Gate operation.
% Readout error. Tensor product type.
% Have a sense of simulation accuracy.
% Why our simulation enables the reliability simulation.
% ]

To obtain high-accuracy results, we further apply three error mitigation techniques, CF, SV, and CMX, to reduce the effects of gate errors. 
% The noise channel of the new circuit is expected to be close to the Clifford analogue circuit as the two-qubit gates which contribute to dominant errors will be unchanged. 
For the observable in the Hamiltonian, we aim to obtain an error-mitigated expectation value by learning a fitting function $f$, which maps a noisy expectation value obtained in experiments to an error-mitigated one.
% To do so, prior to VQE simulation, we first run Clifford analogue circuits of UCC by replacing most of the non-Clifford gates with random Clifford gates. 
To get the function $f$, we construct Clifford-analogue circuits of the UCC ansatz by replacing most of the non-Clifford gates with random Clifford gates. The fitting function of the new circuit instances $\mathcal{S}$ is expected to be similar to the UCC circuit as the two-qubit gates which contribute to dominant errors are unchanged. 
Since Clifford circuits $\mathcal{S}$ are classically simulable, the function $f$ can be learned by fitting the ideal and experimental results, which are obtained by classical calculation and quantum measurements, respectively.
% Since Clifford circuits are classically simulable, we can obtain $o^{\mathrm{noisy}}$ and $o^{\mathrm{ideal}}$ by measuring on circuits $\mathcal{S}$ and by classical calculation, respectively.
% the circuit noise model can be learnt by fitting the ideal and experimental results, which are obtained by classical calculation and quantum measurements, respectively, and generally have a linear dependence.
% maps a noisy expectation value to an error-mitigated one. 
For example, for the XXYY observable in a 4-qubit H$_2$-molecule case, the noisy measurement results and the ideal results have a clear linear dependence, as shown in Fig.~\ref{fig2}c.
With the fitting function that is learned prior to VQE experiments, we can obtain an error-mitigated result from a given noisy measurement outcome, referred to as the CF method.
% \red{This error-mitigation protocol is known as CF.}
During the VQE process, for each observable, we first apply REM to mitigate readout errors and then obtain an error-mitigated observable expectation value by using {CF}.
The symmetry-verified result can be obtained by measuring additional observables $\hat{H}\hat{S}$ with the conserved symmetry $\hat{S}$, after which the final ground-state energy is estimated using CMX.
% It is important that the noise channel stays the same in the training and testing processes. This can be seen from the fact that
% This can be seen from the fact that all102
% non-Clifford gates are single-qubit rotation gates and we randomly replace them with Clifford ones of the same type,103
% i.e. single-qubit rotation with another angle that makes it Clifford. Therefore, with the rotation angles of single-qubit104
% gates altered, the noise channel for the quantum circuit persists
% The two-qubit gates which contribute more errors are Clifford gates, and all non-Clifford gates are single-qubit rotation gates, which will be randomly replaced with Clifford ones. 
As shown in Fig.~\ref{fig2}d, error mitigation is critical in improving calculation accuracy. The combination of REM, CF, and SV clearly improves the energy accuracy along the optimisation iteration for the $\mol{H_2}$ molecule. CMX further enhances the final energy within the chemical accuracy compared to the true ground state energy.
The implementation and experimental comparisons of the error mitigation techniques can be found in Supplementary Section IV. 
% \sun{Yukun: could you add more descriptions about figure 2.? including why QEM works, and so on.}
% \blue{The readout error is mitigated by  REM, which effectively inverses the readout error process by assuming a certain form of the error. The variational algorithms are predominantly affected by incoherent and readout errors but are robust to coherent errors. Thus, we apply the SV and CF methods for suppressing the effect of incoherent errors. As the ideal ground state conserves particle numbers, the SV method suppresses the effect of incoherent error that could compromise the symmetry-preserving properties by projecting the wave function prepared experimentally to the symmetry-preserving sub-manifold. Besides, the CF methods work by first learning the mapping relation between the experimental and idea estimated expectation value of an observable using Clifford circuits that are classically simulatable. Once the mapping function is properly learned, we then apply the function to general cases for predicting the ideal expectation value from experimental results. }

{
Equipped} with the theoretical and experimental improvements, we now show the calculation of the potential energy curves for $\mol{H_2}$ (4 qubits), LiH (6 qubits) and $\mol{F_2}$ 
(12 qubits) molecules. 
{The selection of active spaces and problem encoding strategies are shown in Supplementary Section II.}
Fig.~\ref{fig3} shows a significant decrease  in energy errors with error mitigation (around two orders of magnitude)  for all three molecules. 
The raw experiment results indicate that the simulation becomes less accurate with increasing system size. For instance, the raw errors for LiH and F$_2$ are above $0.1$ Hartree and $1$ Hartree, respectively.
% \cheng{This sentence is a arkward.}
This is because VQE experiments generally entail a large number of gates and measurement shots~\cite{wang2021noise,bittel2021training,tilly2022variational}, which results in difficulties in the demonstration of large molecular systems on current hardware.
In our experiment, we improve {the} simulation accuracy with dedicated device optimisation and error mitigation.
In particular, 
% we achieve chemical accuracy for $\mol{H_2}$ at all  bond distances
 % and  LiH at small bond distances.
{the average absolute energy errors for $\mol{H_2}$ and LiH  are $0.565 \pm 0.542$ milli-Hartree and 
% $1.80 \pm 1.45$ milli-Hartree 
$0.815 \pm 2.526$ milli-Hartree (excluding distances above 2.6 $\rm{\AA}$), respectively, which are within the chemical accuracy threshold.}
The energy error for LiH increases for large bond distances since UCC becomes less accurate in the dissociation regime. This has been alleviated by an optimised initial state preparation with a multi-reference state which is a superposition of the Hartree-Fock state and the dissociated state (see Supplementary Information). 
{To validate the effectiveness of our error mitigation schemes, we compare the error-mitigated energy with the one that is computed numerically  using the experimentally found parameters. The error-mitigated energies are found to be close to the numerical ones, which indicates the reliability of our scheme.}

\begin{figure*}[htb!]
\centering
\includegraphics[width=0.9\textwidth]{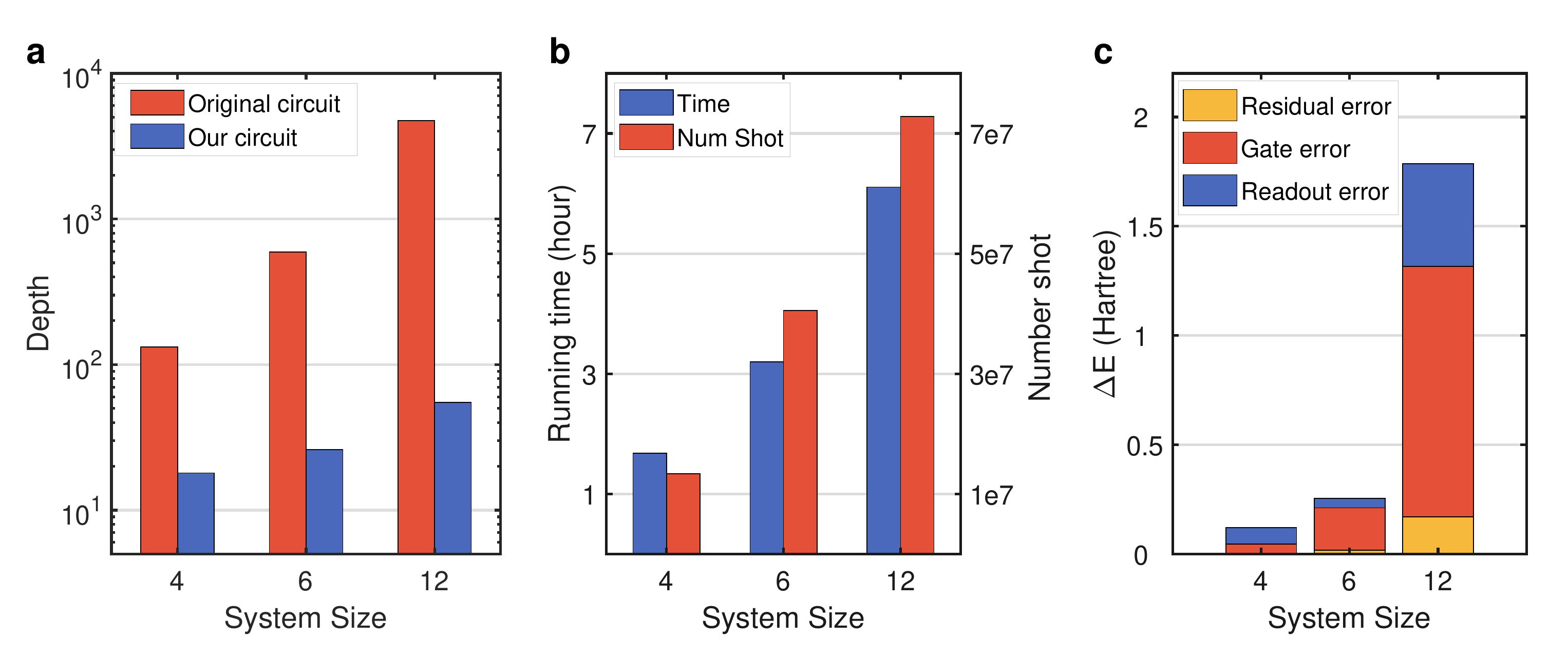}
\caption{
\textbf{Resource estimation and error analysis with increasing system size.}
(a) Circuit depth reduction. The total circuit depth is compared with the original UCCSD circuit ansatz considering single and double excitations.  The total circuit depth~\cite{note_depth} concerning both single-qubit gates and CZ gates for $\mol{H_2}$, LiH and $\mol{F_2}$ are 18, 26 and 55, respectively. The number of CZ  gates for $\mol{H_2}$, LiH and $\mol{F_2}$ are 10, 18 and 50, respectively, and the number of single-qubit gates are 14, 19 and 63, respectively.
(b) The resource estimation in terms of {average} running time and measurement shots {of a single bond distance} for the three molecules.
(c) The analysis of error sources for $\mol{H_2}$ (R = 1.5), LiH (R = 1.5) and $\mol{F_2}$ (R = 2.0). 
Here, we denote the energy before REM, after REM, after {REM and Clifford fitting,}
%the rest error mitigation methods, 
and the numerical result as $E_1$, $E_2$, $E_3$ and $E_4$, respectively. The contributions from readout errors (blue), gate errors (red) and residual errors (yellow) are respectively obtained by $E_2-E_1$, $E_3-E_2$ and $E_4-E_3$.
}
\label{fig4}
\end{figure*}

We further demonstrate the simulation for $\mol{F_2}$.
Owing to the larger system size and deeper circuit, the energy errors for $\mol{F_2}$ are around  $10^{-2}$ to $10^{-1}$, above the chemical accuracy threshold. 
{The average absolute energy error for $\mol{F_2}$  is $0.0174 \pm 0.0417$ Hartree.}
{Nevertheless, if the energy is calculated using the experimentally found parameters,  the energy error could be consistently suppressed below $10^{-2}$. The average absolute error for the numerically calculated energy is 4.81 milli-Hartree, which is 370 times smaller compared to the initial point, i.e., the experimentally measured Hartree-Fock energy. 
This indicates that the parameters that characterise the ground state can be  found approximately, and the variational scheme is still effective, even though the error cannot be fully mitigated for deep circuits. 
The gate and readout fidelities need further improvements to decrease simulation error below the chemical accuracy threshold.
% This indicates that the parameters can be , although the gate and readout fidelities need further improvement to decrease the error below the chemical accuracy.
% Need to be revised
}
% The experimental results indicate that the gate and readout fidelities need further improvement to decrease the error below the chemical accuracy.

% Indeed, compared to the energy of the initial state, {we achieved $0.278\%$, $0.778\%$ ($0.326\%$ except LiH R2.6 and R3.0), and $1.00\%$ relative error for the final state of these three molecules with the combination EM schemes of REM, CF and CMX ($1.56 \times 10^{-7}$, $0.783\%$ ($0.392\%$ except R2.6 and R3.0), and $0.268\%$ for numerical results.}

%Circuit depth reduction. The total circuit depth used in this work are compared with the original UCCSD circuit.  For $\mol{H_2}$, LiH and $\mol{F_2}$, the circuit depth in this work are 10, 18, 50, respectively, and the number of CZ gate with the original UCCSD circuit are 56, 280, 1920, respectively. We note that the number of single-qubit gate(excluding virtual Z gate) in this work are 18, 26, 84, respectively. The circuit depths considered with single-qubit gates and two-qubit gates are 22, 32, 67 layers, respectively.

With the experimental results of the three molecules, we  analyse the resource cost and scalability of VQE in our experiment. 
As shown in Fig.~\ref{fig4}a, our circuit optimisation strategy consistently reduces the depth of UCC circuits for molecules with different system sizes.
The circuit reduction strategy is efficient and crucial for extending our {scheme} to larger molecules.
Since the VQE process essentially extracts the information from the measurement outcome, the number of measurements required is a critical concern in quantum chemistry simulation. 
While previous analyses of measurement resources appeared pessimistic~\cite{Gonthier_Measurements_2022, Wecker_Progress_2015, yen_deterministic_2023}, our optimised grouping measurement strategy significantly reduces the number of measurement bases. For instance, the number of measurements for estimating the energy of $\mol{F}_2$ is $8.7 \times 10^5$, with a running time of about three minutes.
Comparing the three molecules in Fig.~\ref{fig4}b, we observe that the resource cost ({average} running time and measurement shots) increases almost linearly in the system size. 
Nevertheless, we still need to be careful about device imperfections, where the primary error sources include readout and gate errors.
We conduct an analysis to assess their respective contributions to the energy calculation, as shown in Fig.~\ref{fig4}c.
Our findings indicate that readout errors dominate for small molecules, while gate errors become more serious when the system size increases. 
Therefore, the readout and gate fidelities need to be improved systematically to achieve high accuracy for large molecules.

% \sun{
% A recent work \cite{obrien_purification-based_2023} demonstrates purification-based error mitigation but did not test variational algorithms - the circuit is generated by classical computing and thus actual, making the results good but did not provide evidence for whether quantum computers can be used to solve the electronic structure of molecules.
% The actual optimisation process distinguishes VQE from other experiments. This is the challenging part for hardware. This is evident in our experiment which is stable.
% }

 % we address the challenges of solving molecular electronic structures with noisy superconducting processors.
% We systematically improve quantum gate operations and measurement readout fidelities.
% With improved quantum gate operations and algorithm optimisations, we demonstrate an implementation of VQE with UCC for H$_2$, LiH, and F$_2$ molecules from 4 to 12 qubits.----
% Quantum chemistry simulation requires high accuracy. However, the accuracy of chemistry simulation greatly suffered from decoherence and readout errors.
% We realised QEM in mitigating decoherence and readout errors in chemistry by realising QEM.
% , and mitigated the crucial errors in chemistry simulation.
% ------
%\section{Discussion}

% With improved quantum gate operations and algorithm optimisations, we experimentally implement variational quantum eigensolver with the UCC ansatz for quantum computational chemistry. 
% A significant improvement in accuracy of around two orders of magnitude is observed when using error mitigation techniques 
In this work, we demonstrate an efficient and reliable quantum chemistry solution on a noisy intermediate-scale superconducting quantum processor. 
We develop a systemic way to reduce quantum resources in terms of circuit depths and measurement cost, and improve our hardware in gate operations and measurement. We integrate error mitigation techniques for mitigating different types of simulation errors due to decoherence and measurement readout. With hardware optimisation tailored for chemistry simulation, we experimentally implement variational quantum eigensolvers with optimised chemically inspired ansatz and scale the simulation up to 12 qubits.
Our solution enables us to achieve chemical accuracy for $\mol{H_2}$ at all bond distances and  LiH at small bond distances. We demonstrate a significant enhancement of the accuracy by around two orders of magnitude with error mitigation techniques.
We examine the efficacy of error mitigation in mitigating gate and readout errors, which serves as a reference for understanding the potential benefits and limitations of error mitigation approaches, aiding in the design and optimization of future implementations.

% for the three molecules. 
% \blue{The average absolute error is below the simulation accuracy.
% }

% The optimisation process sets VQE apart from other quantum experiments, which demanding for quantum hardware. 
The optimisation process in VQE imposes strict demands on the stability of quantum devices.
We monitored the parameter evolution and observed its stability throughout the optimisation process over several hours, which indicates the stability of our devices as discussed in Supplementary Sections IV.
While owing to non-negligible hardware errors, the current quantum processor is insufficient for larger molecules, the variational scheme {is tested to be} effective and reliable in our experimental implementation.
Compared to the energy of the initial input state, we achieved $0.278\%$, $0.778\%$, 
% ($0.326\%$ except LiH R2.6 and R3.0), 
and $1.00\%$ relative errors for the final output states of the three molecules by applying a combination of REM, CF and CMX.
We further compare the circuit error estimated using different methods to understand the effect of noise with an increasing system size. We observed that the circuit error obtained by fitting the experimental data is much lower than the error computed by multiplying the error of each individual gate, which is discussed in detail in  Supplementary Section~IV.
% , and we refer to Supplementary Information for details.
 % ($1.56 \times 10^{-7}$, $0.783\%$ ($0.392\%$ except R2.6 and R3.0), and $0.268\%$ for numerical results
Our experimental result and resource analysis indicate the potential of simulating larger molecules with improved gate and readout operations.
% \red{Mention Fig.S17:  As shown in Supplementary Fig.S17, the VQE procedure could partially reduce the effects of hardware control errors, and the actual VQE circuit error rate is much lower than the error estimated with individual XEB results when the system size increases.}

% the necessity for improved gate and readout operations in order to achieve chemical accuracy for larger molecules. 

%\vspace{10cm}

\begin{acknowledgments}
    The authors thank the USTC Center for Micro- and Nanoscale Research and Fabrication for supporting the sample fabrication. 
The authors also thank QuantumCTek Co., Ltd. for supporting the fabrication and maintenance of room-temperature electronics.
% \textbf{Funding:}
This research was supported by the Innovation Program for Quantum Science and Technology (Grant No.~2021ZD0300200), Shanghai Municipal Science and Technology Major Project (Grant No.~2019SHZDZX01), Anhui Initiative in Quantum Information Technologies, National Natural Science Foundation of China (Grants No.~11905217, No.~11774326, No.~12175003), Natural Science Foundation of Shandong Province, China (grant number ZR202209080019), and Special funds from Jinan science and Technology Bureau and Jinan high tech Zone Management Committee. H.-L. H. acknowledges support from the Youth Talent Lifting Project (Grant No.~2020-JCJQ-QT-030), National Natural Science Foundation of China (Grants No.~11905294, 12274464), China Postdoctoral Science Foundation, and the Open Research Fund from State Key Laboratory of High Performance Computing of China (Grant No.~201901-01). 
J. Sun acknowledges the Samsung GRC grant for financial support.
M. Gong was supported by the Youth Innovation Promotion Association of CAS (Grant No.~2022460). X.B. Zhu acknowledges support from THE XPLORER PRIZE. 
\end{acknowledgments}

% \section*{Acknowledgements}

\section*{Author contributions}
X.Y, M.G, J.S. initiated the project. J.S. developed the theoretical aspect of the project and built up the source code for numerical simulation with input from X.Y. and Y.Z.. J.S., Y.Z, and K.L carried out the numerical simulation under the supervision of X.Y.. Y.Z. and K.L. generated the measurement bases with the algorithm developed by J.S.. 
S.G., H.Q., M.G., F.C. and S.C. performed the measurements.
S.G., J.S., H.Q., M.G., Y.Z.  analysed the experimental data.
Q.Z., Y.Y., C.Y., F.C. and S.L. designed the processor.
S.C., Y.L., K.Z., S.G., H.Q., T.-H.C., H.R., H.D. and Y.-H.H. fabricated the processor.
M.G., S.W., C.Z., Y.Z., S.L., C.Y., J.Y., D.F., D.W. and H.S. contributed to the construction of the ultracold and low-noise measurement system. 
J.L., Y.X., F.L., C.G., L.S., N.L. and C.-Z.P. developed the room-temperature electronics.
Experiments were performed using a quantum processor that was developed and fabricated with a large effort from the experimental team.
J.S., S.G., Y.Z., X.Y. and M.G. wrote the manuscript with input from all the authors.
All the authors contributed to writing up the manuscript.
X.Y., X.Z., and J.-W.P. supervised the project.\\

\vspace{6pt}
\textbf{Competing Interests}
The authors declare no competing interests.\\

\vspace{6pt}
\textbf{Data and materials availability}
% \textbf{Data and materials availability:} 
The data shown in this paper are available from the corresponding authors upon reasonable request.\\
%\textbf{Code availability:} The code in this paper is available from the corresponding authors upon reasonable request.

\vspace{6pt}
\textbf{Note:} During the preparation of this work, we became aware of two independent works realising the unitary pair-coupled cluster double ansatz with  superconducting qubits~\cite{obrien_purification-based_2023} and trapped ions~\cite{2022arXiv221202482Z}, respectively, for chemical systems with the problem size at most 10 and 12 qubits.  
Ref.~\cite{obrien_purification-based_2023} showcased the efficacy of purification-based error mitigation; however, the circuits in their experiments were produced through classical computation.

\bibliographystyle{unsrt}
\bibliography{ref_chem}

\begin{thebibliography}{10}
\bibitem{note_depth}
The total circuit depth serves as an indicator of the system's performance to
  decoherence, as in our system the time duration for single-qubit and
  two-qubit gates are similar.
\end{thebibliography}

\clearpage

\newpage

\section*{Supplementary Information}

\tableofcontents

{\section{Device and calibration}}
\label{sec:hardware}
\subsection{\blue{Device}}
{The ``Zuchongzhi 2.0'' quantum processor contains 66 qubits and 110 couplers. In order to control and wire such a large-scale quantum processor, the qubit layer and wiring layer are prepared on separate chips, which are then combined using the flip-chip packaging process. High-purity aluminium thin films are grown on a sapphire substrate by molecular beam epitaxy (MBE) for both chips, which can reduce material loss and improve qubit decoherence performance. The wiring layer employs photolithography and etching processes to fabricate readout circuits and control circuits and utilizes a lift-off process for creating airbridges. These airbridges effectively establish grounding connections, which prevent slot line modes, and simultaneously reduce crosstalk between control circuits. This results in superior qubit isolation and enables high-precision gate operations. The qubits and couplers are fabricated on the qubit layer. In this process, the capacitive elements of the qubits and couplers are defined using photolithography and etching. The Josephson junction is created by defining the junction area using electron beam equipment and then prepared using a dual-angle evaporation process followed by a lift-off process. During the fabrication process, we fine-tuned the junction's critical current by adjusting the oxidation conditions and the size of the junction region. This allowed us to achieve the optimal frequency adjustment range for both qubits and couplers. Furthermore, a considerable number of indium bumps as superconducting interconnects are employed to establish a solid ground connection between the qubit layer and the wiring layer, effectively reducing parasitic modes.}

\blue{Following the chip fabrication process, the processor is wire-bonded to a printed circuit board located within a sample box. This sample box is equipped with gold-plated shielding inside and $\mu$-metal shielding outside. Finally, the packaged processor is mounted to the plate of a dilution refrigerator and connected to room-temperature electronics via a series of attenuators, filters, and amplifiers.}

\subsection{System calibration}
Our experiment is carried out on the ``Zuchongzhi 2.0'' quantum processor. We selected up to 12 qubits to run the quantum circuits. We turned off the coupling between used and unused qubits. Then, we optimised and calibrated the performance of the selected qubits, especially the quantum gates and readout performance.

The  optimisation procedure of our quantum processor can be summarised into the following steps:

\begin{enumerate}
    \item \textbf{Select idle frequencies}. Idle frequencies refer to the qubit operating frequencies when the qubits are in idle operations or single-qubit gates. As the qubit performance varies strongly with the frequency, setting the qubits in good idle frequencies is necessary to achieve high-fidelity gates. To arrange the idle frequencies, we need to consider a combination of factors, including coherence time, residual coupling between qubits, the distance between idle frequencies and CZ interaction frequencies (related to $z$ distortions),  two-level systems (TLSs), and XY-crosstalk. XY-crosstalk comes from the interaction between control lines and the distance of the driven frequencies.
    
    \item \textbf{Turn off the coupling between qubits.} In the setting of idle frequencies, the frequency distance between nearest-neighbour qubits is about 100MHz. Turning off the coupling to suppress the interaction between qubits is necessary to decrease parallel-gate errors. After setting qubit idle frequencies, we determine the coupler zero-coupling point by measuring the relative two-qubit conditional phase when varying the flux of the coupler. Then, we turn off the qubit-qubit coupling by adjusting the coupler flux bias to the selected zero-coupling point.  
    
    \item \textbf{Optimise single-qubit gates.} We fine-tune the XY drive pulse amplitudes and apply Derivative Reduction by Adiabatic Gate (DRAG) correction to the pulse envelope with a length of 25 ns.

    \item \textbf{Optimise the Josephson Parametric Amplifiers (JPAs).} The JPA can significantly improve the readout signal-to-noise ratio (SNR) and fidelity~\cite{lin2013single,mutus2014strong}. First, we coarsely tune the JPA direct current (DC) bias, pump signal power and frequency according to our design. Then, these parameters are optimised by the Nelder–Mead algorithm, and the target function is the average SNRs of the qubit readout signals. 
    
    \item \textbf{Optimise the pulse distortion of each qubit.} Distortion or reflection occurs when the pulse is transmitted from a waveform generator to qubits, and it will cause a shift in the qubit frequency which increases the gate error in the quantum circuit. In our experiment, we first measure the effect of pulse distortion on qubit phase accumulation over a fixed time of 400 ns. Then we calculate the waveform after distortion based on the result of phase accumulation and apply a reverse one to compensate for the pulse. \blue{The full protocol is detailed in Ref ~\cite{zhiguangYan2019}}
    
    \item \textbf{Optimise the readout crosstalk.} 
    Readout crosstalk may be generated by factors such as the influence of neighbouring qubits on readout ones, coupling between readout resonators, frequency collision (caused by the overlap of demodulation frequencies, intermodulation of parametric amplifiers, etc.). It can reduce the readout fidelity and make REM unreliable. The readout crosstalk can be well suppressed by optimising the readout parameters, including frequencies and lengths of readout signal and qubits' frequencies for readout. 
    
    \item \textbf{Calibrate the readout performance.}
    After optimising the readout crosstalk, we need to confirm whether residual crosstalk is negligible. Here, we first calibrate the standard readout fidelity with qubits preparing in $\ket{0}^{\otimes n}$ and $\ket{1}^{\otimes n}$. Then, we prepare the qubits in a random state set and calibrate the readout fidelity again. If the fidelities are close in both cases, the residual crosstalk is negligible. 
    
    \item \textbf{Optimise CZ gates.} 
    In the UCC ansatz, the multi-qubit Pauli rotation gate $\hat{V}(\vec \theta)$ is an elementary operator, which is compiled with single-qubit gates and CZ gates. The CZ gate with tunable-coupler structure has been investigated and realised with high fidelity~\cite{Ye_2021,sung2021realization}. In our experiment, CZ gates are realised by utilising the resonance coupling between $\ket{11}$ and $\ket{02}$ states and optimised by the following steps:
\begin{enumerate}
    \item According to the pattern in circuits, The CZ gates and single-qubit gates are divided into different groups. The next steps will be performed individually in these groups.
    \item We collect coherence information (T1 and T2) for a large frequency range ($\sim$100MHz) of qubits and determine the frequency region that is suitable for CZ interaction accordingly.
    
    \item We first prepare the qubits in $\ket{11}$ states. Then, we measure the conditional phase with varying coupling strength (tuning by coupler) for a fixed 50 ns gate time. Next, we measure the leakage of qubit states to $\ket{02}$ or $\ket{20}$ as qubit interaction frequencies vary. Finally, we coarsely determine the interaction frequencies and coupling strengths by minimising the cost function of conditional phase error and leakage error.
    
    \item Select a CZ gate in the group and finely adjust its parameters by optimising the XEB (cross-entropy benchmarking) fidelity with a fixed cycle of 20 and fixed random circuits of 60 while keeping other parameters of the CZ gate unchanged.
    
    \item We repeat step (d) for the CZ gate with the lowest fidelity in the group until the average fidelity no longer improves.
\end{enumerate}
    
   \item \textbf{Calibrate single-qubit gates and CZ gates.} We utilise the SPB and XEB measurements to calibrate single-qubit and CZ gates. For CZ gates, the dynamical phases on both qubits are also determined in the XEB experiment.
    
\end{enumerate}

All steps can be realised automatically as a directed graph~\cite{Kelly2018}, except for step {6}, which needs to be checked manually. We refer to Ref.~\cite{wu2021strong} for more details about the performance calibration of quantum devices. 

\begin{figure}[htb]
\centering
\includegraphics[width=1\textwidth]{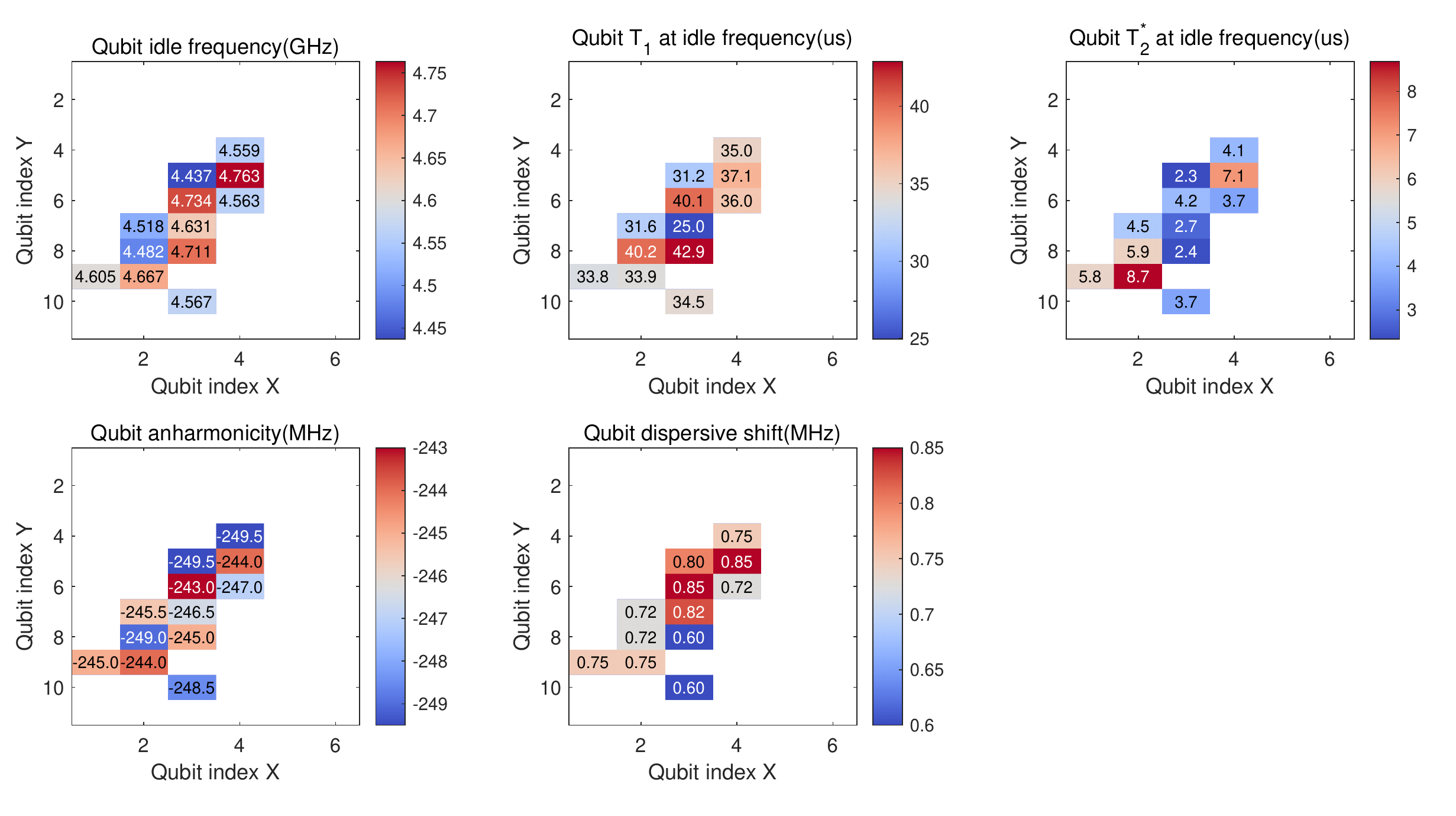}
\caption{\blue{\textbf{The typical distribution of the qubit parameters}. Idle frequency, $T_1$, $T_2^*$, anharmonicity and dispersive shift for the selected 12 qubits.}}
\label{fig:supp_qubit_parameter}
\end{figure}

\begin{figure}[htb]
\centering
\includegraphics[width=1\textwidth]{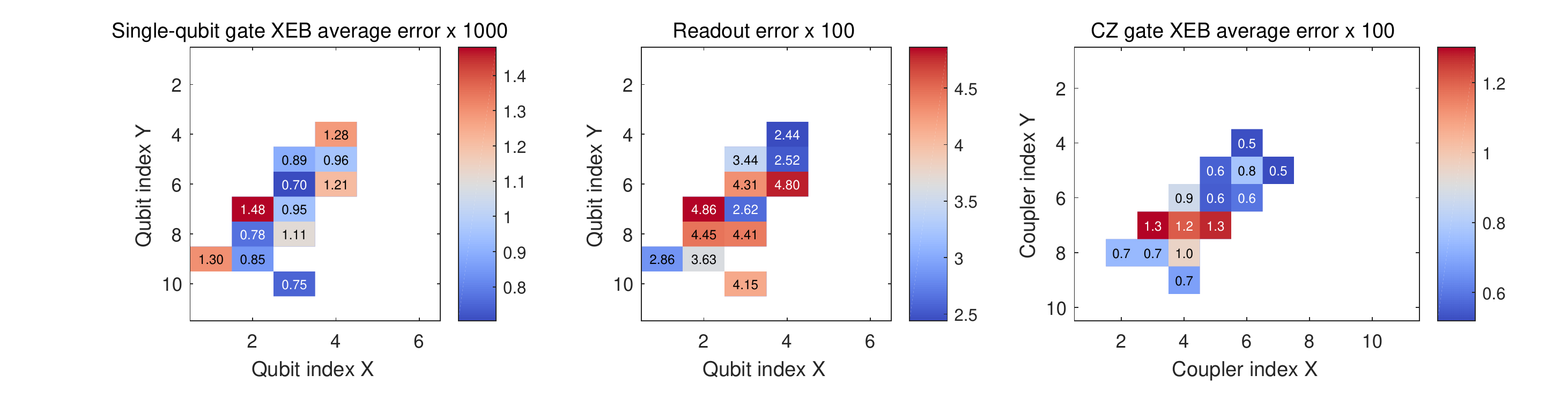}
\caption{
\textbf{The performance of the quantum processor after calibration}.
Error for readout,  single-qubit, and CZ gates for the selected 12 qubits.
}
\label{fig:fidelity_color_map}
\end{figure}

After all the calibrations and optimisations, \blue{we summarised the qubit parameters in \autoref{fig:supp_qubit_parameter}, and the errors for single-qubit gates, two-qubit gates and readout  in Fig.~2a in the main text and \autoref{fig:fidelity_color_map}}. Fig.~2a in the main text shows the accumulated distribution of single-qubit gate, CZ gate and readout errors. \autoref{fig:fidelity_color_map} shows the spatial distribution of those gate errors for selected qubits on the ``Zuchongzhi 2.0'' quantum processor. There is no strong spatial error distribution in our qubit set.
We run the automatic calibration steps {3, 7, 8, and 9} every three hours to maintain the system's performance. 

\subsection{Multi-qubit Pauli rotation gate calibration and optimisation}

The multi-qubit Pauli rotation gate $\hat{V}(\vec \theta)$ is an elementary operator in the UCC ansatz. Consequently, the quality of $\hat{V}(\vec \theta)$ directly determines the accuracy of our VQE experiment. In addition to high-fidelity single-qubit and CZ gates,   two more experimental improvements are made to increase the quality of $\hat{V}(\vec \theta)$.

The Z-pulse distortion on the couplers could lead to a non-ideal coupling strength during the CZ gate and cause a residual wave trailing after the gate. Since  $\hat{V}(\vec \theta)$ is mainly composed of  CZ gates,   a non-ideal transmission characteristic will generate unwanted pulse distortion, thus increasing the error of the following gates. The residual effect is difficult to observe in the two-qubit XEB experiment. However, the trailing signal on the coupler will shift the qubit frequencies and reduce the fidelity of the gates in multi-qubit Pauli rotation. We first correct the Z-pulse distortion on the coupler to address this problem. \blue{We apply the Z waveform to the couplers and measure the phase accumulation of the adjacent coupled qubit. With the relation between coupler Z bias amplitude and qubit frequency, we could calculate the response function of the coupler control line, thereby achieving the correction through reverse waveform compensation. }

Secondly, $\mathrm{R_x}(\vec\theta)$ in $\hat{V}(\vec \theta)$ is a non-Clifford gate, with different rotating angles originally realised with different pulse amplitude by linear adjustment. The error of $\mathrm{R_x}(\vec\theta)$ is significantly larger than a standard single-qubit gate due to a non-linear response nonlinear response \blue{in the signal generation and transformation. 
On the other hand, the DRAG parameters are calibrated using $\pi/2$ gate, a fixed DRAG parameter for an arbitrary rotation gate $\mathrm{R_x}(\vec\theta)$ would lead to additional control error and state leakage.}
Therefore, we replace the $\mathrm{R_x}(\vec\theta)$  with $\mathrm{R_y}(-\pi/2)$-$\mathrm{R_z}(\vec\theta)$-$\mathrm{R_y}(\pi/2)$. In our experiment, $\mathrm{R_z}(\theta)$ is a virtual gate. This replacement reduces the error of $\mathrm{R_x}(\vec\theta)$ and guarantees the performance of Clifford fitting, which highly relies on the control accuracy in $\mathrm{R_x}(\vec\theta)$, as introduced in \autoref{sec:clifford_fitting}.

We benchmark the operation quality with and without the above two improvements by the following protocol:
\begin{enumerate}
    \item Initialise all qubits in the  state $\ket{0}^{\otimes n}$.
    \item Apply a sequence consisting of $m$ sets of $\hat{V} \hat{V}^{\dag}$ pairs with different random-selected rotating angle $\theta$. 
    \item {Perform projection measurements along the $Z$ direction} to obtain the projection probability to $\ket{0}^{\otimes n}$. 
    % expectation value.
    \item Repeat steps 2 and 3 for $k$ times to obtain a stable average probability.
    %\sun{what is k???? why we need k times???}
    \item Change the length $m$ and repeat steps 2, 3 and 4.
    % to build up an exponential decay of the measurement results.
\end{enumerate}
Note that when the noise is sufficiently small, $\hat V\hat V^\dag$ should be close to identity, and hence the projection probability should be close to 1. Therefore, the projection probability value reflects the implemented gate's quality. 
Let us denote the implemented noisy gates as $\tilde{V}$ and $\tilde{V}^{\dag}$.
Here, in order to see a more explicit relation, we make the following assumptions:
    \begin{enumerate}
        \item  The implemented noisy gates $\tilde{V}$ and $\tilde{V}^{\dag}$ are the corresponding noiseless gates ${V}$ and ${V}^{\dag}$ concatenated by an error process $\mathcal{E}(\rho)$.
        % \item The fidelity of $\hat{V}(\vec \theta)$ with different random $\vec \theta$ are the same.
        \item The error process $\mathcal{E}$ of the implemented noisy $\tilde{V}$ is gate-independent, time-independent, and parameter-independent, and can be described by a depolarising channel $\mathcal{E}(\rho) = p \rho + (1-p) I/2^n$ with the identity operation $I$ and the error rate $1-p$.
        
    \end{enumerate}

\noindent Under the above assumptions, the averaged gate fidelity of the noisy $\tilde{V}$ is
\begin{equation}
    F_{\rm avg} = \int d\psi \tr[\mathcal V(\psi)\tilde{ \mathcal{V}}(\psi)] \approx p
\end{equation}
Here $\mathcal{V}(\cdot) = \hat{V}(\cdot)\hat{V}^{\dag}$, $\tilde{\mathcal{V}}(\cdot) = \tilde{V}(\cdot)\tilde{V}^{\dag}$, and we have neglected exponential dependence of $\mathcal O(1/2^n)$. 

Now, consider $m$ sets of noisy $\tilde V\tilde{V}^{\dag}$ on $\rho_0 = \ket{ \bar 0}\bra{\bar 0}$, we have
\begin{equation}
    \rho = \prod_{i=1}^m \mathcal{E}   \mathcal{V}^\dag   \mathcal{E} \mathcal{V}  (\rho_0) = p^{2m} \rho_0 + (1-p^{2m}) I/2^n.  
\end{equation}
  % with $\rho_0 = \ket{ \bar 0}\bra{\bar 0}$, $\mathcal{V}_1(\cdot) = \hat{V}(\cdot)\hat{V}^{\dag}$, and $\mathcal{V}_2(\cdot) = \hat{V}^{\dag}(\cdot)\hat{V} = p^{2m} \rho_0 + (1-p^{2m}) I/2^n$.
The probability to obtain $\ket{ \bar 0}\bra{\bar 0}$ is given by
  $
      \tr( \ket{ \bar 0}\bra{\bar 0} \rho) \approx p^{2m},
  $
where we again neglect exponential small dependence of $\mathcal O(1/2^n)$.  
Considering state preparation and measurement errors (assuming the stochastic noise model), we can show that the expectation value of $Z$ takes the form of $ Ap^{2m} + B$. Therefore, the error rate $p$ can be obtained by fitting the measurement outcomes with different choices of $m$. 

% The state fidelity  is given by 
% \begin{equation}
%    F(m) = \tr(\rho_0 \rho ) = p^{2m} + (1-p^{2m})/2^n.
% \end{equation}

% \sun{Can we use this results??}
% We fit the average sequence fidelity $F$ by $F(m) = Ap^{2m} + B$, where $m$ represents the total number of $\hat{V} \hat{V}^{\dag}$ pairs.

% Finally, we acquire the operation \guo{depolarising error $e_d = (1 - p)$ by the fitting result. We can further acquire the operation depolarising fidelity $F (\hat{V}) = 1 - e_d $ }. 

%Next, we fit the average sequence fidelity $F$ to the equation $F(m) = Ap^{2m} + B$ where $m$ represents the total number of $\hat{V}$ and $\hat{V}^{\dag}$. Finally, we acquire the operation \guo{Pauli error $e_p = (1 - p)(1 - 1 / D ^ 2)$ by the fitting result, where $D = 2 ^ n$ is the dimension of the Hilbert space. We can further acquire the operation fidelity $F (\hat{V}) = 1 - e_p $ }. 

\begin{figure}[htb]
\centering
\includegraphics[width=0.8\textwidth]{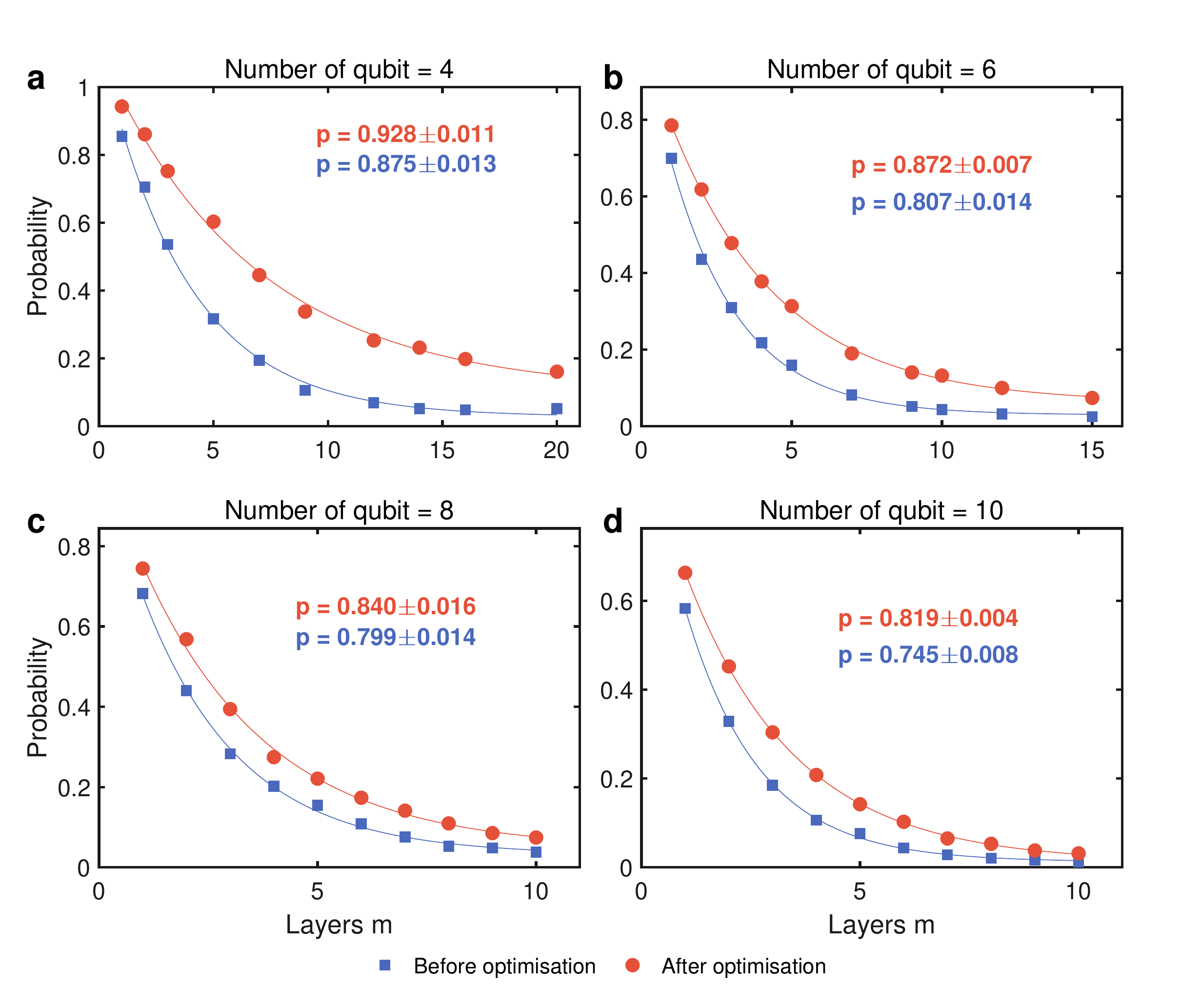}
\caption{
\textbf{Benchmark of multi-qubit Pauli rotation gate.}
The error rate of each sequence, as a characteristic of the averaged fidelity, can be obtained by measuring the probability of $\ket{0}^{\otimes n}$. The data points are the average values at each sequence length. 
}
\label{fig:V_gate_fit}
\end{figure}

In the experiment, we use the fitting curve as a  benchmark of the noisy $\tilde{V}(\vec \theta)$, as shown in \autoref{fig:V_gate_fit}. This figure clearly shows the difference in the quality before and after optimisation for the number of qubits as 4, 6, 8 and 10. After optimisation, there is an average $6\%$ improvement for the $\tilde{V}(\vec \theta)$ gate. 
This experiment demonstrates a fast, scalable, and efficient protocol to benchmark a multi-qubit quantum computational chemistry operation.
% Our results indicate a hi

\begin{figure}[htb]
\centering
\includegraphics[width=0.6\textwidth]{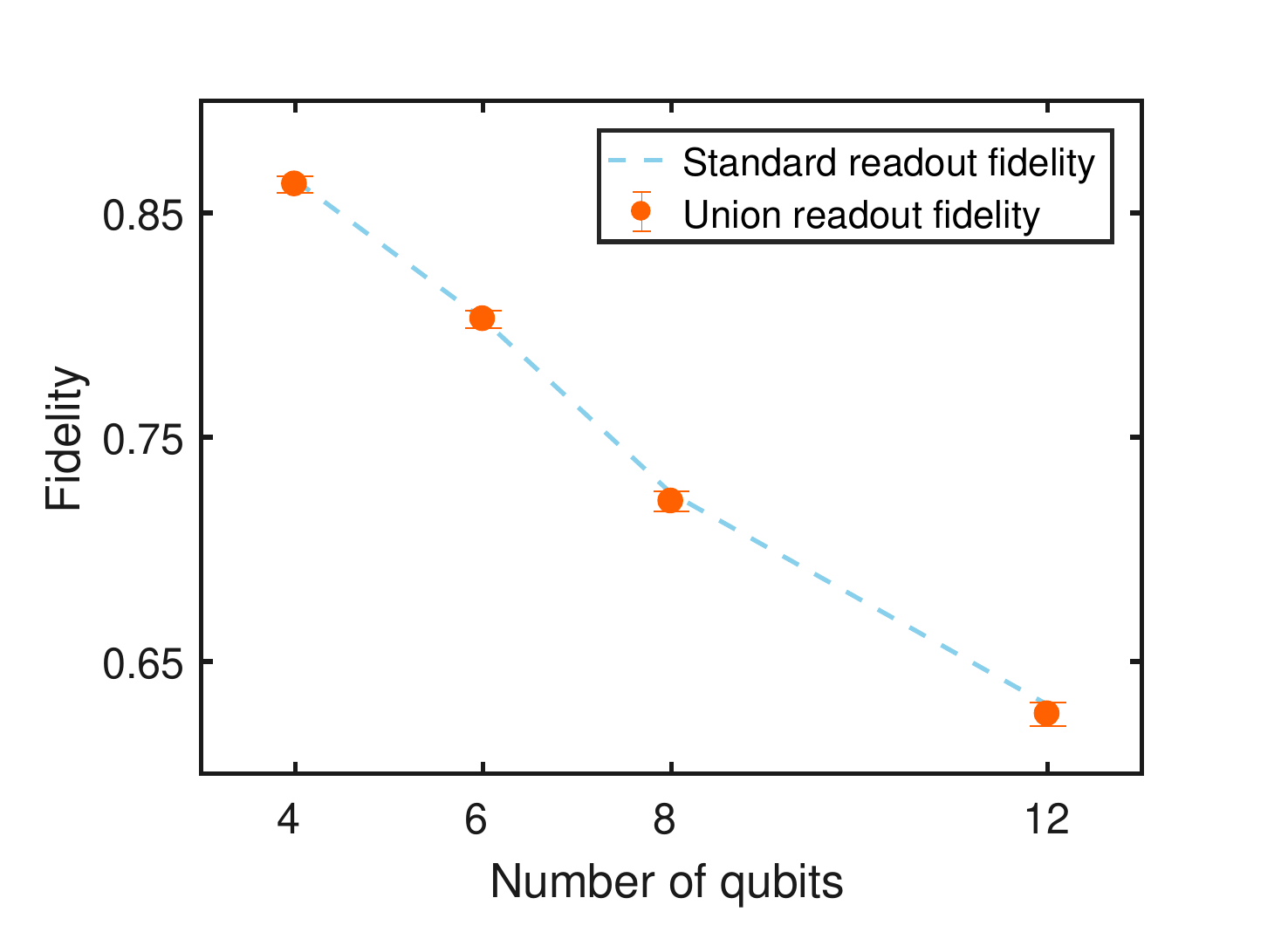}
\caption{
\textbf{Union readout fidelity.}
Comparison of average union readout fidelity (orange dots) and standard readout fidelity (blue dashed line) for four qubits, six qubits, eight qubits and twelve qubits.
}
\label{fig:random_state_fidelity}
\end{figure}

\subsection{Correlated readout error and random state measurements}

Quantum state measurements determine the information that we can extract from quantum experiments. Measurement errors can significantly impact simulation accuracy since they introduce a bias in the expectation value.  Therefore, it is essential to mitigate measurement errors to ensure reliable results.
% Measurement errors can greatly impact simulation accuracy since they introduce a bias in the expectation value. 
% Therefore, it is crucial to suppress measurement readout errors.

To mitigate measurement errors, the first step is to learn the measurement noise matrix $\Lambda$ from readout calibration measurements.
Provided an invertible measurement noise matrix  $\Lambda$, we can mitigate measurement readout errors by applying an inverse matrix $\Lambda^{-1}$ to the noisy measurement outcomes.

% The measurement noise matrix can be learned from the readout calibration measurement.
There are two types of measurement calibration matrices, one takes a tensor product form, a tensor product of local single-qubit calibration matrices, while the other one characterises correlated noise. The latter requires more experimental resources, which is not desirable in experiments. When the correlated readout error is negligible,  noise can be well described by a measurement noise matrix with a tensor product form.

In general, correlated readout error arises from the residual coupling and the spectral overlap of readout resonators. We make two improvements to ensure that the correlations of measurement errors are weak enough to be neglected.

Firstly, we suppressed the residual coupling by adjusting the working point of the coupler delicately. Secondly, we optimised readout parameters simultaneously, including readout pulse frequencies, readout pulse amplitudes, readout pulse length and qubit readout frequencies.

After these improvements, we verified the correlated readout error was suppressed by comparing the results of standard measurement and random state measurement. The verification by  in the following steps:
\begin{enumerate}
    \item Prepare all qubits in $\ket{0}^{\otimes n}$ and $\ket{1}^{\otimes n}$ to acquire $F_{00}$ and $F_{11}$. Here, $F_{00}$ ($F_{11}$) means the probability to get a measure result of $\ket{0} (\ket{1})$ when the qubit was prepared in $\ket{0} (\ket{1})$.
    \item Random state measurements. Randomly prepare the qubit state and use the same discrimination line of the readout signal demodulated IQ clouds to calculate fidelity. Multiply the fidelities for all qubits, and we refer to the multiplication as the union readout fidelity.
    \item With hundreds of repeated experiments, averaged union readout fidelity is computed.
\end{enumerate}
\autoref{fig:random_state_fidelity} 
 compares the average standard and union readout fidelity. These two types of readout fidelity show a good agreement, and the correlated readout error is negligible, which ensures the tensor product form is applicable in this experiment.

\section{Variational quantum state preparation and measurement}
\label{sec:method_VQE}

This section discusses quantum algorithms for quantum chemistry problems, including variational state preparation, optimisation, and quantum state measurement. 
This section is organised as follows.
In \autoref{sec:init_state_prep}, we first demonstrate multi-reference state preparation on a quantum processor and illustrate the corresponding circuit. 
In \autoref{sec:circuit_design}, we introduce quantum circuit design strategies based on unitary coupled cluster ansatz, and propose a  circuit reduction strategy to enable the simulation on current hardware. Quantum circuits for the three molecules are shown in \autoref{sec:circuit_compilation}.
In \autoref{sec:optimisation_SGD}, we introduce an optimisation scheme by analytic stochastic gradient descent and present the analytical form of the gradient estimation.
In \autoref{sec:OGM}, we introduce an efficient quantum state measurement strategy for measuring multiple observables.

\subsection{Problem encoding and qubit reduction}

% \sun{These sentences redudant and not the 'direct'.
% Active space, intuition.
% Stay chemical inert, which is called core orbitals.
% Slater determiant is occupied.
% Occuputation number XXX (in ) to select the orbitals.
% }

% We select the core orbitals as those who  stay chemical inert, that is, their electronic occupation population stays to be two when expanding the ground state of the Hamiltonian as a linear combination of Slater determinants. 

This work considers second-quantised molecular Hamiltonians represented in a discrete basis set in the active space. 
We implement qubit reduction by separating the full space of molecular orbitals into three parts: the core, active, and virtual space.
We select the core orbitals as those that stay chemically inert. That is, their occupation numbers are sufficiently close to two when expanding the ground state of the Hamiltonian as a linear combination of Slater determinants, and  virtual orbitals 
as those whose occupation numbers are close to zero. 
By classifying molecular orbitals into these three categories, we only need to compute the ground state in the selected active space, and thus the number of qubits required is reduced.
As discussed in Ref.~\cite{mcardle2018quantum}, a generic way to determine the classification of molecular orbitals is by computing the one-particle reduced density matrix (1-RDM) using an approximation of the ground state generated by a classical tractable method, such as the coupled cluster method. The 1-RDM is defined as
$
    ^{1}D^i_j=\braket{\hat{a}_i^\dagger\hat{a}_j}.
$ 
By diagonalising the 1-RDM by a basis rotation operation, we obtain the natural orbital occupied numbers (NOONs) for every molecular orbital.  NOON indicates the expected occupation number for the corresponding orbital, based on which we can determine the core, active and virtual spaces by the pre-determined thresholds. 

%\sun{Changsu: By the way, this method works only when we generate 1-RDM with some "correlation" methods beyond HF.}

\begin{figure}[htb]
\centering
\includegraphics[width=1\textwidth]{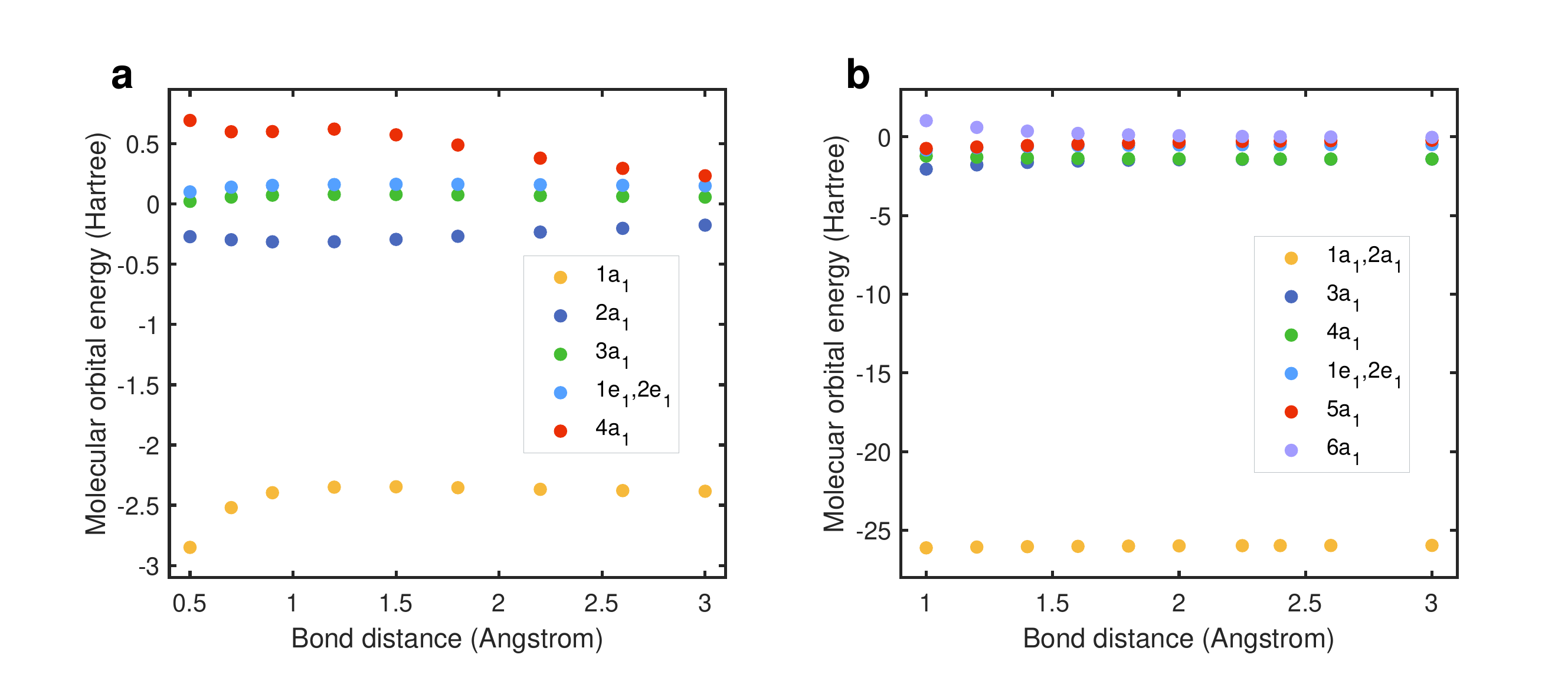}
\caption{
\blue{Hartree Fock} orbital energies for the (a) LiH and (b) $\mol{F_2}$ molecules. 
% Each molecular orbital is indexed by a number along with its irrep. 
Each \blue{Hartree Fock} is represented by a coloured node in the figure, indexed by its irrep. Molecular orbitals with close energy are combined.
}
\label{fig:MO_Irreps}
\end{figure}

We show the irreducible representations (irreps) of orbitals in the selected active space for the molecules that are considered in this work. 
The Hamiltonians for various molecules are generated in the minimal basis set, known as the STO-3G basis, using the PySCF package~\cite{sun2018pyscf}. 
For the $\mol{H_2}$ molecule, we do not implement qubit reduction. 
The irreps of \blue{Hartree Fock} orbitals of the LiH and $\mol{F_2}$ molecules at different bond distances are plotted in \autoref{fig:MO_Irreps}. The point group of the LiH molecule is $C_{\infty v}$. Four \blue{Hartree Fock} orbitals of LiH are of irreps $a_1$ and the other two are of $e_1$. We freeze the lowest orbitals with $1a_1$ and $2a_1$ irreps, and remove the orbitals with $1e_1$ and $2e_1$ irreps. The point group of $\mol{F_2}$  is $D_{\infty h}$. We freeze the four lowest core orbitals with irreps $1a_1$, $2a_1$, $3a_1$ and $4a_1$. The other orbitals of $1e_1$, $2e_1$, $3e_1$, $4e_1$, $5a_1$ and $6a_1$ irreps are selected in the active space. 
% XXX explain the results of FIGURE.

With the selected orbitals in the active space, these molecular Hamiltonian can be expressed in the second quantisation as
\begin{equation}
\label{secondh}
\hat{H}=\sum\limits_{i,j} h_{ij}\hat{a}_i^{\dagger}\hat{a}_j
    	+\frac{1}{2}\sum\limits_{i,j,k,l} g_{ijkl}\hat{a}_i^{\dagger}\hat{a}_j^{\dagger}\hat{a}_{k} \hat{a}_{l},
\end{equation}
where ${\hat{a}_i}^\dag$ and $\hat{a}_i$ denote the fermionic creation and annihilation operators associated with $i$th molecular orbitals \sun{in the active space}, respectively. The coefficients $h_{ij}$ and $g_{ijkl}$ are the one- and two-electron integrals that can be evaluated classically {provided the spin-orbital basis function}.

Under the Jordan-Wignar transformation, the fermionic Hamiltonian can be mapped to a qubit form which is suitable for evaluation on a quantum processor.
The Jordan-Wigner transformation maps the fermionic operators $\hat a_j$ and $\hat a_j^{\dagger}$ on each fermionic mode to the qubit Pauli operators as 
\begin{equation}
    \hat a_{j} \mapsto \frac{1}{2}\left( X_{j}+i Y_{j}\right) \bigotimes_{i=1}^{j-1} Z_{i},~ \hat a_{j}^{\dagger} \mapsto \frac{1}{2}\left( X_{j} - i Y_{j}\right) \bigotimes_{i=1}^{j-1} Z_{i}
\end{equation}
with Pauli operators $X$, $Y$, $Z$ acting on the $j$th qubit.

\subsection{Initial state preparation}
\label{sec:init_state_prep}

As the spin Hamiltonian is constructed according to the $\textrm{SU}(2)$ symmetry, we consider the initial state or the reference state obeying this symmetry by assigning an equal number of electrons to the spin $\alpha$ and $\beta$ sectors. Also, as only closed-shell molecules are considered,  the reference state is restricted to a singlet state. In the first step of generating the molecular Hamiltonians, we adopt the restricted Hartree-Fock (RHF) method, where the ground-state wave function $\Psi_0$ is expressed by a single Slater determinant of molecular orbitals. The energy $E=\langle \Psi_0|H|\Psi_0\rangle $ is then minimised subject to orbital orthogonality. We first consider the initial state {to be a single reference state (i.e., a product state in the computational basis).} Suppose the total number of electrons is  $2 \eta$. The single reference state is constructed in a way that $\eta$ of electrons occupy the lowest energy spin orbitals in the spin $\alpha$ and $\beta$ sectors, respectively, which is referred to as the Hartree-Fock state $\ket{\Psi_\textrm{HF}}$. 

In this work, qubits are arranged in the following way: i) From left to right, each qubit corresponds to a spin-orbital that increases in its molecular-orbital energy; ii) The spin $\alpha$ and $\beta$ sector are arranged separately; the spin $\alpha$ is assigned first in the left part, then followed by the spin $\beta$ part. For example, the HF state of the $6$-qbuit, $2$-electron LiH molecule reads $\ket{100100}$, where the first and last $3$-qubit state $\ket{100}$ correspond to the spin $\alpha$ and $\beta$ parts, respectively.

% $\braket{ \Psi_{\mathrm{GS}} | \hat{H}|\Psi_{\mathrm{GS}}}$
% Given a basis set
% exact  ground state energy  and that calculated from HF state.

% Here the HF state represents a single reference state prepared based on the coefficients from classical HF method.
% Single slater determinant/single reference  

% This is in contrast to the HF energy ca

% HF energy at a mean-field level.
% $\braket{ \Psi_{\mathrm{HF}} | \hat{H}|\Psi_{\mathrm{HF}}}$.

Conventionally, the HF state is a good initial guess when the molecule is close to equilibrium. However, the closer the molecule approaches dissociation, the worse the HF state behaves due to the quasi-degeneration of different configurations. Consequently, the correlation energy, defined as the difference between the ground state energy and HF energy, becomes larger when the bond distance of a bi-atomic molecule stretches. To address this point, we consider the reference state to be a superposition of   multi-configurations,
\begin{equation}
    \ket{\Psi_\mathrm{ref}}=\sum_i c_i\ket{\Psi_i},
\end{equation}
where $\{\ket{\Psi_i}\}$ is the collection of configurations in the reference state with coefficients $\{c_i\}$. The configurations $\ket{\Psi_i}$ are chosen by identifying the fermionic modes that are related to the bond-breaking process and the size $|\{\ket{\Psi_i}\}|$ is a constant as they are only related to a constant number of fermionic modes. Thus, we can optimise efficiently on a classical device to determine the optimal coefficients. This approach is known as a multi-reference method~\cite{mcardle2018quantum} in the quantum chemistry literature. 
We apply the method for the LiH and $\mol{F_2}$ molecule and observe that higher accuracy is achieved compared to the HF state in the dissociation regime. 
As the two molecules we considered here are bi-atomic, the molecule can be seen as two individual atoms with negligible interaction in the large bond-distance limit. Also, the molecular orbitals are constructed as a linear combination of atomic orbitals in a symmetry-adapted way such that two atomic orbitals with the same symmetry are combined in a linear way to two molecular orbitals. For the LiH molecule, the lowest two molecular orbitals in the active space are established from atomic orbitals in this way. Hence, in the dissociation limit, one decouples the molecular orbitals to atomic ones by the equal-superposition state $\frac{1}{\sqrt{2}}\ket{100100}-\frac{1}{\sqrt{2}}\ket{010010}$ where the other configuration is selected as $\ket{010010}$ for the reference. The same logic follows for other molecules.
% and note that the last two orbitals around dissociation are of the same symmetry $A1$ as we depicted in the last section. Thus, the two configurations are chosen as $\ket{11111011110}$ and $\ket{11110111101}$.
As the molecule is in the intermediate state that interpolates between the equilibrium and the dissociation regimes, it is expected the reference state to be a superposition of the above-mentioned two configurations. We elaborate on the compilation of quantum circuits and optimise the coefficients of the reference state in the following.

In this work, we prepare the reference state as 
\begin{equation}
\ket{\Psi_\textrm{ref}} = \frac{1}{\sqrt{1+\beta^2}} (1-\beta \hat{a}^{\dagger}_{\eta+N/2}\hat{a}^{\dagger}_{\eta}   \hat{a}_{\eta+N/2-1}  \hat{a}_{\eta-1}) 
 \ket{\Psi_{\mathrm{HF}}},
 \label{eq:multi_ref_state}
\end{equation} 
which considers the excitations from the highest occupied molecular orbitals of the Hartree-Fock state, and the qubits are numbered from $0$.
When $\beta = 0$, the initial state reduces to the Hartree-Fock state.
The quantum circuit for preparing the multi-reference state  in \autoref{eq:multi_ref_state} is first creating a rotation along the y-axis on the qubit \blue{indexed by $(\eta-1)$} as 
\begin{equation}
    R_y \ket{0}^{\otimes N} =  \frac{1}{\sqrt{1+\beta^2}}  \ket{0}^{\otimes (\eta-1) } \otimes (  \ket{0} - \beta \ket{1}) \otimes \ket{0}^{\otimes N- \eta},
\end{equation}
followed by 3 CNOT gates  and single-qubit X gates, which can be explicitly expressed as
\begin{equation}
 \left( \bigotimes_{i=N/2}^{\eta+N/2-1} X_{i}  \bigotimes_{i=0}^{\eta-1} X_{i} \right)\mathrm{CNOT}_{(N/2+\eta-1) \rightarrow (N/2+\eta)} \mathrm{CNOT}_{\eta \rightarrow (N/2+\eta-1) }  
    \mathrm{CNOT}_{(\eta-1) \rightarrow \eta }.    
\end{equation}
 
\begin{figure}[htb]
\centering
\includegraphics[width=0.28\textwidth]{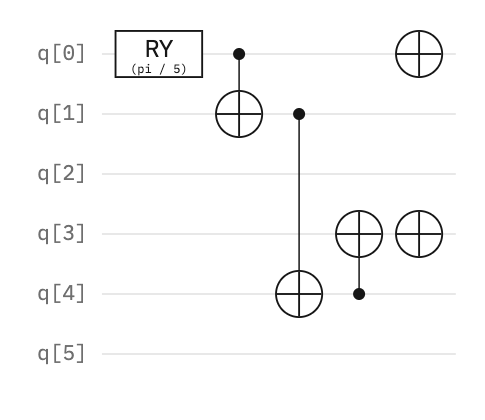}
\caption{
The initial state preparation for LiH molecule $ \ket{\Psi_0} =  
  \frac{1}{\sqrt{1+\beta^2}} ( \ket{100100} - \beta \ket{010010}) $. The  Pauli rotation along the y-axis prepares the state $\mathrm{R_y}\ket{0} = \frac{1}{\sqrt{1+\beta^2}}   (  \ket{0} - \beta \ket{1})$.
  }
\label{fig:LiH_multi_ref}
\end{figure}

% We note that this multi-reference state can be simulated classically.
It is easy to find that compared to the Hartree-Fock state, the state preparation for \autoref{eq:multi_ref_state} only introduces additional $3$ CNOT gates.
For instance, we prepare $ \ket{\Psi_0} =  
  \frac{1}{\sqrt{1+\beta^2}} ( \ket{100100} - \beta \ket{010010}) $ for the $\mol{LiH}$ molecule.
The circuit for initial state preparation is shown in \autoref{fig:LiH_multi_ref}.

\begin{figure}[htb]
\centering
\includegraphics[width=0.6\textwidth]{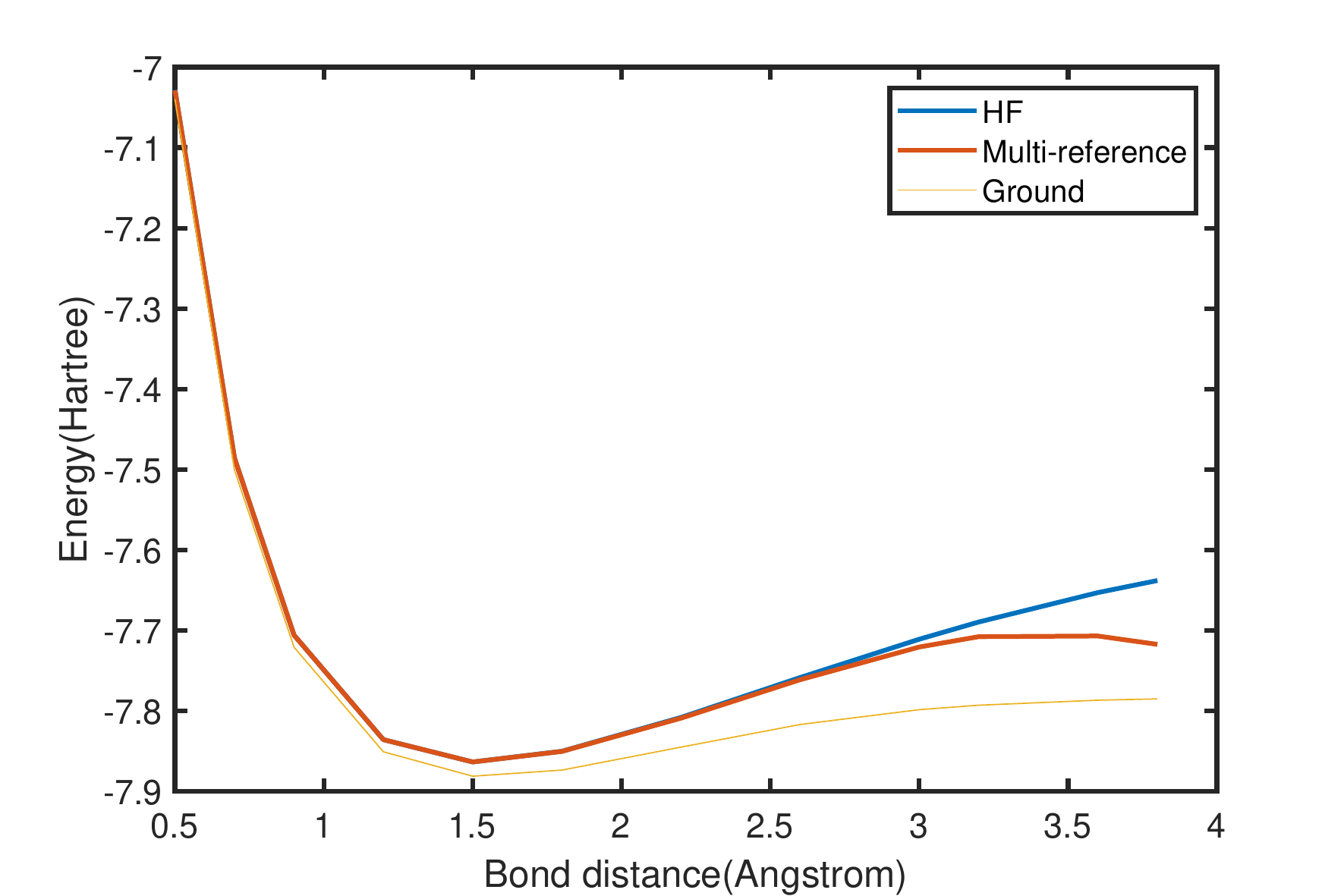}
\caption{
Comparison of different energies associated with the Hartree-Fock state, the multi-reference state, and the ground state. 
}
\label{fig:stabliser_energy}
\end{figure}

The parameter $\beta$ is determined by minimising the energy on the state, i.e.,   $\min_{\beta} \braket{ \Psi_0| \hat{H} | \Psi_0}$. It is easy to check that this minimisation problem can be computed efficiently.

A comparison of the initial state energy associated with the multi-reference state, Hartree-Fock state energy, and the true ground state energy is shown in  \autoref{fig:stabliser_energy}. The multi-reference state outperforms the Hartree-Fock state at a large bond distance and thus serves as a better initial state in this case.

\subsection{Quantum circuit ansatz design and compilation}
\label{sec:circuit_design}

\subsubsection{Unitary coupled cluster circuit ansatz}

% Selection of 9

% The ansatz derived from the unitary coupled-cluster~(UCC) method~\cite{hoffmann1988unitary,bartlett1989alternative,mcardle2020quantum,shen2017quantum,liu2021variational} is one of the most popular choices.

One crucial step for simulating large molecular systems by variational quantum algorithms is to design an appropriate quantum circuit that can effectively represent the ground state of the target system, and meanwhile, circuit depth is shallow. 
The unitary coupled cluster (UCC) ansatz has been widely used in the context of quantum computational chemistry~\cite{ucc-2}. 
As shown in the main text, the quantum state 
  can be constructed as
\begin{equation}
|\Psi(\Vec{\theta})\rangle=e^{\hat{T}({\Vec{\theta}})-\hat{T}^{\dagger}({\Vec{\theta}})}|\Psi_0\rangle,
\end{equation}
where $\hat{T}(\vec \theta)$ is the coupled-cluster excitations, and $|\Psi_0\rangle$ is the initial state.
The coupled-cluster excitations are usually truncated to single and double excitations, which are also called unitary coupled-cluster singles-and-doubles (UCCSD). 
The cluster operator truncated at single- and double-excitations has the form
\begin{equation}
\label{eq:UCCSD_operator}
    \hat T(\vec{\theta}) = \sum_{\substack{p\in vir\\i\in occ}} {\theta_{pi} \hat{T}_{pi}} + \sum_{\substack{p,q\in vir\\i,j\in occ}} {\theta_{pqji} \hat{T}_{pqji}},  
\end{equation}
where the one- and two-body terms are defined as
$
    %\label{eq8}
    \hat{T}_{pi} = \hat{a}^{\dagger}_{p} \hat{a}_{i}
$ and
$
    %\label{eq9}
    \hat{T}_{pqji} = \hat{a}^{\dagger}_{p} \hat{a}^{\dagger}_{q} \hat{a}_{j} \hat{a}_{i}
$, respectively.
As the molecular Hamiltonian conserves particles with definite spins, the ansatz is designed to conserve {this symmetry} as well, whereas the spin degree of freedom is abbreviated for simplicity.

To implement the VQE circuit of UCC ansatz on a quantum device, we apply the first-order Trotter-Suzuki expansion. and the circuit ansatz becomes 
\begin{equation}
\label{eq:UCCSD_Trotter}
%|\Psi(\Vec{\theta})\rangle  \approx 
|\Psi(\Vec{\theta}) \rangle = 
\prod_{\hat{T}_i \in \mathcal{O} } e^{\hat{T}_i - \hat{T}_i^{\dagger}}|\Psi_{0}\rangle,
\end{equation} 
where we denote the operator pool as $\mathcal{O} := \{ \hat{T}_i \}$ with $\hat{T}_i$ representing the excitation operators in \autoref{eq:UCCSD_operator}.
Here, it is worth remarking that the order of $\hat{T}_i$ will result in a different performance of the circuit.

As can be found from \autoref{eq:UCCSD_operator} and \autoref{eq:UCCSD_Trotter}, the original UCCSD  ansatz considering the single- and double-excitations is composed of a large number of operators scaling as $\mathcal{O} ( (N-\eta)^2 \eta^2)$ with electron number $\eta$, and the circuit depth with respect to the two-qubit entangling gates scales as $\mathcal{O} ( N(N-\eta)^2 \eta^2)$, As shown in the main text, the CZ gate counts for $\mol{H_2}$, $\mol{LiH}$, and $\mol{F_2}$ are 56, 280, 2920, respectively. Note that the multi-qubit rotation gate is complied into a  ladder of CZ gates and a single-qubit Pauli rotation gate, $\exp (-i \theta \hat \sigma/2)$, with $\hat \sigma = \{X,Y,Z \}$ being the single-qubit Pauli operator.
The high demand for the CZ gates constraints practical implementation of the original UCC ansatz.

\subsubsection{Point group symmetry constraint}

There has been considerable progress in reducing the number of operators in the circuit ansatz, such as using classical pre-calculation to screen out the terms that are small enough~\cite{romero2018strategies} or only choosing the important excitation operators~\cite{wecker2015progress}, for instance, paired double excitation operators in ansatz~\cite{lee2018generalized}. 
% \textcolor{blue}{However, it should be noted that the strategy may not be significant to reduce circuit depth when it applies to larger molecules when it is less symmetric.}
In the following, we discuss the circuit reduction strategy using the point group symmetry constraint.

To reduce the circuit depth without sacrificing accuracy, we consider  using the point group property to   reduce the UCC ansatz operators. 
Point group symmetry (spatial symmetry) is a fundamental property of   molecular systems that characterises  molecules' geometric and electronic structures.  The amplitude associated with the one- and two-electron excitations in both coupled cluster and unitary coupled cluster ansatz is proven to  be zero unless the corresponding term preserves the symmetry of the initial state~\cite{stanton1991direct}. 
This indicates that the expansion of the unitary coupled cluster wavefunction will only contain the terms with the same irreducible representation (irrep) as the reference wavefunction~\cite{cao2021larger}.
For the simple closed-shell molecules, the excitation operations should preserve the symmetry, and thus the irrep of the ground state  should be $\rm{A_g}$, which is a totally symmetric irreducible representation, $\rm{A_g}$, of which the characters are all 1.
This  applies to   all the cases considered in this work.
% For instance, if the initial state is the Hartree-Fock state, the excitation operations should preserve the symmetry, which is totally symmetric irreducible representation, $\rm{A_g}$, of which the characters are all 1.

% (~\cite{CCsym_1,CCsym_2}).

% Symmetric properties in molecular systems such as the particle number conservation ($U(1)$ symmetry), the fermionic parity conservation ($Z_2$ symmetry), are used to be restrictions in VQE~\cite{bravyi2017tapering, Gard2019, Greene-Diniz2021}. 

 % \subsection{Symmetry reduction}

% The excitation operator in UCC is allowed only when it preserves the symmetry of the reference Hartree-Fock state~\cite{cao2021larger}.
% So the targeted unitary coupled cluster solution $|\Psi\rangle$  constructed from $\left|\Psi_{0}\right\rangle$ is of the same irrep $\rm{A_g}$ as well, and the expansion of $|\Psi\rangle$ to Hartree-Fock Slater determinants will contain only the terms of $\rm{A_g}$ irrep. 

In our implementation, we filter out the operators that break the symmetry of the reference Hartree-Fock state when constructing the UCC ansatz.
Only the terms excite $\left|\Psi_{0}\right\rangle$ to $A_g$ states remain, and the rest of the terms in the exciting operator are excluded in constructing the cluster operator. 
As such, we select the operators $\hat{T}_i$ that satisfy the constraint of symmetry, that is,
\begin{equation}
\mathcal{O} := \left \{ \hat{T}_i \mid
D\left( e^{\hat{T}_i - \hat{T}_i^{\dagger}}\left|\Psi_{0}\right\rangle\right) = D\left(\left|\Psi_{0}\right\rangle\right)  \right \}
\label{eq:symmetry_constrait}
\end{equation}
%%%%%% The benefit of the method %%%%%
where $D$ is the irrep of the corresponding wavefunction. 

% Let us take molecule $\mol{F_2}$ as an example. 
The point group of $\mol{LiH}$ is $\rm{C_{\infty v}}$, which has three irreducible representations \textit{D}, i.e. $D{\in}\{A_g, E_{1x}, E_{1y}\}$.  
All the UCCSD excitation operators satisfy the symmetry constraint in \autoref{eq:symmetry_constrait}, and thus there is no circuit reduction for $\mol{LiH}$.
However, for large molecules, quantum circuits can be reduced without sacrificing accuracy.
The point group of $\mol{F_2}$ is $\rm{D_{\infty h}}$. $\mol{F_2}$ has \textcolor{blue}{six} irreducible representations \textit{D},  $D{\in}\{A_{1g}, E_{1gx}, E_{1gy},  A_{1u }, E_{1uy}, E_{1ux}\}$.   In total, there are ten single excitation operators and 25 double excitation operators. Only nine excitations have the same irreducible representation of $A_{1g}$ and need to be included in the circuit ansatz. 

However, we note that the strategy based on Point group symmetry constraints may not yield significant reductions in circuit depth for larger molecules which tend to be less symmetric. Below, we introduce a circuit reduction strategy by selecting the dominant excitation operations.

\subsubsection{Selection of excitation operators}

The above circuit reduction exploits intrinsic information of the molecules.
We can further construct a more compact quantum circuit  by selecting operators that contribute dominantly to the energy decrease from a reference state, which has been proposed in \cite{fan2021circuit}.   
We start from the initial state $\ket{\Psi_0}$, which is a Hartree-Fock state or a multi-reference state. The excitation operators are selected from the  UCCSD operator pool $\mathcal{O}$  in \autoref{eq:UCCSD_operator}  composed of the possible single- and double-excitation operators $\hat{T}_{pi}$ and $\hat{T}_{pqji}$ that satisfy the symmetry constraints.

This relies on the fact that the quantum circuit with a single operator can be efficiently simulated.  
The quantum state under one excitation  becomes 
\begin{equation}
    \ket{\Psi} = e^{ \theta ( \hat{T_i}-\hat{T_i}^\dagger) } \ket{\Psi_0},
    \label{eq:ESVQE_state}
\end{equation}
where $\hat T_i \in \mathcal{O} $ is a one-body or two-body excitation operator in the operator pool $\mathcal{O}$ with the form   $\hat{T}_i =\hat a_p^{\dagger } \hat a_i$ or $\hat a_p^{\dagger } \hat a_q^{\dagger } \hat a_j \hat a_i$.
The VQE optimisation iteration is carried for each operator $\hat{T}_{i} \in \mathcal{O}$ for $E_i$.
The minimised energy is obtained by the variational principle 
\begin{equation}
    E_i = \min_{\theta_i} \braket{\Psi_{0}|e^{-\theta_i (\hat T_i - \hat T_i^\dagger)}\hat{H}e^{\theta_i (\hat T_i - \hat T_i^\dagger)}|\Psi_{0}}. 
    \label{eq:minimisation_single_Xop}
\end{equation}
The importance of the operator is evaluated by calculating the energy difference with respect to the initial state energy, $\Delta E_{i} = |E_0- E_{i}|$, with $E_0 =   \braket{\Psi_{0}| \hat{H} |\Psi_{0}} $.
The list $\{(\Delta E_{i}, \hat{T}_{i})\}_{\textrm{sorted}}$ is recorded. The operators are sorted according to the energy difference in descending order. The  operators with contributions above a threshold $|\Delta E_{i}| > \varepsilon_{\mathrm{thres}}$ are picked out and used to perform the VQE optimisation. In what follows, we show that this selection procedure is classically efficient.

% \subsubsection{YUKUN complete}

The circuit operator $e^{\theta (\hat{T}_i-\hat{T}_i^{\dagger})} $ can be expanded by the Taylor series as
\begin{equation}
\begin{aligned}
e^{\theta (\hat{T}_i-\hat{T}_i^{\dagger})} &= \sum_{n = 0}^{\infty} \frac{\theta^n (\hat{T}_i-\hat{T}_i^\dagger )^n}{  n!} \\
&= I + \sum_{n = 1}^{\infty} (-1)^n \theta^{2n} \frac{ (\hat{T}_i \hat{T}_i^{\dagger})^n + (\hat{T}_i^{\dagger} \hat{T}_i )^{n}}{  (2n)!}  + \sum_{n = 0}^{\infty} (-1)^n \theta^{2n+1} \frac{ -\hat{T}_i^{\dagger} (\hat{T}_i \hat{T}_i^{\dagger})^n + \hat{T}_i (\hat{T}_i^{\dagger} \hat{T}_i )^{n}}{  (2n+1)!}  \\
&=  \cos( \theta \sqrt{\hat{T}_i \hat{T}_i^{\dagger}}) + \cos(\theta \sqrt{\hat{T}_i^{\dagger} \hat{T}_i}) - 
\frac{\hat{T}_i^{\dagger} }{ \sqrt{\hat{T}_i \hat{T}_i^{\dagger}} }  \sin( \theta\sqrt{\hat{T}_i \hat{T}_i^{\dagger}}) + \frac{\hat{T}_i }{\sqrt{\hat{T}_i^{\dagger} \hat{T}_i}
} \sin(\theta \sqrt{\hat{T}_i^{\dagger} \hat{T}_i}) 
\label{eq:UCCSD_exponential}
\end{aligned}
\end{equation}
where we have used $\hat T_i^2 =  (\hat T_i^{\dagger})^2 = 0$.
From \autoref{eq:UCCSD_exponential}, the circuit operator comprises four terms only, and each term can be simulated classically~\cite{2021FQE}.
Let us take $\cos( \theta\sqrt{\hat{T}_i \hat{T}_i^{\dagger}}) $ for example.
Without loss of generality, $\hat{T}_i$ can be expressed as  $\hat{T}_i = \hat{a}^{\dagger}_3\hat{a}^{\dagger}_1 \hat{a}_2\hat{a}_0$, and then
\begin{equation}
    \begin{aligned}
\hat{T}_i \hat{T}_i^{\dagger} & =\hat{a}_3^{\dagger} \hat{a}_1^{\dagger} \hat{a}_2 \hat{a}_0 \hat{a}_0^{\dagger} \hat{a}_2^{\dagger} \hat{a}_1 \hat{a}_3  =  \hat{n}_3 \hat{n}_1\left(1-\hat{n}_2\right)\left(1-\hat{n}_0\right).
\end{aligned}
\end{equation}
Since the action of the number operator gives 0 or 1 and $\hat{T}_i \hat{T}_i^{\dagger}$ is idempotent,  the square root of $\hat{T}_i \hat{T}_i^{\dagger}$ can be defined as
\begin{equation}
    \sqrt{\hat{T}_i \hat{T}_i^{\dagger}} \equiv  \hat{n}_3 \hat{n}_1\left(1-\hat{n}_2\right)\left(1-\hat{n}_0\right),
\end{equation}
and we have
\begin{equation}
    \cos(\theta \sqrt{\hat{T}_i \hat{T}_i^{\dagger}}) =   \cos\theta  \sqrt{\hat{T}_i \hat{T}_i^{\dagger}} 
    =  \cos\theta  \hat{n}_3 \hat{n}_1\left(1-\hat{n}_2\right)\left(1-\hat{n}_0\right).
\end{equation}
The energy expectation $ \langle \Psi_0| e^{-\theta(\hat T_i-\hat T_i^\dagger)} \hat{H} e^{\theta(\hat T_i-\hat T_i ^\dagger)} | \Psi_0 \rangle $ contains   $16$ terms, each of which can be simulated classically. Therefore, the optimisation process described by \autoref{eq:minimisation_single_Xop} can be simulated classically.

The above discussion holds for any operators, and hence their actions on any states.
The expression can be further simplified when we consider the action of the operator on the Hartree-Fock state. 
Define $\hat G := -i (\hat T_i - \hat T_i^{\dagger})$. It is easy to check that $G$ is Hermitian $\hat G  = G^{\dagger}$. For $T$ contains only one excitation, the action of $G$ on the HF state $\ket{\Psi_{\mathrm{HF}}}$ is
\begin{equation}
   \hat G^2 \ket{\Psi_{\mathrm{HF}}} = - (\hat T_i^2 + (\hat T_i^{\dagger})^2 - \hat T_i  \hat T_i^{\dagger}  -  \hat T_i^{\dagger} \hat T_i)   \ket{\Psi_{\mathrm{HF}}} = \ket{\Psi_{\mathrm{HF}}}.
\end{equation}
Therefore, we have
\begin{equation}
    \ket{\Psi} =  e^{i\theta  \hat G} \ket{\Psi_{\mathrm{HF}}} = (\cos  \theta \hat I + i \sin \theta \hat G) \ket{\Psi_{\mathrm{HF}}}.
\end{equation}
The energy can now be calculated as
\begin{equation}
    \langle \Psi_{\mathrm{HF}}| e^{-\theta(\hat T_i-\hat T_i^\dagger)} \hat{H} e^{\theta(\hat T_i-\hat T_i ^\dagger)} | \Psi_{\mathrm{HF}} \rangle = E_{0} \cos ^2 \theta + \sin^2 \theta \braket{\Psi_{\mathrm{HF}} | \hat{G} \hat{H} \hat{G} | \Psi_{\mathrm{HF}}}
\end{equation}
with $E_0 = \bra{\Psi_{\mathrm{HF}}}\hat H \ket{\Psi_{\mathrm{HF}}}$.
It is straightforward to see that this optimisation process with one excitation operator acting on the Hartree-Fock state can be simulated classically.

Suppose we have selected the operator set $\{  {\hat{T}}_i \}$ in its qubit form. We map   ${\hat{T}}_i$ to its qubit form under the Jordon-Wigner transformation. Since each term is symmetric and the contribution of each term  is the same, we only choose one term in the qubit form to further reduce the gate count and hence circuit depth.
In the next section, we display the quantum circuits for the three molecules.

% This ansatz can be regarded the heuristic hardware efficient ansatz that mimics the UCCSD ansatz.  
% The total gate count is reduced by two orders of magnitude.  

\subsubsection{Quantum circuit compilation}
\label{sec:circuit_compilation}

In this section, we present the dominant operators and quantum  circuits for the three molecules after circuit compilation. 
The dominant excitation operators for $\mol{H_2}$ are 
$   a_3^{\dagger }a_{1}^{\dagger} a_2  a_0 
$.
The dominant excitation operators for $\mol{LiH}$ are 
$ a_5^{\dagger} a_1^{\dagger} a_3   a_0  $,
$a_4^{\dagger} a_2^{\dagger} a_3   a_0 $, and
$a_5^{\dagger}  a_2^{\dagger}  a_3  a_0$.
The dominant excitation operators for $\mol{F_2}$ are 
$ a_{11}^{\dagger} a_5^{\dagger}  a_6  a_0  $,
$a_{11}^{\dagger} a_5^{\dagger}  a_7    a_1 $,
$a_{11}^{\dagger} a_5^{\dagger}  a_8    a_2$,
$a_{11}^{\dagger} a_5^{\dagger}  a_9    a_3 $, and
$a_{11}^{\dagger} a_5^{\dagger}  a_{10}   a_4$.

Next, we optimise the quantum circuits with respect to  the circuit depth of CZ gates.
We exploit the 2D geometry of   the qubit layout of the quantum processor to encode the spin-ups and spin-downs, and we optimise the depth by  parallelisation of CZ gates.
The parameterised quantum circuits for $\mol{H_2}$, $\mol{LiH}$ and $\mol{F_2}$ are
\begin{equation}
    U_{\mol{H_2}} = \exp(i \theta X_0Y_1X_2X_3)\exp(-i \theta Y_0 X_1X_2X_3) ,
\end{equation}
\begin{equation}
    U_{\mol{LiH}} =  \exp(-i \theta_3 Y_0Z_1X_2 X_3 Z_4 X_5)    \exp(-i \theta_2 Y_0 Z_1X_2X_3X_4) \exp(-i \theta_1 Y_0 X_1 X_3Z_4X_5),
\end{equation}
and 
\begin{equation}
\begin{aligned}
    U_{\mol{F_2}} =  &\exp(-i \theta_5 X_4 X_5 Y_{10} X_{11}) \exp(-i \theta_4 Y_3Z_4 X_5 X_9Z_{10} X_{11})  \exp(-i \theta_3 Y_2Z_3Z_4 X_5 X_8Z_9Z_{10} X_{11}) \\  & \exp(-i \theta_2 Y_1Z_2Z_3Z_4 X_5 X_7Z_8Z_9Z_{10} X_{11})  \exp(-i \theta_1 Y_0Z_1Z_2Z_3Z_4 X_5 X_6Z_7Z_8Z_9Z_{10} X_{11}),
\end{aligned}
\end{equation}
        respectively.

\begin{figure}[htb]
\centering
\includegraphics[width=0.6\textwidth]{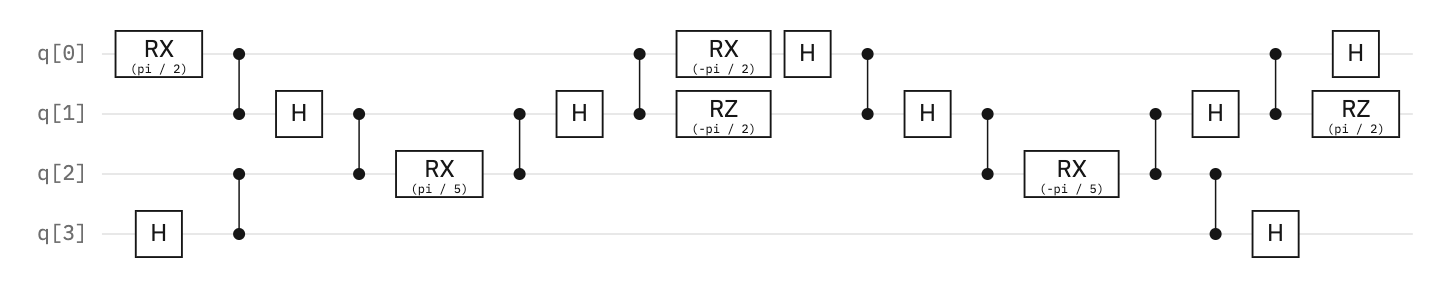}
\caption{
The compiled quantum circuit ansatz for $\mol{H_2}$.  
}
\label{fig:circuit_H2}
\end{figure}

\begin{figure}[htb]
\centering
\includegraphics[width=0.9\textwidth]{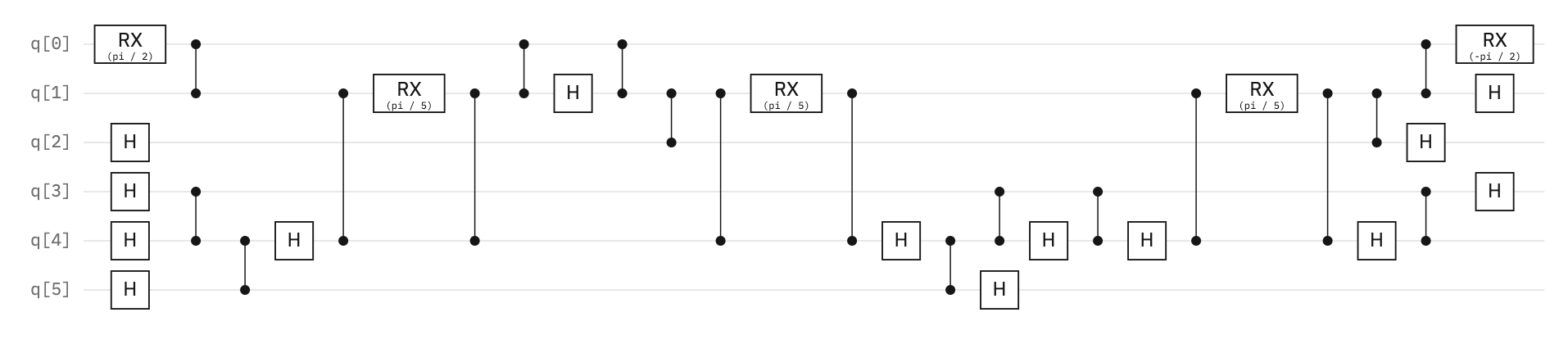}
\caption{
The compiled quantum circuit ansatz for LiH 
}
\label{fig:circuit_LiH}
\end{figure}

\begin{figure}[htb]
\centering
\includegraphics[width=1.0\textwidth]{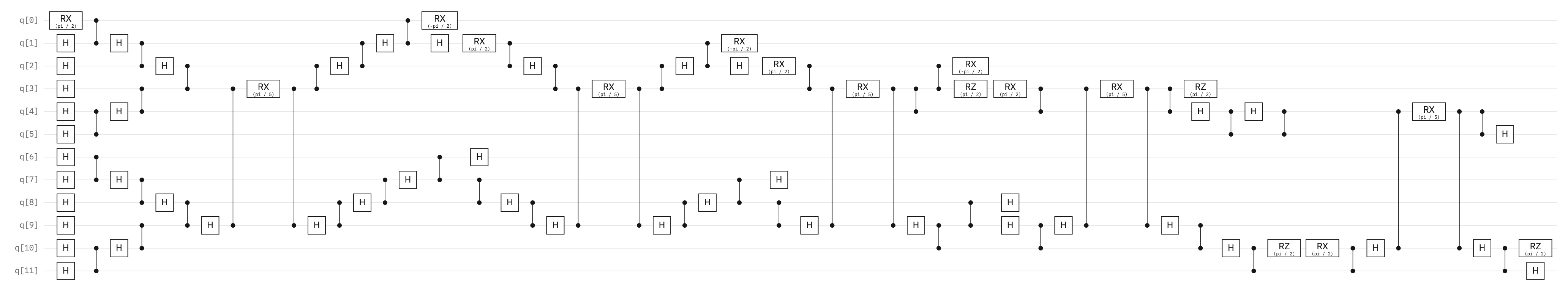}
\caption{
The compiled quantum circuit ansatz for $\mol{F_2}$ 
}
\label{fig:circuit_F2}
\end{figure}

The compiled quantum circuits (excluding the initial state preparation and measurement) for $\mol{H_2}$, LiH, and $\mol{F_2}$ are shown in \autoref{fig:circuit_H2}, \autoref{fig:circuit_LiH}, and \autoref{fig:circuit_F2}, respectively.
Here, the parameters are chosen as $\pm \pi/5$ as an example.
The circuit for initial preparation has been discussed in \autoref{sec:init_state_prep}, and the circuit for LiH is shown in \autoref{fig:LiH_multi_ref}.

\subsection{Optimisation by analytic stochastic  gradient descent}
\label{sec:optimisation_SGD}

% In VQE, the ground state is approximated by solving $\mathcal{L} =  \min_{\vec\theta}\braket{\Psi_0 | U(\vec \theta )H U(\vec \theta ) \Psi_0}_{\vec\theta}$ using a  classical optimiser.

Now, we show how to  analytically estimate the gradient of expectation values of the observable $\hat{O}$.
In VQE, the ground state is approximated by solving $\mathcal{L}(\vec \theta ) =  \min_{\vec\theta}\braket{\Psi_0 | U^\dagger(\vec \theta ) \hat{O} U(\vec \theta ) \Psi_0}_{\vec\theta}$ with $\hat{O} = \hat{H}$ using a  classical optimiser.
The circuit that is constructed using the methods developed in \autoref{sec:circuit_design} takes the form of 
\begin{equation}
    U(\vec \theta) = \prod_j \prod_{ l_j }^{L_j} V_{j, l_j}(\vec \theta_j),
\end{equation}
where each gate operation $V_{j, l_j}$ is a multi-qubit Pauli rotation $V_{j, l_j} = \exp( -i \vec \theta_j P_{j, l_j}/2)$ with $ P_{j, l_j} \in \{ I,X,Y,Z \}^{\otimes N}$ being the tensor product of single-qubit Pauli operators over $N$ qubits.
Prior works usually consider $L_j = 1$, that is, each parameter is associated with only one term. However, in   quantum chemistry experiments, the parameters are often associated with many terms.
The objective is to get the first-order derivative with respect to the $j$th parameter, i.e., $g_j (\vec \theta ) = {\partial \mathcal{L}(\vec \theta ) } / { \partial \vec \theta_j}$.
Using the Leibniz rule, the gradient is given by
\begin{equation}
 g_j = \sum_{l_j}  \braket{\Phi_{j,l_j} | \frac{ \partial \bar{V}_{j,l_j} (\vec \theta_{j, l_j})}{ \partial {\vec \theta_{j, l_j}} } | \Phi_{j,l_j} }
\end{equation}
where  we introduce $\vec \theta_{j,l_j}$  to represent the parameter associated with the operator $V_{j,l_j}$ with $\vec \theta_{j,l_j} = \vec \theta_j$, and  we have
\begin{equation}
   \ket{\Phi_{j,l_j} }  = \prod_{k < l_j} V_{j, k} (\vec \theta_{j,k}) \prod_{k < j} \prod_{l_k} V_{k, l_k} (\vec \theta_{k,l_k}) \ket{\Psi_0}
\end{equation}
and 
\begin{equation}
\bar{V}_{j,l_j} (\vec \theta_{j,l_j}) =V_{j, l_j}  (\vec \theta_{j, l_j}) ^{\dagger}\Bar{O} V_{j, l_j}  (\vec \theta_{j, l_j})
\end{equation}
with
\begin{equation}
\Bar{O}=  \left(\prod_{k > j} \prod_{l_k} V_{k, l_k} (\vec \theta_{k, l_k}) \prod_{k > l_j }   V_{j, k}  (\vec \theta_{j, k}) \right)^{\dagger} O \prod_{k > j} \prod_{l_k} V_{k, l_k} (\vec \theta_{k, l_k}) \prod_{k > l_j }   V_{j, k}  (\vec \theta_{j, k}).
\label{eq:V_bar_j_lj}
\end{equation}
% \begin{equation}
% \bar{V}_{j,l_j} (\vec \theta_{j,l_j}) =  \left(\prod_{k > j} \prod_{l_k} V_{k, l_k} (\vec \theta_{k, l_k}) \prod_{k \geq l_j }   V_{j, k}  (\vec \theta_{j, k}) \right)^{\dagger} O \prod_{k > j} \prod_{l_k} V_{k, l_k} (\vec \theta_{k, l_k}) \prod_{k \geq l_j }   V_{j, k}  (\vec \theta_{j, k})
% \end{equation}
Since $P_{j, l_j}^2 = I$, each of the gate can be written as \begin{equation}
     V_{j, l_j} (\vec \theta_{j, l_j}) = \cos( \vec \theta_{j, l_j} /2 ) I - i \sin  (\vec \theta_{j, l_j}/2) P_{j, l_j},
\end{equation}
and thus $ \bar{V}_{j,l_j} (\vec \theta_{j, l_j})$  can   be written as
\begin{equation}
   \bar{V}_{j,l_j} (\vec \theta_{j, l_j}) = A + B \cos(\vec \theta_{j, l_j}) + C \sin (\vec \theta_{j, l_j})
   \label{eq:V_theta_A_B_C}
\end{equation}
where $A$, $B$, and $C$ are operators that rely on ${V}_{j,l_j}$ and $\Bar{O}$ yet independent of the parameters $\vec \theta_{j, l_j}$
From \autoref{eq:V_theta_A_B_C}, we can show that the summand in the gradient is related to a parameter-shifted $\bar{V}_{j,l_j} (\vec \theta_{j, l_j} \pm s)$ with a shift  $s$, which is given by
\begin{equation}
    \frac{ \partial \bar{V}_{j,l_j} (\vec \theta_{j, l_j})} {\vec \theta_{j, l_j}}=  \frac{\bar{V}_{j,l_j} (\vec \theta_{j, l_j} + s) - \bar{V}_{j,l_j} (\vec \theta_{j, l_j} -s )}{ 2 \sin s}.
\end{equation}
The gradient is thus given by
\begin{equation}
\begin{aligned}
	g_j &= \frac{\sum_{l_j} \left(  \braket{ \Phi_{j,l_j}  | \bar{V}_{j,l_j} (\vec \theta +  s \vec e_{j, l_j})   | \Phi_{j,l_j}} - \braket{ \Phi_{j,l_j}  | \bar{V}_{j,l_j} (\vec \theta -  s \vec e_{j, l_j})   | \Phi_{j,l_j}} \right)} {2 \sin(s)} \\
    &= \frac{\sum_{l_j}  \mathcal{L} (\vec \theta +  s \vec e_{j, l_j})  - \mathcal{L}  (\vec \theta -  s \vec e_{j, l_j}) }{2 \sin(s)}
    \label{eq:gradient_g_j}
\end{aligned}
\end{equation}
with unit vector $\vec e$.
\autoref{eq:gradient_g_j} indicates that we can get the gradient of the energy expectation analytically by measuring the energy expectation by shifting   parameters, termed the parameter-shift rule~\cite{mitarai2018quantum,mari2021estimating}. 
% Note that the gradient in \autoref{eq:gradient_g_j} is different from the one derived by finite-difference methods, in which the denominator is $s$ and we

In the experiment, the gradient is estimated by measuring the two parameter-shifted quantities. Due to statistical fluctuation, the gradient can only be estimated up to certain precision.
The estimator of $g_j$ is given by
\begin{equation}
    \hat g_j =\frac{ \sum_{l_j}  \hat {\mathcal{L}} (\vec \theta +  s \vec e_{j, l_j})  - \hat {\mathcal{L}} (\vec \theta -  s \vec e_{j, l_j}) }{2 \sin(s)}.
\end{equation}
The estimation by the parameter-shift rule has an unbiased estimation of the gradient as 
\begin{equation}
	\mathbb E \hat g_j = g_j.
\end{equation}
In what follows, we discuss the variance of the estimator.
Assume that energy estimation has the same variance as \begin{equation}
	\var [\hat {\mathcal{L}}] = \var [ \hat {\mathcal{L}} (\vec \theta +  s \vec e_{j, l_j}) ],~\forall j, ~\forall l_j.
\end{equation} 
Then the variance of the gradient estimation is given by
\begin{equation}
	\var [\hat g_j ] =  \frac{L_j}{ \sin^2(s)} \var [\mathcal{\hat L}] \geq L_j  \var [\mathcal{\hat L}] . 
	\label{eq:var_g_j}
\end{equation}
It is straightforward  that the variance of the estimator is minimised when $s = \pi/2$.

The  gradient estimation by the parameter-shift rule shows a clear advantage over the finite difference method where the gradient is estimated as 
\begin{equation}
	\hat g_j^{\textrm{FD}} = \frac{ \sum_{l_j}  \hat {\mathcal{L}} (\vec \theta +  \delta  \vec e_{j, l_j})  - \hat {\mathcal{L}} (\vec \theta -  \delta  \vec e_{j, l_j}) }{2  \delta }.
\end{equation}
The estimation is unbiased when the step size approaches zero, $\delta  \rightarrow 0$. To make the estimation close to the true gradient, the step $\delta $ should be taken to be a small number. However, the variance of the estimation is
\begin{equation}
	\var [\hat g_j^{\textrm{FD}}] = \frac{L_j}{ \delta^2} \var [\mathcal{\hat L}],
\end{equation}
which can be much larger than the variance in \autoref{eq:var_g_j}.

To make it more efficient, we adopt a stochastic gradient descent strategy.
For the $k$th iteration with parameters $\vec \theta^{k}$, 
we obtain the gradient element $g_j^k(\vec \theta^{k})$ by the parameter shift rule
\begin{equation}
    \hat g_j^{\textrm{SGD}, k}(\vec \theta^{k}) = \frac{ \hat{\mathcal{L}}(\vec\theta^+)-\ \hat{\mathcal{L}}(\vec\theta^-)}{2}
    \label{eq:gradient_SGD}
\end{equation}
where we define
\begin{equation}
	 \hat{\mathcal{L}}(\vec\theta^{\pm}) = \sum_{l_j} \hat{\mathcal{L}} (\vec \theta \pm  \pi \Delta_{j} \vec e_{j, l_j} /2).  
	 \end{equation}
  \label{eq:gradient_SGD_L_pm_def}

% , and
% 
% we measure the expectation value of the Hamiltonian for two parameters $\braket{H}_{\vec\theta^+}$ and $\braket{H}_{\vec\theta^-}$.
% with 
% the learning rate $\alpha^k$.

% and the variance of gradient estimation is minimised  when $c^k = \frac{\pi}{2}$. 
In experiments,    the parameters to be optimised  are  chosen stochastically   by $\Delta^k_j = 0$ or $1$ and satisfying $|\Delta^k| >  N_{p}/2$ to reduce the sample cost for the gradients. 
% with $N_p  $ the number of parameters and set $\Delta_0$ around $N_p / 2$ in our experiment.
We update the parameters by
$
    \vec \theta^{k+1}= \vec \theta^{k} + \alpha^{{k}} \vec g^k(\vec \theta^{k})
$ with an appropriate learning rate $\alpha^k$.
The estimation of the gradient is also unbiased  
\begin{equation}
	\mathbb E \hat g_j^{\textrm{SGD}, k}(\vec \theta^{k}) = g_j^k(\vec \theta^{k}) .
\end{equation}

\subsection{Quantum state measurement and classical post-processing}
\label{sec:OGM}

The power of variational quantum algorithms is at the expense of measurements. This becomes a severe problem in  molecular systems, in which the number of measurements at each iteration scales up to $\mathcal{O} (N^4)$ with the problem size $N$.
However, as many of the observables have small coefficients and are qubit-wise compatible (hence could be  measured simultaneously), we could exploit more efficient measurement schemes to alleviate the measurement cost. 
In this section, we discuss how to measure the observable efficiently by exploiting the qubit-wise compatibility of the target observable.

\subsubsection{Overlapped grouping measurement}
The observables in our experiment include the Hamiltonian $\hat{H}$, {the projected Hamiltonian in the symmetry verified space}  $\Pi \hat{H } \Pi$ with the projector $\Pi$, and Hamiltonian moments $\hat{H }^k$ for $k \geq 2$.
The target observable $\Omat$ can be decomposed into the Pauli basis as 
\begin{equation}
    \Omat = \sum_l \alpha_l \Olmat,
    \label{eq:observ_decomp}
\end{equation}
with $\Olmat \in \{I,X,Y,Z\}^{\otimes n}$ being the tensor product of single-qubit Pauli operators.
The main idea is that if two observables $\Omat_l$ and $\Omat_{l'}$ are qubit-wise compatible, we can measure the two observables simultaneously in one measurement basis. 
Formally, for a multi-qubit Pauli operator $\Qmat = \otimes_{i=1}^n Q_i$ with $Q_i \in \{I, X,Y,Z\}$ being a single-qubit Pauli operator acting on the $i$th qubit, its expectation value can be obtained by measurements in any Pauli basis $\Pmat=\otimes_{i=1}^n P_i$ whenever $Q_i = P_i$ or $Q_i = I$ for any $i$. We refer to this circumstance as $\Pmat$ hits $\Qmat$ and denote by $\Qmat\triangleright\Pmat$.  We say that $\Omat_l$ and $\Omat_{l'}$ are compatible with each other when $\Omat_l \triangleright \Pmat$ and $\Omat_{l'} \triangleright \Pmat$, i.e., they are hit by the same basis $\Pmat$.
 
To estimate $\tr( \rho \Omat) $ for an $n$-qubit unknown quantum state $\rho$, we randomly select the measurement basis $\Pmat$ and generate an estimation of  $\tr( \rho O) $ by measuring in the measurement basis $\Pmat$  with probability $\Kcal(\Pmat)$.
An estimator of the target observable $\Omat$ is expressed as
\begin{equation}
\hat{\omat}(\Pmat) = \sum_l \alpha_l  f(\Pmat,\Olmat,\Kcal )\mu(\Pmat, \textrm{supp}(\Olmat))
\label{eq:estimator_OGM}
\end{equation}
where $\mu(\Pmat,\textrm{supp}(\Olmat)) = \prod_{i\in \textrm{supp}(\Olmat)}\mu(P_i)$ with $\mu(P_i)$ being the single-shot outcome of measurement $P_i$ on the $i$th qubit, $\textrm{supp}(\Qmat)=\cbra{i|Q_i\ne I}$.
Suppose we have chosen the measurement basis set $\{\Pmat_j)\}$ and the associated
probability distribution $\{\Kcal(\Pmat_j)\}$, the function $f$ is defined as
\begin{equation}
    f(\Pmat,\Qmat, {\Kcal}) =  \chi(\Qmat)^{-1} \delta_{\Qmat \triangleright \Pmat},
\label{eq:DefFuncOGM}
\end{equation}
where $\chi(\Qmat) = \sum_{\Pmat:\Qmat \triangleright \Pmat} \Kcal(\Pmat)$ represents the probability that $\Qmat$ is effectively measured with the measurement basis $\Pmat$.
\autoref{eq:estimator_OGM} gives an unbiased estimation
% $\mathbb{E}[\hat{\omat}] = \tr(\Omat \rho)$, 
\begin{equation}\label{eq:unbiasedestimation}
    \mathbb{E}[\hat{\omat}] = \tr(\Omat \rho), 
\end{equation}
where the average is over the probability distribution $\Kcal(\Pmat)$.  
The variance of the estimator in \autoref{eq:estimator_OGM} for a single sample could be calculated by the definition as
\begin{equation}
\textrm{Var}[\hat{\omat}] = \mathbb{E}_{\Pmat} \sum_{ l,l' } \alpha_{l} \alpha_{l'} \tr(\rho \Olmat \Olpmat)  f(\Pmat, \Olmat, \Kcal) f(\Pmat, \Olpmat, \Kcal) - \tr(\rho \Omat)^2,
\label{eq:Var_general}
\end{equation}
where we use the equality $\mathbb{E}_{\mu(\Pmat)} \mu(\Pmat, \operatorname{supp}(\Olmat)) \mu(\Pmat, \operatorname{supp}(\Olpmat))=\mathbb{E}_{\mu(\Pmat)} \mu(\Pmat, \operatorname{supp}(\Olmat\Olpmat) )=\tr(\rho \Olmat \Olpmat)$. The detailed proof can be found in Refs.~\cite{wu2021overlapped,hadfield2022measurements}.

% We show in Algorithm~\autoref{alg_overlapSet} our strategy to determine the overlapped sets $\bm e_1, \ldots, \bm e_s$ and the associated probability $\Kcal_j:=\Kcal(\Pmat_j)$.
% %{The main idea is that we give priority to observables with larger absolute weights. Under the premise of covering all the objective observables, we only account for sets with large total weights.}
% The main idea is that we give priority to observables with larger absolute weights.

% The following questions are to determine the measurement basis set and the optimised probability distribution.
We assign the observables $\{ \Omat_l \}$ into overlapped grouping sets, 
\begin{equation}
    \mathcal{S} = \{\bm e_1, \ldots, \bm e_s\},
\end{equation}
where each set $\bm e_j$ is composed of compatible observables that can be measured in  basis $\Pmat_j$ and it satisfies 
\begin{equation}
    \cup_j \bm e_j = \mathcal O,~\bm e_j = \{ \Qmat \triangleright \Pmat_j \},\forall \Qmat\in \bm e_j.
\end{equation}
The strategy to determine the  overlapped grouping sets is that we add an observable $\Omat_l$ that has not been accessed into a new set and all compatible observables  into this set. 
Observables with larger absolute weights are given high  priority    since they have more contributions to the variance, as can be found from \autoref{eq:Var_general}. 
The algorithm generates the measurements $\{\Pmat\}$ with non-optimised probabilities $\{\Kcal\}$.

\subsubsection{Optimisation process}

% we first generate the overlapped groups $\Scal$, the measurements $P_1, \ldots, P_s$, and initialized distribution $\vec\Kcal=(\Kcal_1,\ldots, \Kcal_s)$ with Algorithm~$\rm{alg_overlapSet}$.
% To 
% to speed up the optimisation over $\vec\Kcal$,  

The probability distribution associated with set $\bm e_j$ by default is the weight of the set.
We choose to use the strategy proposed in \cite{wu2021overlapped} to optimise the probability distribution $\{\Kcal\}$ such that   the  variance of the estimator can be further minimised.    We consider a diagonal approximation of {$ \var\pbra{\hat{ \omat }}$}, which is explicitly expressed as
\begin{equation}
    l(\vec \Kcal)= \sum_{l} \frac{\alpha_l^2}{\chi({\Olmat, \vec \Kcal})},
    \label{eq:diag_var}
\end{equation}
where $\chi(\Olmat, \vec \Kcal) = \sum_{\Pmat:\Olmat \triangleright \Pmat} \Kcal(\Pmat)$ and $\vec \Kcal := \pbra{\Kcal_1, \cdots, \Kcal_s}$ represents all the corresponding probabilities.  
% \comments{ 
% There are several advantages of using the diagonal approximation $l(\vec \Kcal)$ instead of the actual variance~---~(1)   independence of the quantum state, (2) fast classical evaluation, (3) including dominant contribution to the variance since $\tr\pbra{\rho \Olmat\Omat_k}< \tr\pbra{\rho \Olmat\Olmat} = 1$ when $j\ne k$.  Therefore, we could instead regard $l(\vec \Kcal)$ as the cost function and minimize it by optimising over $\vec \Kcal$. }
% From the expression of $l\pbra{\vec \Kcal}$ in \autoref{eq:diag_var}, we see the cost function is not convex in $\vec \Kcal$, and hence there is no closed minimum solution.
% An estimation could be generated by searching for a local minimum solution of the cost function in \autoref{eq:diag_var}.  
% For the optimisation process of the OGM method, 
% we will further speed it up by adaptively deleting the groups that have very small initial probabilities 
% % and thus fewer contributions for the estimator, 
% until the cost function stops decreasing with the disturbance.
In our experiment, since changing to another measurement basis requires a certain initialization time,  we attempt to reduce the total number of basis sets to as few as possible. We delete   observable sets with small weights.
As such, we consider the final cost function as
\begin{equation}
    l(\vec \Kcal) = \sum_{\Olmat\in \Scal} \frac{\alpha_l^2}{\chi\pbra{\Olmat}} + \sum_{\Olmat\not\in \Scal }\alpha_l^2 N_s,
\label{eq:lfFinal}
\end{equation}
where $N_s$ is the total number of samples, $\Olmat\in \Scal$ if there exists a set $\bm e$ such that $\Olmat\in \bm e$, and $\sum_{\Olmat\not\in \Scal}$ is the penalty caused by deleting some sets. 
The selection of the final cost function in \autoref{eq:lfFinal} is inspired by the linear dependence  of the variance and the number of samplings, which is indicated by Chebyshev inequality $N_s\geq \text{Var}(\hat{\omat})/\pbra{\delta\varepsilon^2}$.The regularised term $\alpha_j^2 N_s$   compensates the initial error $\varepsilon_0 |$ for excluding the observable $\Olmat$ in the estimation of $\Omat$.
The initial error, which can be expressed as $\varepsilon_0 = | \sum_{j:\Olmat\not \in \Scal} \alpha_l \tr\pbra{\rho \Olmat} |$, implies a bias of our estimation since $\Olmat$ will not be covered by any set.

% average number shots for H2, LiH, F2: 1.3e5, 3.7e5, 8.7e5
% average running time for H2, LiH, F2: 0.4min, 1.2min, 2.9min

\blue{
It's important to note that we optimise the overlapped grouping strategy which was possibly used for comparison in \cite{yen_deterministic_2023} whose results appear pessimistic regarding the measurement cost. Specifically, we prioritise the measurement bases with large coefficients while discarding those with very small coefficients. This initial selection may introduce some level of error into our computations, termed initial errors. However, we can set a tolerance threshold and ensure the errors due to insufficient measurement bases are below such a threshold. This approach, while introducing certain errors, can significantly reduce the total number of measurements, and thus make it possible to run the algorithm on our device.}

% We will show that by means of the OGM framework, our local minimum is better than other existing algorithms in the later numerical results.

In our experiment, we reduce the total number of basis sets by decreasing the default number of samples $N_s$ and cut down sets with a weight smaller than a threshold. This strategy will generate a small number of sets and could reduce the running time for measurement. We could search for an optimised $N_s$ in a real experiment with a small-scale input size with an initial $N_{s0}$.
Since the cost function in \autoref{eq:lfFinal} is not convex, we could find a local minimum solution using the nonconvex optimisation methods. 

% The major
% which might introduce a large initial error. We need to check if it is above a threshold. When generating the basis set, we need to ensure the initial error is small, and the total number of basis sets should be smaller compared to other methods, such as grouping by the largest degree first (LDF) scheme.

Since our overlapped grouping measurement strategy assumes measurements drawn from the probability distribution, the measurement accuracy may fluctuate.  In our experiment, we derandomise the scheme by fixing the  measurement basis   $\Pmat$. 
Suppose that we have determined the measurement basis sets $\Scal =\{ \bm e_1, \ldots, \bm e_s \}$ and the optimised probability distribution $\{(\Pmat,\Kcal (\Pmat))\}$ using the above strategy. 
% In practical computation, we usually have constraints on the maximum allowed number of measurements. 
% In what follows, we provide a partially derandomized strategy with a given number of measurements $N_s$.
For the $j$th measurement $\Pmat_j$ with sampling probability $\Kcal_j$, we choose the number of measurements for $\Pmat_j$ to be $\floor{N_s \Kcal_j}$, and select an additional one $\Pmat_j$ with probability $N_s\Kcal_j - \floor{N_s\Kcal_j}$, where the total number of measurements is $N_s$. 

\blue{With these improvement, in our experiment, the average number of shots for measuring the energy of $\mol{H_2}$, LiH, and $\mol{F_2}$ is 1.3e5, 3.7e5, and 8.7e5, respectively, and the
average running time for $\mol{H_2}$, LiH, and $\mol{F_2}$ is 0.4min, 1.2min, and 2.9min, respectively.}

\section{Error-mitigated observable estimation}
\label{sec:QEM}

The VQE process requires the evaluation of gradient and energy. However, due to gate and measurement errors, these values unavoidably deviate from the ideal ones.
It is crucial to mitigate the errors in order to get reliable results.
In this section, we show how to obtain error-mitigated expectation values for observables.

\subsection{ Readout error mitigation}

% Quantum state measurements determine the amount
% of information that we can extract from the quantum experiments, and thus it is essential
% to mitigate the measurement error to ensure the results are reliable

In experiments, we extract quantum state information by performing projective measurements on a computational basis and analysing the measurement outcomes.
The quantum state $\ket{\Psi}$ collapses onto the    computational basis $\ket{i} \in \{ \ket{0}, \ket{1}\}^{\otimes N}$ with probability $p_i = |\braket{i|\Psi}|^2$ by Born’s rule, which can be obtained by  sampling the classical measurement outcomes.
Getting unbiased probability $\{p_i\}$ is central for us to estimate the property of the quantum state.
However, due to readout noise, the probability distribution deviates from the ideal one, which results in an error in the estimation.
In what follows, we show how to mitigate the measurement readout errors. 

The measurement error can be described as a classical random process, where any given ideal probabilistic measured result is transformed into a noisy one. Thus, the error can be fully described by a size $2^n\times 2^n$ transition matrix, also known as the calibration matrix $\Lambda$. The noisy measurement outcomes  $\tilde{\vec{b}}_{\mathrm{noisy}} = \{\tilde b_1, \cdots,\tilde b_n\}$ is related to the  ideal measurement outcomes  $\vec{b}_{\mathrm{ideal}} = \{ b_1, \cdots, b_n\}$ via the calibration matrix by
\begin{equation}
    \tilde{\vec{b}}_{\mathrm{noisy}}  = \Lambda \vec{b}_{\mathrm{ideal}}. 
\end{equation}

To circumvent the intractability of acquiring the complete calibration matrix, one can model the measurement error by taking tensor products of each local single-qubit calibration matrix of size $2\times 2$. The underlying assumption is that the measurement crosstalk in the quantum device is small enough that it can be completely omitted. 

In our experiment, we optimise qubit-qubit residual coupling and readout parameters and check the correlated readout errors by random state measurement experiments.
As shown in \autoref{fig:random_state_fidelity}, correlated readout errors are negligible.
Therefore, it is reasonable to assume that the measurement calibration matrix $\Lambda $ admits a tensor product form as
\begin{equation}\label{eq:calibration_matrix}
    \Lambda = \bigotimes_{j=1}^N \Lambda_j,
\end{equation}
where $\Lambda_j $ is the calibration matrix for the $j$th qubit, and it is expressed by
\begin{equation}
\Lambda_{j}=\left(\begin{array}{ll}
p(0 | 0) & p(0 | 1) \\
p(1 | 0) & p(1 | 1)
\end{array}\right)=:\left(\begin{array}{cc}
1-\epsilon_{j} & \gamma_{j} \\
\epsilon_{j} & 1-\gamma_{j}
\end{array}\right),
\label{eq:Lambda_j}
\end{equation}
where $p(x|y)$ represents the probability of the measured basis is $\ket{x}$ given the prepared state $\ket{y}$ under computational bases $\{|0\rangle , |1\rangle\}$.
With \autoref{eq:Lambda_j},  $ 2n$ parameters, $\{\varepsilon_j, \gamma_j \}$ for $j \in [1,n] $, can be used to describe the measurement error.

Next, we conduct the calibration process to learn the tensor product model parameters. We first select a set of product states $\mathcal{T}$. For each state $x\in \mathcal{T}$, we prepare the state on the quantum device and collect the measured product state $y$ at each time for $N_c$ times. We denote $m(y,x)$ as the number of times for y is observed when we prepare the state x. Then, as shown in Ref.~\cite{bravyi2021mitigating} the estimated two parameters $\epsilon_j$ and $\gamma_j$ for the $j$th qubit is given by
\begin{equation}
    \begin{aligned}
\epsilon_j & =\frac{\sum_{x, y} m(y, x)\left\langle 1 | y_j\right\rangle\left\langle x_j | 0\right\rangle}{\sum_{x, y} m(y, x)\left\langle x_j | 0\right\rangle}, \\
\gamma_j & =\frac{\sum_{x, y} m(y, x)\left\langle 0 | y_j\right\rangle\left\langle x_j | 1\right\rangle}{\sum_{x, y} m(y, x)\left\langle x_j | 1\right\rangle}.
\end{aligned}
\end{equation}
Remark that to effectively learn the parameters, we are supposed to let the selected, prepared product state set $\mathcal{T}$ contain at least once $x_j=0$ and $1$ for each qubit $j$.

The REM process essentially applies the inverse of \autoref{eq:calibration_matrix} to the noisy measurement result for estimating the expectation value of an observable. Given an arbitrary Hermitian observable in the form of \autoref{eq:observ_decomp}, and each $\Omat_l$ can be expressed in a tensor product form,
$    \Omat_l = \bigotimes_{k=1}^N \Omat_{l}^k,
$
where $\Omat_{l}^k$ acts on the $k$th qubit.
% For simplicity, we suppose each $\Omat_{l}^i$ is diagonal in the computational basis.
The estimator of the expectation value of $\Omat_l$ after REM is given by 
\begin{equation}
\hat{\omat}_l  = 
% \sum_x O(x)\left\langle x\left|A^{-1}\right| s^i\right\rangle \\
\frac{1}{M}    \sum_{m=1}^{M} \prod_{k=1}^N\left\langle
e\left|\Omat_{l}^k \Lambda_k^{-1}\right| s_k^m\right\rangle,
\label{eq:estimator_o_l_REM}
\end{equation}
where $M$ is the measurement shots assigned for observable $\Omat_l$, $\ket{e}=\ket{0}+\ket{1}$, and $s^m_k\in \{0,1\}$ is the obtained measurement outcome for the $k$th qubit in the $m$th measurement.

% As such, the estimator of the expectation value of observable $\Omat$ can be expressed by
% \begin{equation}
%     \hat \omat = \sum_j \sum_{l:\Olmat \in \mathcal{S}_j } \frac{\alpha_l N_j}{s_l}   \hat{o}_{l,j}.
%     \label{eq:estimate_derand_sum}
% \end{equation}

Now, we discuss the estimator of the observable within the grouping measurement framework.
Suppose we have determined the measurement basis set $\{\Pmat_j\}$ and the associated measurement times $N_j$.
Similarly to the grouping method, we denote $\mathcal{S}_j$ containing all $\Olmat$ hitted by $\Pmat_j$ (element-wise commute with $\Pmat_j$). 
The estimator of the expectation value of the observable $\Omat$ can be expressed by
\begin{equation}
\hat{\omat} = \sum_l \alpha_l  \hat{\omat}_l,
 \label{eq:estimate_derand_sum}
\end{equation}
where $\hat{\omat}_l $ denotes the estimator of $\Olmat$ measured in basis $\Pmat_j$.
% , and it can  be expressed as
% \begin{equation}
%     \hat{\omat}_l = 
%     \frac{\sum_{j: \Omat_l  \triangleright \mathcal{S}_j}  N_j \hat{o}_{l,j}}{\sum_{j: \Omat_l  \triangleright \mathcal{S}_j} N_j }.
% \end{equation}
In our experiment, we estimate the expectation value of the observable $\Olmat$ by
\begin{equation}
    \hat{\omat}_l = 
    \frac{\sum_{j: \Omat_l  \triangleright \mathcal{S}_j}    \hat{o}_{l,j}}{s_l }.
\end{equation}
Here, the denominator $s_l = \sum_{j} \delta_{\Omat_l  \triangleright \mathcal{S}_j}$ is the total number of times $\Olmat$ being hit by $\mathbf{P}_j$.
As such, the variance of $\hat \omat$ with a single measurement can be estimated as
\begin{equation}
    \var [ \hat{\omat}] = \sum_l \alpha_l^2 \frac{\sum_{j: \Omat_l  \triangleright \mathcal{S}_j}  \var[\hat{o}_{l,j}] }{s_l^2}.
\end{equation}

\blue{ Without REM, when the observable $\Olmat$ is measured by $\mathbf{P}_j$, it yields two potential outcomes: either $+1$ or $-1$. Let ${t}_{l,j}$ represent the count of measurements resulting in an outcome of $+1$. It is straightforward to see that the measurement result (i.e. estimator) associated with the measurement $\Pmat_j$ for observable  $\Omat_l$ is $\hat{o}_{l,j} = 2{t}_{l,j}/N_j - 1$.}

With REM, $\hat{o}_{l,j} $ is given by \autoref{eq:estimator_o_l_REM}. For later convenience, we give the explicit form of the observable estimation as
\begin{equation}
\hat{\omat}_l^{\mathrm{REM}}  = 
\frac{1}{N_j}    \sum_{j: \Omat_l  \triangleright \mathcal{S}_j}   \sum_{m=1}^{N_j} \prod_{k=1}^N\left\langle
e\left|\Omat_{l}^k \Lambda_k^{-1}\right| s_k^m\right\rangle.
\label{eq:estimator_o_l_OGM_REM}
\end{equation}
It is thus straightforward that the readout error mitigated estimator of the expectation value of observable $\Omat$ can be explicitly expressed as 
\begin{equation}
\hat{\omat}^{\mathrm{REM}} =   \sum_l \frac{\alpha_l}{s_l}  \sum_{j: \Omat_l  \triangleright \mathcal{S}_j}   \sum_{m=1}^{N_j} \prod_{k=1}^N\left\langle
e\left|\Omat_{l}^k \Lambda_k^{-1}\right| s_k^m\right\rangle.
 \label{eq:estimate_O_REM}
\end{equation}
The readout error mitigated estimation in \autoref{eq:estimate_O_REM}
is unbiased if every observable is assigned at least one measurement basis $N_j \geq 1$, otherwise, the estimation has an initial error.

\subsection{Gate error learning by Clifford  fitting \label{sec:clifford_fitting}}

\subsubsection{Methods}

We consider Clifford fitting (also called Clifford data regression)~\cite{czarnik2021error} to mitigate the dominant CZ gate error.
The key idea is to obtain an error-mitigated expectation value by fitting a regression function $f^{\mathrm{CF}}$ which maps a noisy expectation value to an error-mitigated one as $$o^{\mathrm{ideal}} = f^{\mathrm{CF}}(o^{\mathrm{noisy}}).$$
The regression function is obtained in a learning style using data pairs of ideal and noisy results generated from classically simulable (near)-Clifford circuits and noisy quantum circuits. Here, we choose the fitting function to be linear,  $$f^{\mathrm{CF}}(\cdot)=a(\cdot)+b,$$ 
where $a$ and $b$ are the coefficients to be fitted. 

For Clifford fitting in quantum chemistry simulation, which relies on a parameterised quantum circuit, we choose a collection of quantum circuits by randomly replacing the parameterised gates with Clifford gates, and the remaining parameterised gates are assigned  with a rotating angle choosing independently, uniformly from $[0,2\pi]$. 
Here, we denote $n_p$ as the total number of parameterised gates in a parameterised quantum circuit, $K$ as the number of gates not being substituted, and $\textit{S}_u=\{u_i\}$ as the set of the randomly assigned circuit. Hence, the remaining $n_p-K$ gates are replaced with single-qubit Clifford gates. 
% We choose the $n_p-K$ replaced Clifford gates to be the identity gates $I$ for experimental concerns. 

For a given Pauli observable $\Omat_j \in P^{\otimes n}\setminus I^{\otimes n}$ with $P\in \{I,X,Y,Z \}$, we aim to establish a training data set consists of data pairs of ideal expectation value simulated on a classical computer and noisy value measured from the quantum device with the set of quantum circuits we construct above. For efficient classical simulation, it is crucial that they are ``nearly Clifford'' in the sense that the stabiliser rank, i.e. the number of non-Clifford gates $K$ in the quantum circuit, of each quantum circuit is a constant. In this case, these quantum circuits can be simulated classically~\cite{bravyi2019simulation}. 
% The Clifford fitting method is especially suitable for our experiment in that the scalability is guaranteed once the training (fitting) process is done. 
Compared to other error mitigation methods, such as zero noise extrapolation approaches, the Clifford fitting method does not require stretching the implementation time of the gates. 
\blue{Besides, the effect of noise on the training and corresponding testing (experimental) circuit on the expectation values of observables is expected to remain the same as only single-qubit rotation gates are replaced by Clifford gates through changing the rotation angle and the two-qubit CZ gates are unchanged.}

% To understand why a linear fitting function   works, let us consider the incoherent noise and assume that the underlying noise process is time- and gate-independent. Two different error rates, $\xi_1$ and $\xi_2$, represent the single- and double-qubit error rates are the key gradients to describe the noise process. Note that for the superconducting qubit system, $\xi_2$ is often one order greater than $\xi_1$. The noise process is thus expressed by a complete positive trace-preserving (CPTP) map acting on the ideal density matrix of the quantum circuit. Since the CPTP map is linear, the transformation from the ideal to the noisy expectation value for an observable will be an affine transformation. As the Clifford fitting does not change the arrangement of the quantum circuit, e.g., only the single-qubit gates are being substituted, the same affine transformation holds for different quantum circuits we used for fitting.
% It is thus valid for us to generalise to unseen cases once the training is done.

\subsubsection{Experimental procedure}

% We enumerate the chosen $K$ for different molecules in the following table.

For our goal to estimate the energy, we run the Clifford fitting process for each observable in the Hamiltonian. The number of quantum circuits could be enormous and impractical for experimental realisation. 
Here, we discuss how to reduce instances of the circuit to be implemented for Clifford fitting. 

We observe that in practical computation, some observables' expectation values remain unchanged or change in a negligible small range for different parameters in different quantum circuits, called unchanged observables. It is, therefore, unnecessary to run the algorithm for these unchanged observables on quantum hardware,  upon evaluating the energy of the quantum system. 
% Instead, we only need to record its ideal value.
% The criterion for determining if an observable $\Omat_j$ is whether changing or not follows: i) We simulate classically the expectation value $\braket{\Omat_j}_{\phi_i}$ of the observable with respect to each $\ket{\phi_i}$ in $\textit{S}_\phi$; ii) Compute the range value $\sigma_j$ and the variance $\mu_j$ of each expectation value; iii) If both $\sigma_j$ and $\mu_j$ are greater than the threshold we chose, then $\Omat_j$ is changing, or it is unchanging otherwise.
We elaborate on the complete protocol in the following.
\begin{enumerate}
    \item In the first step, we identify and record all the changed and unchanged observables in the Hamiltonian. For each Pauli observable $\Omat_j$ in the Hamiltonian $\hat{H}$, randomly assign all the parameters uniformly from $[0,2\pi]$ and obtain a $L$-size quantum circuit instance set $\mathcal{S}^{(Z)}_u=\{{u_i}\}_{i=1}^L$, where the upper-script $Z$ stands for the number of non-Clifford gates in the circuit, and ${u_i}$ denotes the $i$th quantum circuit instance obtained from the random-parameter assignment. The size $L$ needs to be sufficiently large to create enough quantum circuit instances to determine if the expected value of an observable changes. Compute  the expectation value $\{\braket{\Omat_j}_{u_i}\}_{i=1}^L$ classically for each circuit instance in $\mathcal{S}^{(Z)}_u$ for each observable.  Compute the range value $\sigma_j$ and the variance $\mu_j$ of the expectation values. If both $\sigma_j$ and $\mu_j$ are greater than the thresholds $\sigma_T$ and $\mu_T$ we chose, then $\Omat_j$ is labelled as a changed observable, otherwise, an unchanged observable. We record the invariant observables, which will not be measured on quantum hardware. Record its value for estimating the energy in the final step.
    \item Let $M$ be the total number of parameterised gates. Randomly choose $K$ out of $M$ parameterised gates and assign value independently to each parameter from $[0,2\pi]$ to construct the quantum circuits set $\mathcal{S}^{(K)}_u$ of size $R$. Set the left $M-K$ gates to the identity gate. $R$ is chosen to be sufficient for fitting a line for the linear regression, such as $10$.
    \item We refer to the set of changed observables as $\{O_C\}$, run step 4 for each observable in $O_C$. Then go to step 5.
    \item For a given observable $\Omat_j$, calculate the ideal expected value $\{\braket{\Omat_j}_{u_i}\}_{i=1}^R$ with respect to state prepared by each $u_i$. Here, we want the ideal values to be as spread out between $[-1,1]$ as possible. The purpose is that at least these different instances will appear in different intervals so that the values are dispersed enough for better linear regression. To this end, we classify different expected values according to their intervals, such as selecting instances in the intervals of [-1,0.9], [-0.9,0.8],…,[0.9,1]. 
    \item Set $\{O_S\}=\emptyset$. Update $\{O_S\}$ by adding all the observables that satisfy the dispersion condition in step 4. Repeat step 4 for multiple times for observables in $\{O_C\}\cap\overline{\{O_S\}}$ until $\{O_C\}\cap\overline{\{O_S\}}$ is empty. At each repetition, add qualified observables to $\{O_S\}$, and record the quantum circuit set $\mathcal{S}^{(K)}_u$ as well.
    % \item  Determine if $\{O_S\}$ is not equivalent to $\{O_C\}$. If not, repeat steps 4-6 to obtain multiple sets of parameters. 
    \item For each recorded quantum circuit run on the quantum device to get the noisy expectation value for all its corresponding observables. For each observable $O$, feed the ideal and corresponding noisy value to fit the linear regression function and optimise the loss function
    \begin{equation}
        C=\sum_{\phi_i \in \mathcal{S}_\psi^{(K)}}\left(\braket{O}_{u_i}^{\text {exact }}-\left(a \braket{O}_{u_i}^{\text {noisy }}+b\right)\right)^2.
    \end{equation}
    Therefore we get different linear mapping functions for different observables.
    \item Use the fitting function to mitigate the error in expectation value for the corresponding observable.
    % The number of sets should not be many.
\end{enumerate}

The estimation of an observable $\Omat$ can be obtained by \autoref{eq:estimate_derand_sum}, and is formally expressed as
\begin{equation}
\hat{\omat}^{\mathrm{CF}} = \sum_l \alpha_l \hat{\omat}_l^{\mathrm{CF}},
    \label{eq:CF_estimator}
\end{equation}
where the estimator for $\Omat_l$ after REM and CF is  given by
\begin{equation}
    \hat{\omat}_l^{\mathrm{CF}} = f_l^{\textrm{CF}}(\hat{\omat}_l ^{\mathrm{REM}} ),
    \label{eq:CF_estimator_o_l}
\end{equation}
and $f_l$ is the fitting function associated with $\Omat_l$.

\subsection{Symmetry verification}
Error mitigation by symmetry verification takes advantage of prior knowledge of the physical subspace where we search for the solution to the problem. 
There are two types of symmetries in the molecular problems considered here, which are the conservation of particle number and singlet state for the ground state, need to be satisfied. The particle number, hence the parity, of the ground state should be conserved throughout the simulation process. In addition, it is known in the quantum chemistry community that for closed-shell molecules, the ground state is bound to be a spin singlet state.

Below, we briefly review error mitigation by symmetry verification introduced in \cite{PhysRevA.98.062339}. 
For each symmetry verification operator $\hat{S}_i$, we have,
\begin{equation}
\hat{S}_i|\psi\rangle=s_i|\psi\rangle,
\end{equation}
where the $s_i\in{\pm1}$. The density matrix, which is projected into the subspace that conserves the symmetry, can be expressed as,
\begin{equation}
\rho_{s_i} = \frac{\Pi_{s_i} \rho \Pi_{s_i}}{\tr(\Pi_{s_i}\rho)},
\end{equation}
with $\Pi_{s_i}=\frac{I+s_i \hat{S}_i}{2}$. 
Specifically, as a result of the closed shell molecule that we deal with, the trial wave function we pick is in an even-parity sub-manifold for total particle number and spin. Therefore, the eigenvalues of parity check operators for the correct symmetry are always one, i.e., $s_i = 1$. 
We project the target molecular Hamiltonian $H$ to the even-parity subspace. For this purpose, we have
\begin{equation}
\tr(\hat{H}\rho_{s_i}) = \frac{\tr(\hat{H}\rho)+\tr(\hat{H}\hat{S}_i\rho)   }{1+\tr(\hat{S}_i\rho)},
\label{eq:SV_expect}
\end{equation}
which effectively conserves the symmetry of the quantum subspace.
This is achieved by measuring $\tr(H\rho)$, $\tr(H\hat{S}_i\rho)$, and $\tr(\hat{S}_i\rho)$.

We have the particle number or parity conservation for the molecular problems considered here, i.e., $[\hat H, \hat S] = 0$ with $\hat S = \bigotimes_i \hat{Z}_i$. 
This symmetry verification expands the to-be-measured observable set, which is composed of the elements in the decomposition of $H$ and $H \hat{S}$.
The  symmetry verification EM has three ingredients $\braket{ \hat H}$, $\braket{ \hat H \hat{S}}$, and $\braket{ \hat{S}}$.
Each of them is estimated by measuring the predetermined basis, introduced in \autoref{sec:OGM}.  The final error mitigated results by SV are  computed classically by \autoref{eq:SV_expect} as
\begin{equation}
    \hat{\omat} = \frac{ \braket{\hat{H}}^{\mathrm{CF}} + \braket{\hat{H} \hat{S} }^{\mathrm{CF}} }{ 1 +  \braket{\hat{S}}^{\mathrm{CF}}}.
\end{equation}
Here, $\braket{\hat{H}}^{\mathrm{CF}} $, $\braket{\hat{H }\hat{S} }^{\mathrm{CF}}$, and $ \braket{\hat{S}}^{\mathrm{CF}}$ are the estimator obtained by sequentially applying REM  in \autoref{eq:estimator_o_l_OGM_REM} and CF in \autoref{eq:CF_estimator}, and classical post-processing by \autoref{eq:estimate_derand_sum}.

In our experiment, we found that a combination of  Clifford fitting and symmetry verification might not perform well. This is because, in our experiment, symmetry verification requires measuring many large-support observables, which demands high-precision measurement for these observables.
In \autoref{fig:symmetryVerification}, we study the performance of symmetry verification and explain why it results in large fluctuations in the energy estimation, which will be elaborated on below.

\subsection{Connected moments expansion\label{sec:connectedMX}}

The connected moments expansion (CMX) method~\cite{kowalski2020quantum} is derived from the imaginary time evolution (ITE) approach that can be expressed in the following formula
\begin{equation}\label{eq:CMX-ITE}
    E(\beta)=\frac{\langle\Phi|H e^{-\beta \hat H}| \Phi\rangle}{\langle\Phi|e^{-\beta \hat H}| \Phi\rangle}=\sum_{k=0}^{\infty} \frac{(-\beta)^k I_{k+1}}{k !} ,
\end{equation}
where $\beta$ is the evolution time of the ITE method, and we expand the formula as the expansion of the connected moments $I_k$ taking advantage of the Taylor expansion. The mechanism of the ITE method is to project the state $\ket{\Psi}$ towards the ground state of $H$, and \autoref{eq:CMX-ITE} gives the exact ground state energy as $\beta$ goes to infinity as long as the state $\ket{\Psi}$ is non-orthogonal to the ground state. 
The evolution time $\beta$ in practice is chosen to be a finite number and is essentially related to the overlap between the state $\ket{\Psi}$ and the ground state. 

The ITE operator $e^{-\beta \hat H}$ is nonunitary and thus generally hard to implement on a quantum device. The CMX method provides a way to circumvent this problem through measuring the higher moments of $\hat H $ in \autoref{eq:CMX-ITE}. Specifically, the connected moment is defined recursively by
\begin{equation}\label{eq:CMX_expansion}
I_{k}=\langle\Phi|\hat{H}^{k}| \Phi\rangle-\sum_{i=0}^{k-2}\left(\begin{array}{c}k-1 \\ i\end{array}\right) I_{i+1}\left\langle\Phi\left|\hat{H}^{k-i-1}\right| \Phi\right\rangle.
\end{equation}
The analytical form for exact energy as $\beta$ goes to infinity
can be represented by 
\begin{equation}
E_{ \mathrm{ground}}=I_{1}-\frac{S_{2,1}^{2}}{S_{3,1}}\left(1+\frac{S_{2,2}^{2}}{S_{2,1}^{2} S_{3,2}} \ldots\left(1+\frac{S_{2, m}^{2}}{S_{2, m-1}^{2} S_{3, m}}\right) \ldots\right),
\label{eq:En_CMX}
\end{equation}
where $S_{k,1} = I_k$, $S_{k,i+1} = S_{k,1} S_{k+2,i} - S_{k+1,i}^2$ and it is straightforward to have
$
S_{2,1} = I_2 = \braket{\hat{H}^2} - \braket{\hat{H}}^2$ and
$S_{3,1} = \braket{\hat{H}^3} - 3 \braket{\hat{H}}\braket{\hat{H}^2} + 2 \braket{\hat{H}}^3$.  For practical concerns, one truncates the expansion up some order $k$. In our experiment, we measure the higher-order moment $\langle\Phi|H^{k}| \Phi\rangle$ for $k \leq 3 $ using the grouping measurement schemes introduced in \autoref{sec:OGM}.
% It has been shown numerically that the measurement cost for measuring the higher-order moment will saturate for sufficiently large orders~\cite{vallury2020quantum}.
It is reasonable for our choice of $k\leq3$ in that we expect the application of various quantum error mitigation (QEM) methods to be able to reproduce most parts of the ground state energy, but in some cases, it may not be able to reach the chemical-accuracy requirement. As such, the state we prepared in the energy estimation step corresponds to an effective state which has a fairly good overlap with the ground state, but is not exactly the same. Assuming to perform the ITE approach in this case, it follows that the evolution time $\beta$ of the ITE method can be chosen to a relatively small one. Thus one can equivalently truncate the expansion \autoref{eq:CMX_expansion} up to some constant order as the expansion is monotonically decreasing. Truncating the expansion to a low order also makes the method affordable experimentally. Note that   CMX is not a variational method, and thus the energies obtained by CMX may be lower than the true ground state energy.

In the experiment, after the VQE procedure, we would obtain the final circuit parameters. We then execute the circuit under different measurement bases to measure the expectation values of different observables ($\hat H$, $\hat H^2$ and $\hat H^3$). Finally, the ground state energy is estimated by \autoref{eq:En_CMX}.

The variance of the energy estimation using the connected moments' expansion up to the second order is calculated by
\begin{equation}
    \var [E_{ \mathrm{ground}}] = \left( \frac{\partial E_{ \mathrm{ground}}}{ \partial a }  \right)^2 \var[a] +\left( \frac{\partial E_{ \mathrm{ground}}}{ \partial b }  \right)^2 \var[b] + \left( \frac{\partial E_{ \mathrm{ground}}}{ \partial c }  \right)^2 \var[c],
\end{equation}
and it takes the form of
\begin{equation}
\var [E_{ \mathrm{ground}}] =  \left( 1+ \frac{4a S_{2,1}}{S_{3,1}} - \frac{S_{2,1}^2(-3b+6a^2)}{S_{3,1}^2} \right)^2 \var[a] + \left(\frac{2 S_{2,1}}{S_{3,1}} + \frac{3 S_{2,1}^2 a}{S_{3,1}^2}\right)^2 \var[b] + \frac{S_{2,1}^4}{S_{3,1}^4} \var[c]
\label{eq:CMX_var}
\end{equation}
where  we have defined $a := \braket{\hat{H}}$,  $b := \braket{\hat{H}^2}$ and  $c := \braket{\hat{H}^3}$. It is worth noting that this calculation for the variance is based on the variance propagation, and the actual variance could be smaller. The error bar for the energy estimation using CMX is calculated according to \autoref{eq:CMX_var}.

\subsection{Experimental procedure for observable estimation} 
\label{sec:observ_estimation}

In the experiment, according to the target observable set, $\{ \Omat_l \}$, including the Hamiltonian $\hat{H}$, the symmetry verified Hamiltonian $\hat{H}\hat{S}$, parity $\hat{S}$, and the Hamiltonian power $\hat{H}^k$,  the measurement basis $\{ \Pmat \}$ and the associated measurement shots are generated using the method introduced in \autoref{sec:OGM}. The experimental procedure for obtaining error-mitigated observable estimation in order is summarised as follows.

\begin{enumerate}
    \item Experimental benchmark for measurement noise. Test whether the measurement errors admit a tensor product type by random state measurement.  Obtain the calibration matrix $\Lambda$ by measurement error calibration.
    \item Generate a series of random quantum circuits for Clifford noise learning. Measure the state prepared by Clifford circuits in the predetermined basis $\{ \Pmat \}$ and obtain the fitting function $f^{\rm CF}$ for each observable.
    \item During the VQE process, measure the observable on the state in basis $\{ \Pmat \}$. For each $\Omat_l$, apply readout error mitigation by \autoref{eq:estimator_o_l_OGM_REM}.
    Get the observable estimation after CF in \autoref{eq:CF_estimator} or \autoref{eq:CF_estimator_o_l}.
    \item Get symmetry verified results using  \autoref{eq:SV_expect}.  In our experiment, we apply symmetry verification to correct the energy for $\mol{H_2}$ and LiH. 
    \item The final ground state energy is estimated using the CMX  in \autoref{eq:En_CMX}.
\end{enumerate}

\begin{figure}[htb]
\centering
\includegraphics[width=1\textwidth]{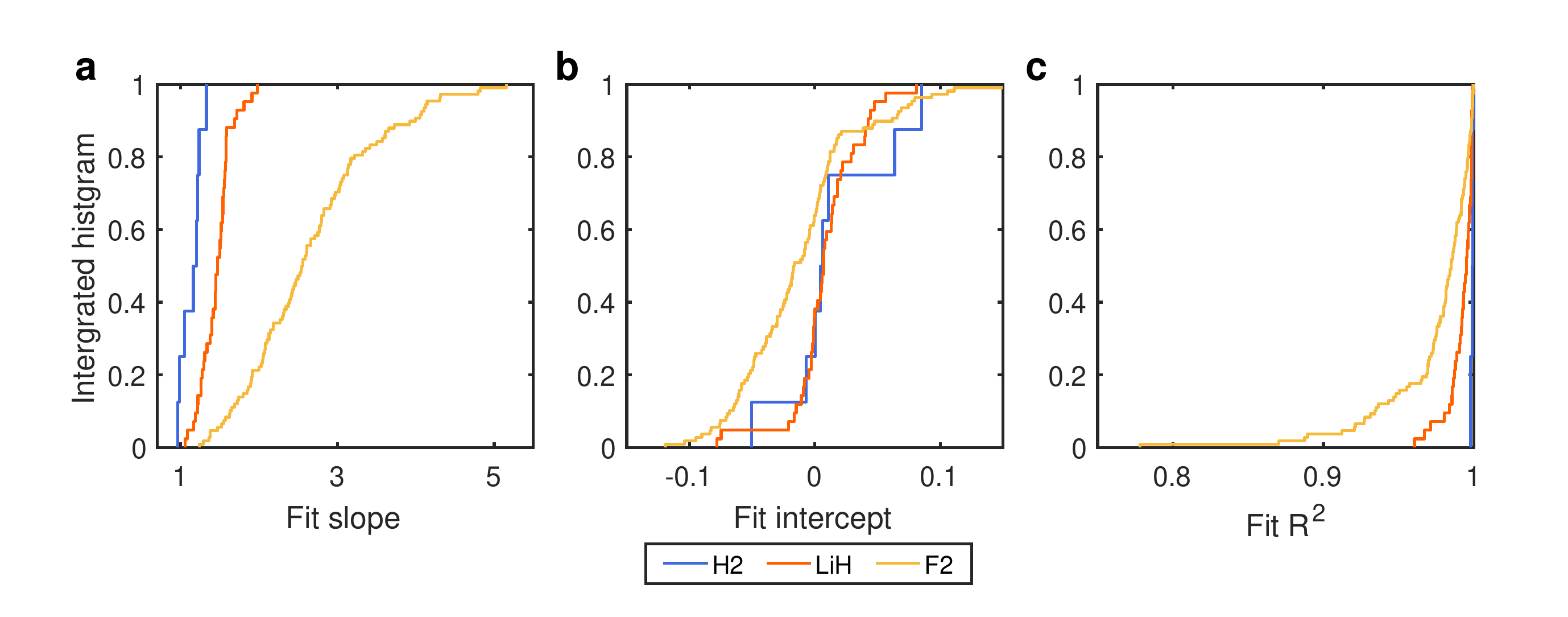}
\caption{
\textbf{Distribution of Clifford fit parameters.}
The accumulated distribution of slope, intercept and goodness of fit for $\mol{H_2}$ (R=2.6), LiH (R=1.5) and $\mol{F_2}$ (R=2.0) are shown in the diagram. With the system size and circuit depth increasing, we can observe that (a) the slopes of Clifford fit become larger due to the impact of decoherence. (b) the variance of intercept becomes larger, and the mean value is off zero, which may be caused by gate correlated errors. (c) the goodness of fit becomes worse. Two main issues may cause it. One is gate-correlated errors that make our model nonlinear. The other is insufficient fit data, which will increase our experiment resources. 
}
\label{fig:cf_distribution}
\end{figure}

\begin{figure}[htb!]
\centering
\includegraphics[width=1\textwidth]{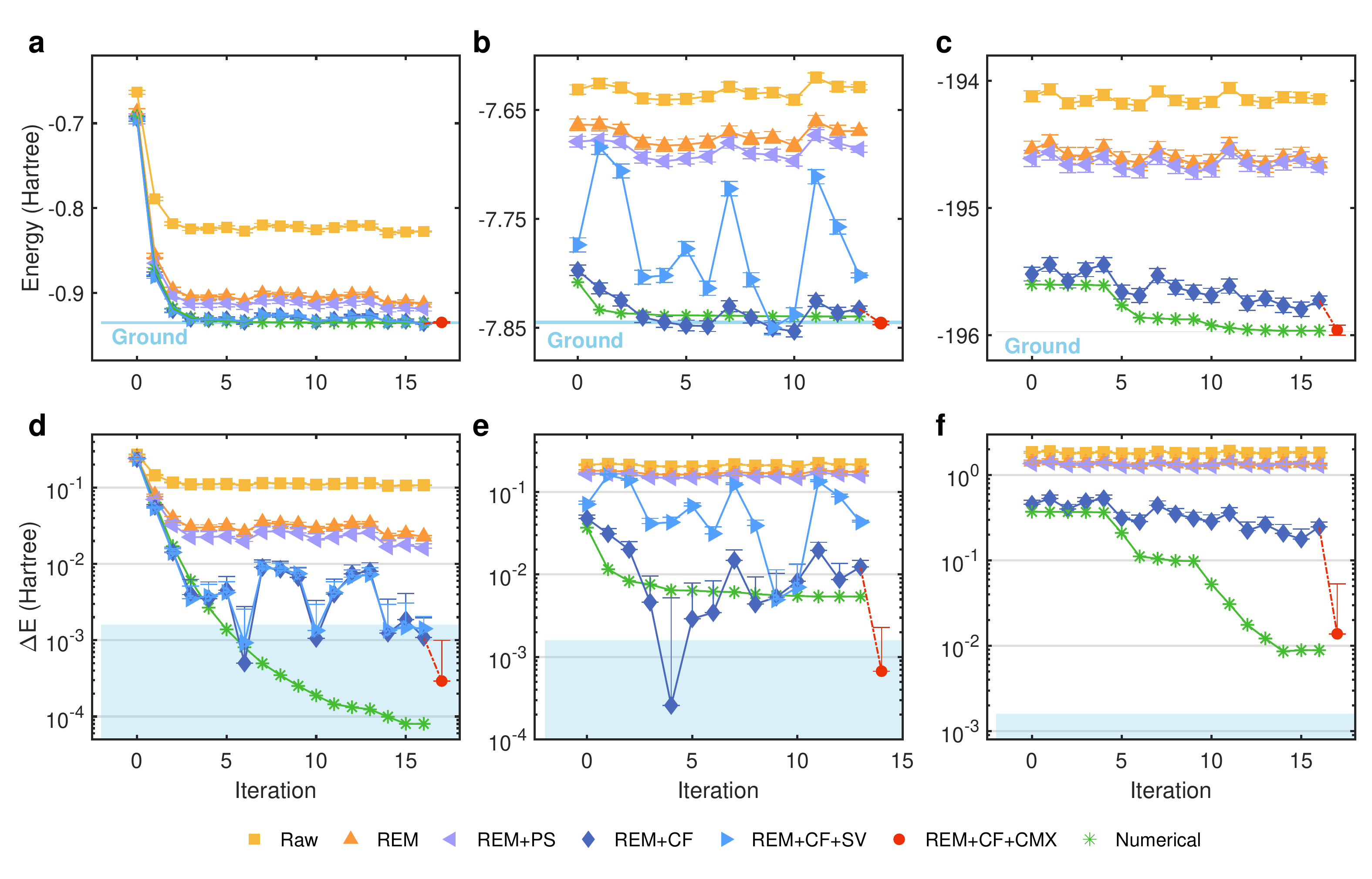}
\caption{
\textbf{Optimisation procedure.}
\blue{The optimisation procedure traces of energy and abs energy error for $\mol{H_2}$ (R=2.6), LiH (R=2.2) and $\mol{F_2}$ (R=2.6) are shown in the picture.} Here in (a)(b)(c), we add energy with REM + PS (purple leftward triangles), while the others are the same as in Figure 2b. \blue{Figure (d)(e)(f) show the traces of abs energy error during the optimisation procedure. }
}
\label{fig:supp_iteration}
\end{figure}

\begin{figure}[htb]
\centering
\includegraphics[width=1\textwidth]{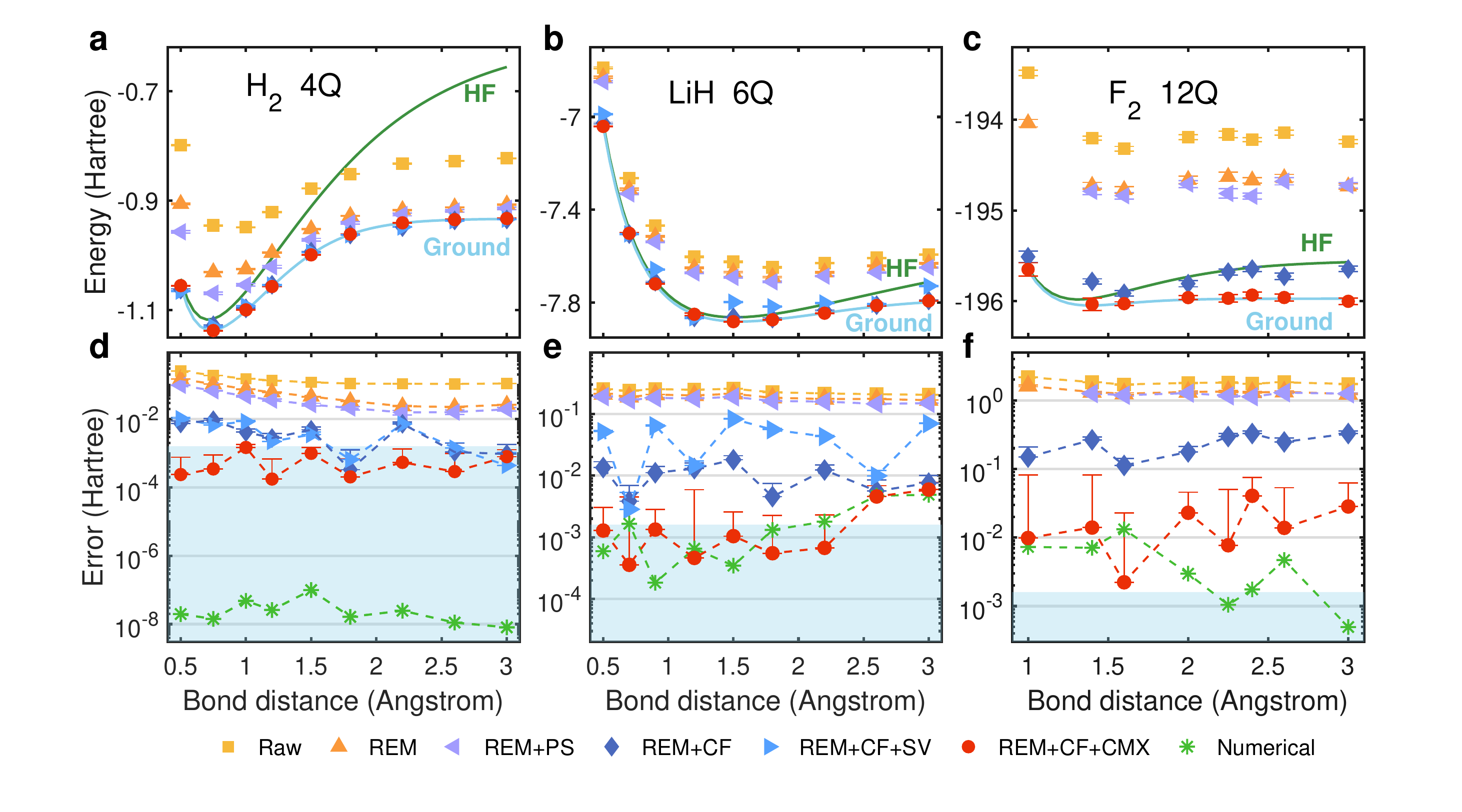}
\caption{
\textbf{The VQE simulations for potential energy curves for different molecules.} Similar to Figure 3, more results with different error mitigation strategies are shown here. (a-c) Potential energy curves as functions of the bond distance for $\mol{H_2}$ (4 qubits), LiH (6 qubits) and $\mol{F_2}$ (12 qubits) molecules with various error mitigation strategies. (d-f) Absolute errors are compared to the FCI results. We compare the raw data (yellow squares) with the application of REM (orange triangles), REM and PS (purple triangles), REM and CF (deep blue diamonds), SV (blue triangles) and CMX (red circles). The results marked by green asterisks are energies calculated by classical simulation with the parameters searched in the experiment. The ground state energy with chemical accuracy (blue regime) is calculated theoretically as a reference.
}
\label{fig:supp_bond}
\end{figure}

\begin{figure}[htb]
\centering
\includegraphics[width=0.8\textwidth]{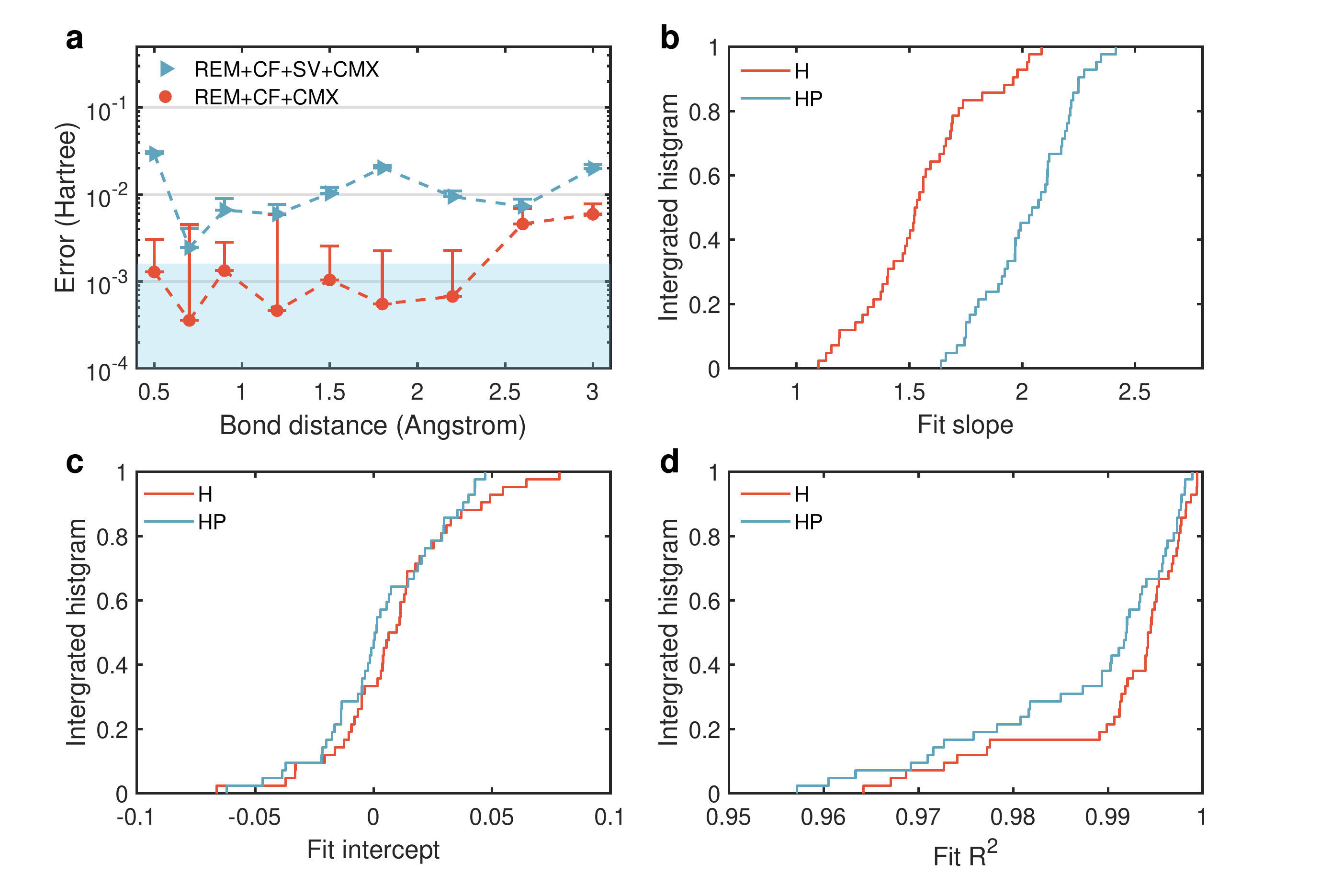}
\caption{
\textbf{Symmetry Verification.} (a)The comparison of the abs energy error with different error mitigation setups for LiH. The red dot is the energy with REM + Clifford fitting + CMX, and the grey-blue triangle is the energy with REM + Clifford fitting + CMX + SV. (b)(c)(d) are the slope, intercept and $R^2$ of the LiH(R = 2.2), respectively. ``H'' refers to the original Hamiltonian $\hat{H} $ and ``HP'' refers to the symmetry verified Hamiltonian $\hat{H}\hat S$.
}
\label{fig:symmetryVerification}
\end{figure}

\begin{figure}[htb]
\centering
\includegraphics[width=1\textwidth]{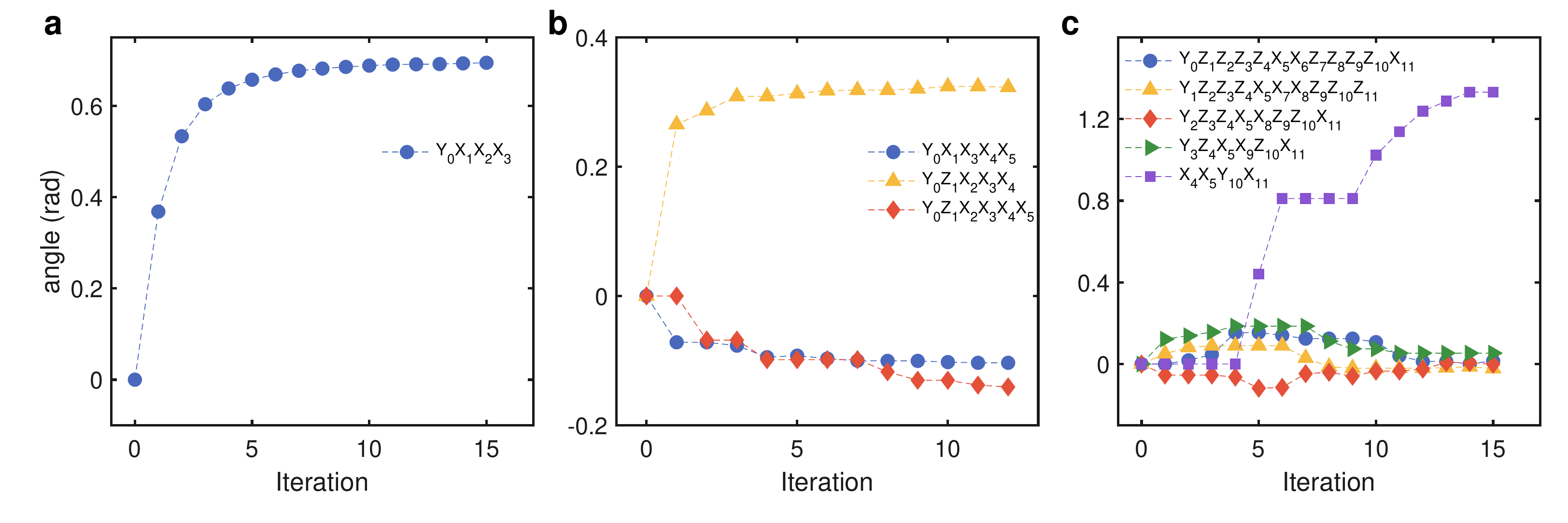}
\caption{
\textbf{Parameter optimisation processes.}
The evolution of variational parameters in optimisation processes for $\mol{H_2}$ (a), LiH (b) and $\mol{F_2}$ (c), respectively. For each iteration, we select N/2 parameters to update, and the parameter converges within 15 iterations for all the molecules.
}
\label{fig:params_iteration}
\end{figure}

\section{Experimental implementation and results}

% Having optimised the Zuchongzhi processor with high-quality qubits and gates and 

In   \autoref{sec:hardware}, we show how to optimise the Zuchongzhi processor to obtain high-fidelity gates and readout.
In \autoref{sec:method_VQE}, we discuss quantum algorithms for the ground state search and quantum state measurement with fewer quantum resources. Subsequently, in \autoref{sec:QEM}, we show how to obtain error-mitigated ground state energy estimation.
With these experimental and theoretical advances, we study systematically the ground state energy estimation for three molecules, $\mol{H_2}$, $\mol{LiH}$, $\mol{F_2}$. 
Finally, we point out the challenges in experimental quantum computational chemistry.

% using the  quantum algorithms developed.

\subsection{Experimental procedure}

\textbf{Benchmarking}: 

Prior to experiments, we first calibrate the qubits and couplers set used in our experiment to obtain the performance with high enough gate fidelity and readout fidelity, as well as to ensure there are no correlated readout errors. The calibrations contain the following steps.
\begin{enumerate}
    \item Calibrate the qubit frequency and Rabi amplitude, update the readout IQ clouds discrimination lines and the readout error mitigation (REM) matrix by standard readout experiment. The REM matrix for single-qubit is described as \autoref{eq:Lambda_j}, where $p(x|y)$ represents the probability of the measured basis is $\ket{x}$ given the prepared state $\ket{y}$ under computational bases $\{|0\rangle, |1\rangle\}$. We prepare the qubit in $\ket{0}$ and $\ket{1}$ in the experiment and select the perpendicular bisector of the two IQ clouds centres as the discrimination line to separate the $\ket{0}$ and $\ket{1}$ as well as acquire $p(x|y)$. \blue{This step runs prior to each VQE experiment for a molecule with a single bond length, as well as before every complete experiment for calibration and benchmark.}
    \item Benchmark single-qubit and CZ gates, and update the control parameters in CZ gates. {Count} the fidelities for all single-qubit gates and CZ gates, and ensure the fidelities are high enough and stable. This step will be run every 3 hours.
    
    \item Check the correlated readout error. Comparing the average union readout fidelity and standard readout fidelity with the random state readout experiment to check whether the corrected readout error is negligible. This step will also be run every 3 hours.
\end{enumerate}

\textbf{Experiments}: 

% The VQE experiment can be divided into two main categories:
% % optimisation and 

% essential
% the Clifford fitting and Imaginary evolution. Clifford fitting experiment characterise the decoherence noise and export the error mitigation function for every observables used in experiment. Then the main VQE optimisation process is executed in Imaginary evolution experiment to get the final error mitigated result. 

A crucial step in optimisation is to get the gradient with respect to a parameter, which is estimated by measuring the parameter-shifted quantum state, as indicated by \autoref{eq:gradient_SGD}.
We apply error mitigation techniques, including REM, CF, and SV, to obtain an error-mitigated gradient and energy.
In the following, we elaborate on the experimental procedures.
The experimental Clifford fitting is composed of the following steps:
\begin{enumerate}
    \item Determine a set of Clifford circuits with the random parameters generated using the method described in \autoref{sec:clifford_fitting}. For each Pauli measurement basis $\Pmat_j$ and the associated observable set $\bm e_j = \{ \Omat_l | \Omat_l \triangleright \Pmat_j \}$,  record instances of quantum circuit $\mathcal{C}_j$. Note that  $\mathcal{C}_j$ could be identical to make the experiments friendly.
    \item 
    Run noisy Clifford circuits $\{ \mathcal{C}_j \}$ with the 
 pre-determined random parameters. Measure in Pauli basis $\{\Pmat\}$ generated using the overlapped grouping methods in \autoref{sec:OGM} and record the noisy expectation value for each observable $\Omat_l$.
    \item Simulate  Clifford circuits $\{ \mathcal{C}_j \}$ classically and record the ideal expectation value for every observable.
    \item Fitting the ideal and noisy expectation value with a linear function \blue{$f_l^{ \mathrm{CF} }(\cdot)=a(\cdot)+b$} for every observable, we get the error mitigation function. Record those fitting functions $f_l^{ \mathrm{CF} }$, and we could use them to mitigate the errors in the observable estimation by \autoref{eq:CF_estimator_o_l}.   
\end{enumerate}

In the VQE experiment, we first prepare the initial state as a Hartree-Fock state or multi-reference state which is described in \autoref{sec:init_state_prep}.
Then, we measure the Hamiltonian to get the initial state energy $E_{0}^{\mathrm{expm}}$. 
In our experiment, the observables are measured in Pauli bases, which are generated using the overlapped grouping strategy introduced in \autoref{sec:OGM}.
We search the ground state by optimising the parameters in the circuit. 
A key component in our VQE experiment is to get the error-mitigated expectation value for observables. To do so, we apply the above error mitigation schemes for each measurement outcome.
The experimental procedures for error mitigation have been shown in \autoref{sec:observ_estimation}

% \begin{enumerate}
%     Get the REM experiment result for all the Pauli basis with the tensor product form readout error mitigation (REM) matrix.
%     \item Compute the expectation value for all observables by experiment result and commute relation.
%     \item Correct the expectation value with the Clifford fitting(CF) error mitigation (EM) function for the corresponding observable.
%     \item Computation of the energy $\braket{H}$ by the EM expectation value and the observables' Hamiltonian coefficient.
%     \item Compute the $\braket{HP}$ and $\braket{P}$ with the same experimental result and different observable set.
% \end{enumerate}

The VQE optimisation is summarised as follows:    
\begin{enumerate}
    % \item Prepare the initial state.
    \item { Gradient estimation and optimisation}. Determine $\vec \theta$ associated with the single-qubit Pauli-rotation gate, in which we select $N/2$ parameters to update. For each selected parameter, the gradient is estimated using \autoref{eq:gradient_SGD}, which requires measuring the Hamiltonian on the parameter-shifted state.
    Get the gradient for all selected parameters and update the parameters. 
    Here, we obtain the error-mitigated measurement outcome $\mathcal{L}(\vec \theta ^{\pm})$ defined in \autoref{eq:gradient_SGD_L_pm_def} by  REM and CF error mitigation schemes.
    \item For each iteration,  we measure the energy with the updated parameters and hence record the optimisation trace.  
    \item Run the 1 and 2 steps to update the parameters until convergence. For all the molecules, the energy converges within 15 iterations.

    % We measure the energy again with more sample number to decrease the result fluctuation,
    % At last, we acquire the energy with REM + CF + CMX error mitigation strategies.
    
\end{enumerate}

Finally, we measure the state with the optimised parameters. We correct the ground state energy using connected moments expansion (CMX), which requires measuring  $\braket{\hat{H}^2}$ and $\braket{\hat{H}^3}$, as described in \autoref{sec:connectedMX}. 
\blue{The expectation value of $\braket{\hat{H}^k}$ for $k \leq 3$ will be measured and then error mitigated in a similar fashion to that of $\braket{\hat{H}}$.
With the error mitigated results, we calculate the CMX energy using equations \autoref{eq:CMX_expansion} and \autoref{eq:En_CMX}.}

\subsection{Results}

% \textbf{Comparison of different error mitigation schemes}

This section shows experimental results for the ground state search and ground state energy estimation.

We select at most 12 qubits on the Zuchongzhi 2.0 quantum processor. Several patterns of 12 qubits are available in our processor. Different pattern choices could lead to significantly different results in our experiment. Since learning the gate error is crucial in our experiment, we run the Clifford fitting experiment with different qubit patterns to characterise the performance and finally select the pattern based on the goodness of fit and the fit slope.

We first select three bond distances for different molecules, including  $\mol{H_2}$ (R=2.6), LiH (R=2.2) and $\mol{F_2}$ (R=2.6), to study the performance of the Clifford fitting.  \autoref{fig:cf_distribution} shows the performance of the Clifford fitting for the three molecules at the selected bond distances by the accumulated distribution of the slope, intercept, and the $R^2$ value.
Through the comparison of a,b, and c, we find that the fitting becomes worse with the system size increases, which may be attributed to the correlated gate errors and the fluctuations of the data.
% the centring of the data

% The correlated gate errors may be
% It may caused by two main issue. 
% There are two major 
% One is gate correlated errors that make our model nonlinear. 
% The other is not sufficient fit data, adding which will make our experiment resource increasing.

Then, we show the ground state searching with the variational algorithms for the three molecules.
% Again, we select three  bond distances for different molecules, including  $\mol{H_2}$ (R=2.6), LiH (R=2.2) and $\mol{F_2}$ (R=2.6), to show the optimisation processes.
We show the optimisation processes for  $\mol{H_2}$ (R=2.6), LiH (R=2.2) and $\mol{F_2}$ (R=2.6) in \autoref{fig:supp_iteration}. The figure presents error-mitigated energy estimations with different error mitigation schemes, including REM, PS, CF, SV, CMX, and their combinations. \autoref{fig:supp_iteration} shows that although measured energies (yellow square) have fluctuations, energies after applying the combination of REM and CF clearly improve the energy accuracy along the optimisation.  Error-mitigated energies tend to decrease consistently in successive iterations.

% In our experiments, we also compare the results with post-selection and symmetry verification.

Next, we show the calculation of potential energy curves in \autoref{fig:supp_bond}. 
\autoref{fig:supp_bond}a, b, c show the energy iteration during the VQE optimisation process for $\mol{H_2}$ (R=2.6), LiH (R=1.5) and $\mol{F_2}$ (R=2.0), respectively.  These results provide a much more detailed comparison of the error mitigation schemes used in the experiment.
In the following, we discuss the performance of these error mitigation schemes.

We first discuss error mitigation with symmetry verification.
As can be found in \autoref{fig:supp_iteration} ad \autoref{fig:supp_bond}, SV results in a large fluctuation in the energy estimation.
In particular, as illustrated in \autoref{fig:supp_bond}(b), the energy with REM, CF and SV (grey right triangles) have a greater fluctuation than REM + CF during the optimisation procedure. 
The energy with REM, CF and SV show a larger error than REM and CF for all bond distances, as can be found in \autoref{fig:symmetryVerification}(a).
This can be understood as follows. The fitting coefficient for the symmetry verified Hamiltonian $\hat{H}\hat S$ is much larger than $\hat{H} $ and thus results in large fluctuation after applying the combination of CF and SV.
This is verified by the slope in \autoref{fig:symmetryVerification}b and the fitting performance in \autoref{fig:symmetryVerification}d.

These results indicate the combination of SV and CF could bring a large fluctuation. Therefore,
we mainly focused on the readout error mitigation, Clifford fitting and ground state energy correction by using connected moment expansion in the main text, while we did not show the results energy of LiH and $\mol{F_2}$ using symmetry verification.

In addition, we add the energy with REM and post-selection (PS). In this scheme, the measurement result of $Z^{\otimes n}$ Pauli basis is post-selected based on the conservation of particle numbers, and other Puali bases are still error-mitigated with REM.
The REM and post-selection (PS) energy perform better than single REM yet worse than REM and CF. With an increasing system size, a combination of REM and PS shows diminishing energy accuracy improvements, implying PS is not a suitable error mitigation strategy for scalable quantum computational chemistry.

\begin{figure}[htb]
\centering
\includegraphics[width=1\textwidth]{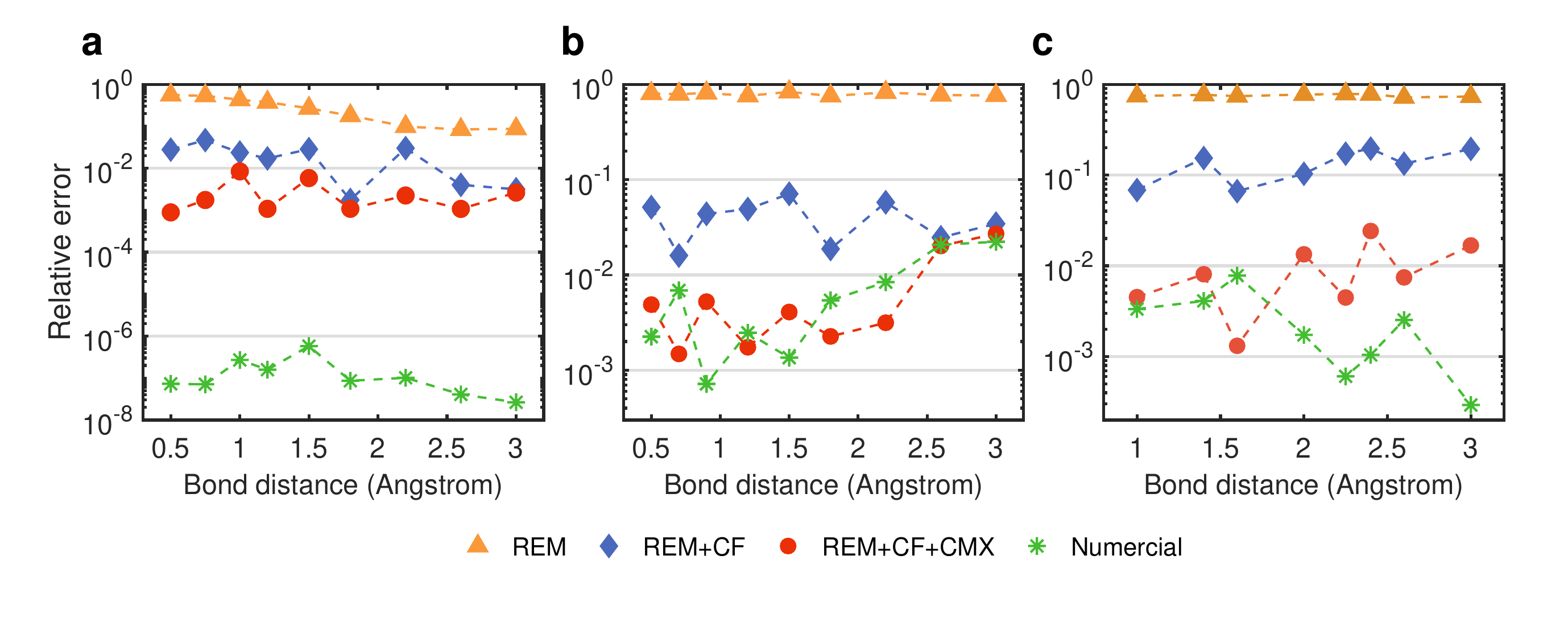}
\caption{
\textbf{Relative error compared to the initial state energy.} Relative errors for $\mol{H_2}$, LiH and $\mol{F_2}$ with different error mitigation schemes, only REM (yellow triangles), REM and Clifford fitting schemes (blue diamonds), REM and Clifford fitting and CMX (red circles), the numerical result (green stars).
}
\label{fig:relative_error}
\end{figure}

Finally, we show the performance of VQE after the optimisation process.
\autoref{fig:relative_error} shows the relative error compared to the initial energy for the three molecules at different bond distances. 
The relative error compared to the initial state energy is calculated by 
\begin{equation}
    \frac{E_i - E_{\rm ground}}{ E_{0}^{\mathrm{expm}}  - E_{\rm ground}},
    \label{eq:rel_error}
\end{equation}
where $E_i$ is the error-mitigated energy or the energy, which is computed using the parameters that are found until convergence, $E_{0}^{\mathrm{expm}}$ is the energy measured on the initial state,  $E_{\rm ground}$ is the ideal ground state energy. 
This relative error characterises the improvement with respect to the initial state energy by VQE.

We can see from \autoref{fig:relative_error} that compared to the initial state energy, the relative average energy errors with only REM for the three molecules are $28.98\%$, $78.86\%$ and $75.42\%$.

The results are improved by $0.278\%$, $0.778\%$ ($0.326\%$ except R2.6 and R3.0), and $1.00\%$ for molecules with REM+CF+CMX EM schemes, and by $1.56 \times 10^{-7}$, $0.783\%$ ($0.392\%$ except R2.6 and R3.0), and $0.268\%$ for numerical result.
The average absolute error associated with different methods, including REM, REM + PS, REM+CF+SV, REM+CF, REM+CF+CMX, and numerical, are displayed in \autoref{table:avg_error}.

\begin{table}[h!]
\centering
\begin{tabular}{c | c | c | c | c | c | c | c} 
 \hline
\hline
Molecules        & Raw  & REM    &  REM + PS     & REM+CF+SV      & REM+CF    & REM+CF+CMX        & Numerical \\
\hline
$\mol{H_2} $ (milli-Hartree)      & $143.661 \pm 0.807$ &  $60.580 \pm 1.015$ & $38.650 \pm 2.367$ & $4.552 \pm 0.740$ & $4.254 \pm 0.997$ & $0.565 \pm 0.542$ &  $2.96\times 10^{-5}$     \\
$\mol{LiH}$  (milli-Hartree)    & $233.564 \pm 2.704$ & $191.358 \pm 3.313$ & $169.987 \pm 3.110$ & $43.888 \pm 2.383$ & $9.965 \pm 2.923$ & $1.802 \pm 2.423$ & $1.794$   \\
$\mol{F_2 }$ (Hartree)    & $1.8315 \pm 0.0264$ & $1.3419 \pm 0.0387$  & $1.2231 \pm 0.0382$ & -- & $0.2378 \pm 0.0354$ & $0.0174 \pm 0.0417$ & $0.0048$     \\
\hline
\hline
\end{tabular}
\caption{Comparison of average absolute energy error for $\mol{H_2}$, LiH and $\mol{F_2}$ using different error mitigation schemes. The first two rows use milli-Hartree.   }
\label{table:avg_error}
\end{table}

\subsection{Parameters iteration}
In \autoref{fig:params_iteration}, we present the parameters iteration for $\mol{H_2}$ (R=2.6), LiH (R=1.5) and $\mol{F_2}$ (R=2.0). For $\mol{H_2}$, LiH and $\mol{F_2}$, there are  1, 3 and 5 parameters in total to iterate respectively. All parameters are optimised from zero. 1, 2 and 3 parameters are randomly selected to optimise during each iteration for $\mol{H_2}$, LiH and $\mol{F_2}$, respectively. All optimisation runs for 5 to 15 iterations.

\subsection{Noise analysis}

In this section, we analyse the noise effect in our chemistry simulation.
We assume noise in the circuit can be described by a depolarising channel
\begin{equation}
    \mathcal{E} (\rho) = p \rho + (1-p) \frac{I}{2^N}, 
\end{equation}
where $1-p$ characterises the error rate, and $p$ characterises the fidelity of the quantum state.
Under depolarising noise, the energy  is given by
\begin{equation}
    E^{\mathrm{noisy}} = \tr (\rho^{\mathrm{ideal}} \hat{H}) = p \tr (\rho^{\mathrm{ideal}}) = p E^{\mathrm{ideal}}.
\end{equation}
Here, the identity operator in the Hamiltonian has been excluded so that $\tr(\hat{H}) = 0$.
We can get the error rate $p$ by fitting the measured energies $E^{\mathrm{noisy}} $ and the corresponding ideal energies $E^{\mathrm{ideal}} $.

\begin{figure}[htb]
\centering
\includegraphics[width=1\textwidth]{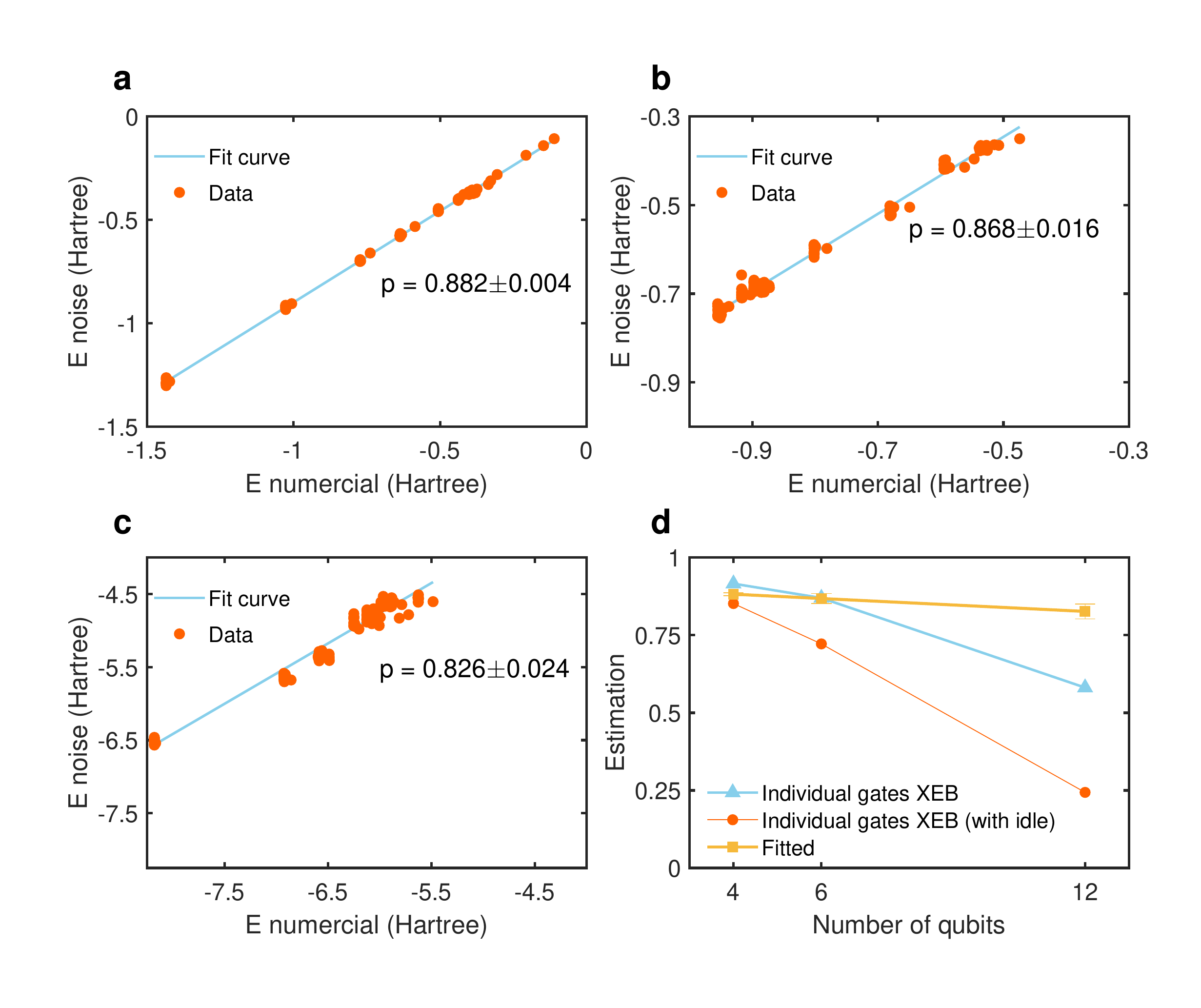}
\caption{
\textbf{Fitted results using a depolarising noise model.} 
% The data contain all the iteration result in VQE procedure for all the molecule bond distance, the curve use a linear fit.
The figures show the ideal energies and measured energies along the optimisation iterations for $\mol{H_2}$ (a), LiH (b) and $\mol{F_2}$ (c) at all the bond distances. \blue{The noisy energies represent the measurement energies only with Readout Error Mitigation (REM).} 
The ideal energies and measured energies have a linear dependence. The slope of the fitting curves characterises the depolarising fidelity $p$.  (d) The circuit fidelity estimation by individual XEB results with and without idle time decay, and the fitted depolarising fidelity obtained by fitting the data points in (a)(b)(c).
}
\label{fig:error_pk}
\end{figure}

We show the ideal energies and measured \blue{noisy} energies in \autoref{fig:error_pk}
along the optimisation iterations for $\mol{H_2}$ (a), LiH (b) and $\mol{F_2}$ (c) at all the bond distances.
\blue{The measured noisy energies represent the measurement energies only with Readout Error Mitigation (REM).}
We observe that the ideal energies and measured energies have a well-linear dependence. 
The slope of the fitting curves gives us the depolarising fidelity $p$, which decreases with a deeper circuit.
% To further reveal the r
 
To further our understanding of noise with an increasing circuit size, we compare the circuit fidelity estimated using different methods. 
As a reference, we calculate the circuit fidelity by multiplying the fidelity of each individual gate, which is obtained by individual XEB.
Specifically, we first calculate the fidelity of each layer using the Pauli fidelity, and calculate the whole circuit fidelity using the depolarising fidelity of different layers.
Since dynamical decoupling is used in our experiment to avoid dephasing during the idle time, we also take account of the idle time decay.
The idle time fidelity is calculated by the fidelity of a single-qubit gate with an equal time length.

\autoref{fig:error_pk}d shows the comparison of the estimation using individual XEB results with (red) and without idle time decay (blue), and the fitted depolarising fidelity by fitting the ideal energies and measured energies (orange). Remarkably, we can find that the actual circuit error rate is much lower than the error estimated with individual XEB results when the system size increases. This result implies the potential for the simulation of larger molecules. A detailed investigation will be an interesting future work.

%\red{add why actual error lower than XEB results 1. over-rotation and under-rotation VQE can 2. XEB fidelity is state fidelity, not energy fidelity}

% In what follows, we briefly discuss the simulation accuracy in relation to the error rate.
% The absolute value of the energy $E_{\ideal}$ of $\mol{F}_2$ is large. Consequently, compared to small molecules, it requires $p$ being much closer to 1 to obtain a high-accuracy result.
% However, this is not an issue if we consider the relative error.  

% The fitted fidelity is higher than estimya

% 
%\subsection{Discussion}

%We note that even if the energy converges,  the parameters does not necessarily converge.
%The fluctuation may appear if certain parameters are redundant.
%\autoref{fig:params_iteration} shows the evolution of variational parameters. This indicates that the parameters also %converge after iteration, verifying the convergence of the VQE process.

%In \autoref{fig:params_iteration}, we present the parameters iteration for $\mol{H_2}$ (R=2.6), LiH (R=1.5) and $\mol{F_2}$ (R=2.0). For $\mol{H_2}$, LiH and $\mol{F_2}$, there are  1, 3 and 5 parameters in total to iterate respectively. All parameters are optimised from zero. 1, 2 and 3 parameters are randomly selected to optimise during each iteration for $\mol{H_2}$, LiH and $\mol{F_2}$, respectively. 

%All optimisation runs for 5 to 15 iterations.

\subsection{Challenges}

As we have discussed here and in the main text, there are many challenges in VQE experiments. 
Compared to dynamics simulation or random circuit sampling, VQE experiments are more challenging. The former examples only require running  one instance on a quantum processor. In contrast, in the VQE experiment, the final results are highly related to the results that are obtained in the prior incidents.
In spite of these challenges, our experiments for the first time achieved chemical accuracy for the six-qubit VQE simulation of molecular systems. Good control of hardware and efficient quantum algorithms enable such high-precision simulation.

For future large-scale quantum computational chemistry experiments, it is crucial to suppress gates and measurement readout errors in order to achieve chimerical accuracy with quantum hardware.
% However, we also wish to remark that challenges do exist in VQE experiments.
We point out these challenges as follows.

\textbf{Stability during experiments.} For large-scale problems, VQE experiments could take several hours or days to obtain the ground state. Therefore, the performance of VQE highly depends on the stability of the experimental system. In our experiment, the readout error mitigation and Clifford fitting schemes are applied. These techniques require maintaining the system feature during the REM matrix extraction experiment, Clifford fitting experiment and VQE experiment. Only when the system float is suppressed can we make the error mitigation schemes work. The fluctuation mainly comes from the following:
\begin{enumerate}
    \item \textbf{Electronic outcome fluctuation.} The electronic outcome determines the \blue{system conditions and manipulation accuracy}, such as the qubit frequencies, coupler flux bias and the control pulse amplitude. The fluctuation of electronics directly affects gate fidelities as well as the results of the quantum circuit. The outcome accuracy of the electronic itself could be suppressed by increasing the samples, but the average outcome amplitude is greatly influenced by ambient temperature.
    \item \textbf{Unstable Two Level System (TLSs).} When the unstable TLSs near close to qubit operating frequency, the coherence of qubits would decrease abruptly. The noisy model changed which make the Clifford fitting fail.
\end{enumerate}

\textbf{Errors and samples.} 
In our experiment, as we pointed out, we require the expectation values of observables for the optimisation and energy evaluation, instead of single-shot measurements. To suppress the statistical errors less than $\epsilon$, the number of samples scales as $1/ \epsilon^2$. 
With the increase of qubit number, 
coherent errors become a main error source, which leads to a large slope for the Clifford fitting function $f^{\mathrm{CF}}$, as shown in \autoref{fig:cf_distribution}, and hence a large variance. 
As a result of fitting, more samples are required for a larger system to suppress the fluctuation.
Nevertheless,   this can be overcome by further improving hardware performance. This is because a smaller error rate will lower the inherent expectation value variance, and hence fewer samples are required. 
% \blue{There is a trade-off between unmitigated-mitigated bias and variance. The large slope in Clifford fitting could mitigate the expectation with a larger bias but with a large variance.} 
In addition, the number of observable increases with the system size, and thus more samples are needed in a larger VQE experiment. This requires a long time to run experiments, and it could be hard to stabilise the system for a long time.

% \sun{SHAOJUn: Unstable for searching. Fluctuation. Disadvantage. Which methods have the worst Fluctuation.}

% We have summarised two challenges in VQE experiment, for the future large scale quantum computational chemistry experiment, the key work is still to suppress the errors in gates and readout. The smaller error rate would lower the inherent expectation value variance which admit less sample number. 

%Finally, we give some outlook for VQE experiment.
%\sun{If the gate error and readout error can be decreased to XXX and XXX,  respectively, we can achieve chemical accuracy for $\mol{F_2}$.}

%As mentioned in main scripts Sec.~III, we summarised the error with readout error, gate error and residual error. The readout error could be partially mitigated by REM, and the gate error also could be partially mitigated by Clifford fitting. The residual error refers to the parts that cannot be mitigated by these two schemes, and it could be partially mitigated by CMX.

% \bibliography{ref_chem}

\end{document}